\newif\iffigs\figstrue
\documentclass[12pt]{article}
\iffigs
   \input{epsf}
\else
\fi
\newcommand{\eqn}[1]{(\ref{#1})}

\newsavebox{\uuunit}
\sbox{\uuunit}
    {\setlength{\unitlength}{0.825em}
     \begin{picture}(0.6,0.7)
        \thinlines
        \put(0,0){\line(1,0){0.5}}
        \put(0.15,0){\line(0,1){0.7}}
        \put(0.35,0){\line(0,1){0.8}}
       \multiput(0.3,0.8)(-0.04,-0.02){12}{\rule{0.5pt}{0.5pt}}
     \end {picture}}

\def\IP{\relax{\rm I\kern-.18em P}}

\def\Es{\bf E_{7(7)}}
\begin{document}
%
\font\cmss=cmss10 \font\cmsss=cmss10 at 7pt
\def\twomat#1#2#3#4{\left(\matrix{#1 & #2 \cr #3 & #4}\right)}
\def\inbar{\vrule height1.5ex width.4pt depth0pt}
\def\IC{\relax\,\hbox{$\inbar\kern-.3em{\rm C}$}}
\def\IG{\relax\,\hbox{$\inbar\kern-.3em{\rm G}$}}
\def\IB{\relax{\rm I\kern-.18em B}}
\def\ID{\relax{\rm I\kern-.18em D}}
\def\IL{\relax{\rm I\kern-.18em L}}
\def\IF{\relax{\rm I\kern-.18em F}}
\def\IH{\relax{\rm I\kern-.18em H}}
\def\II{\relax{\rm I\kern-.17em I}}
\def\IN{\relax{\rm I\kern-.18em N}}
\def\IP{\relax{\rm I\kern-.18em P}}
\def\IQ{\relax\,\hbox{$\inbar\kern-.3em{\rm Q}$}}
\def\bfzero{\relax\,\hbox{$\inbar\kern-.3em{\rm 0}$}}
\def\IK{\relax{\rm I\kern-.18em K}}
\def\IG{\relax\,\hbox{$\inbar\kern-.3em{\rm G}$}}
 \font\cmss=cmss10 \font\cmsss=cmss10 at 7pt
\def\IR{\relax{\rm I\kern-.18em R}}
\def\ZZ{\relax\ifmmode\mathchoice
{\hbox{\cmss Z\kern-.4em Z}}{\hbox{\cmss Z\kern-.4em Z}}
{\lower.9pt\hbox{\cmsss Z\kern-.4em Z}}
{\lower1.2pt\hbox{\cmsss Z\kern-.4em Z}}\else{\cmss Z\kern-.4em
Z}\fi}
\def\bfone{\relax{\rm 1\kern-.35em 1}}
\def\dop{{\rm d}\hskip -1pt}
\def\real{{\rm Re}\hskip 1pt}
\def\trace{{\rm Tr}\hskip 1pt}
\def\ii{{\rm i}}
\def\diag{{\rm diag}}
\def\sch#1#2{\{#1;#2\}}
\def\bfone{\relax{\rm 1\kern-.35em 1}}
\font\cmss=cmss10 \font\cmsss=cmss10 at 7pt
\def\a{\alpha} \def\b{\beta} \def\d{\delta}
\def\e{\epsilon} \def\c{\gamma}
\def\G{\Gamma} \def\l{\lambda}
\def\L{\Lambda} \def\s{\sigma}
\def\cA{{\cal A}} \def\cB{{\cal B}}
\def\cC{{\cal C}} \def\cD{{\cal D}}
\def\cF{{\cal F}} \def\cG{{\cal G}}
\def\cH{{\cal H}} \def\cI{{\cal I}}
\def\cJ{{\cal J}} \def\cK{{\cal K}}
\def\cL{{\cal L}} \def\cM{{\cal M}}
\def\cN{{\cal N}} \def\cO{{\cal O}}
\def\cP{{\cal P}} \def\cQ{{\cal Q}}
\def\cR{{\cal R}} \def\cV{{\cal V}}\def\cW{{\cal W}}
\newcommand{\be}{\begin{equation}}
\newcommand{\ee}{\end{equation}}
\newcommand{\bea}{\begin{eqnarray}}
\newcommand{\eea}{\end{eqnarray}}
\let\la=\label \let\ci=\cite \let\re=\ref
%
%
%
\def\crr{\crcr\noalign{\vskip {8.3333pt}}}
\def\tilde{\widetilde}
\def\bar{\overline}
\def\us#1{\underline{#1}}
\let\shat=\hat
\def\hat{\widehat}
\def\hyp{\vrule height 2.3pt width 2.5pt depth -1.5pt}
\def\square{\mbox{.08}{.08}}
\def\Coeff#1#2{{#1\over #2}}
\def\Coe#1.#2.{{#1\over #2}}
\def\coeff#1#2{\relax{\textstyle {#1 \over #2}}\displaystyle}
\def\coe#1.#2.{\relax{\textstyle {#1 \over #2}}\displaystyle}
\def\half{{1 \over 2}}
\def\shalf{\relax{\textstyle {1 \over 2}}\displaystyle}
\def\dag#1{#1\!\!\!/\,\,\,}
\def\to{\rightarrow}
\def\notin{\hbox{{$\in$}\kern-.51em\hbox{/}}}
\def\shdot{\!\cdot\!}
\def\ket#1{\,\big|\,#1\,\big>\,}
\def\bra#1{\,\big<\,#1\,\big|\,}
\def\equaltop#1{\mathrel{\mathop=^{#1}}}
\def\Trbel#1{\mathop{{\rm Tr}}_{#1}}
\def\inserteq#1{\noalign{\vskip-.2truecm\hbox{#1\hfil}
\vskip-.2cm}}
\def\attac#1{\Bigl\vert
{\phantom{X}\atop{{\rm\scriptstyle #1}}\phantom{X}}}
\def\exx#1{e^{{\displaystyle #1}}}
\def\del{\partial}
\def\delbar{\bar\partial}
\def\nex#1{$N\!=\!#1$}
\def\dex#1{$d\!=\!#1$}
\def\cex#1{$c\!=\!#1$}
\def\eg{{\it e.g.}} \def\ie{{\it i.e.}}
\def\IE{\relax{{\rm I\kern-.18em E}}}
\def\cE{{\cal E}}
\def\rt{{\cR^{(3)}}}
\def\IGam{\relax{{\rm I}\kern-.18em \Gamma}}
\def\IGa{\IA}
\def\ii{{\rm i}}
\begin{titlepage}
\begin{center}
{\LARGE $E_{7(7)}$ Duality, BPS Black--Hole Evolution and Fixed Scalars
 $^*$ }\\
\vfill
{\large Laura Andrianopoli$^1$,
Riccardo D'Auria $^2$, Sergio Ferrara$^3$,
\\
\vskip 0.2cm
Pietro Fr\'e$^4$
 and  Mario Trigiante$^5$   } \\
\vfill
{\small
$^1$ Dipartimento di Fisica Universit\'a di Genova, via Dodecaneso 33,
I-16146 Genova\\
and Istituto Nazionale di Fisica Nucleare (INFN) - Sezione di Torino, Italy\\
\vspace{6pt}
$^2$ Dipartimento di Fisica Politecnico di Torino, C.so Duca degli Abruzzi,
24,
I-10129 Torino\\
and Istituto Nazionale di Fisica Nucleare (INFN) - Sezione di Torino, Italy\\
\vspace{6pt}
$^3$ CERN, Theoretical Division, CH 1211 Geneva, 23,
Switzerland\\
\vspace{6pt}
$^4$ Dipartimento di Fisica Teorica, Universit\'a di Torino, via P. Giuria 1,
I-10125 Torino, \\
 Istituto Nazionale di Fisica Nucleare (INFN) - Sezione di Torino, Italy \\
\vspace{6pt}
$^5$ International School for Advanced Studies (ISAS), via Beirut 2-4,
I-34100 Trieste\\
and Istituto Nazionale di Fisica Nucleare (INFN) - Sezione di Trieste, Italy\\
\vspace{6pt}
}
\end{center}
\vfill
\begin{center}
{\bf Abstract}
\end{center}
{\small We study the general equations determining BPS Black Holes
by using  a Solvable Lie Algebra representation for the homogenous
scalar manifold  U/H  of extended supergravity. In particular we focus  on the N=8
case and we perform a general group theoretical analysis of the
Killing spinor equation enforcing the BPS condition.
Its solutions parametrize the U--duality orbits of BPS solutions that are characterized by
having $40$ of the $70$ scalars fixed to constant values. These
scalars belong to hypermultiplets in the $N=2$ decomposition of the
$N=8$ theory. Indeed it is shown that those decompositions of the Solvable Lie algebra
into appropriate subalgebras which are enforced by the existence of BPS black holes are the
same that single   out consistent  truncations of the N=8 theory to intereacting
theories with lower supersymmetry. As an exemplification of the
method we  consider the simplified case where the only non-zero
fields are in the Cartan subalgebra ${\cal H} \subset Solv(U/H)$ and
correspond to the radii of string toroidal compactification. Here we
derive and solve the mixed system of first and second order
non linear differential equations obeyed by the metric and by the scalar fields. So doing we
retrieve the generating solutions of heterotic black holes with two charges.
Finally, we show that the general $N=8$ generating solution
is based on the 6 dimensional solvable subalgebra
$Solv \left[ \left(SL(2,\IR) /U(1) \right)^3 \right ]$.

}
\vspace{2mm} \vfill \hrule width 3.cm
{\footnotesize
 $^*$ Supported in part by   EEC  under TMR contract
 ERBFMRX-CT96-0045, in which L. Andrianopoli are associated to Torino University
 and S. Ferrara and M. Trigiante are associated to Frascati and by DOE grant DE-FG03-91ER40662.}
\end{titlepage}
\section{Introduction}
\label{introgen}
Interest in the extremal black hole solutions of $D=4$ supergravity
theories has been quite vivid in the last couple of years
\cite{malda,gensugrabh} and it is just part of a more general interest in the
$p$--brane classical solutions of  supergravity theories in all dimensions
$4 \le D \le 11$ \cite{kstellec,duffrep}.
This interest streams from the interpretation of the classical solutions of
supergravity that preserve a fraction of the original supersymmetries
as the BPS non perturbative states necessary to complete the
perturbative string spectrum and make it invariant under the many
conjectured duality symmetries \cite{schw,sesch,huto2,huto,vasch}.
This identification has become quite circumstantial with the advent of
$D$--branes \cite{DbranPolch1} and the possibility raised by them of
a direct construction of the BPS states within the language of
perturbative string theory extended by the choice of Dirichlet
boundary conditions \cite{Polchtasi}.
\par
Basic feature of the non--perturbative states of the string spectrum
is that they can carry Ramond--Ramond charges forbidden at the perturbative level.
On the other hand, the important observation by  Hull
and Townsend \cite{huto,huto2} is that at the level of the low
energy supergravity lagrangians all fields of both the
Neveu-Schwarz,Neveu-Schwarz (NS-NS) and the Ramond-Ramond (R-R) sector are unified
by the group of duality transformations $U$ which is also the
isometry group of the homogenous scalar manifold ${\cal
M}_{scalar}=U/H$. At least this is true in theories with sufficiently
large number of supersymmetries that is with $N\ge 3$ in $D=4$ or,
in a dimensional reduction invariant language with
$ \# \mbox{supercharges} \, \ge \, 12 $.
This points out that the distinction between R-R and
NS-NS sectors is just an artifact of perturbative string theory. It
also points out to the fact that the unifying symmetry between the
perturbative and non-perturbative sectors is already known from
supergravity, namely it is the $U$-duality group. Indeed the basic
conjecture of Hull and Townsend is that the restriction to integers
$U(\ZZ)$ of the $U$ Lie group determined by supergravity should be an
exact symmetry of non-perturbative string theory.
\par
These observations raise the question of the group-theoretical basis
for the separation of the string spectrum, in particular the light modes
yielding the field content of supergravity, into a NS-NS and a R-R sector.
This question was answered by the present authors
\cite{noialtri,noialtri2} using the solvable Lie algebra representation
of the scalar field sector.
\par
In the present paper we want to consider the application of
solvable Lie algebras to the derivation of the differential
equations that characterize BPS states as classical supergravity solutions.
\par
From an abstract viewpoint BPS saturated states are characterized by the
fact that they preserve, in modern parlance, $1/2$ (or $1/4$, or $1/8$) of the original
supersymmetries. What this actually means is that there is a suitable
projection operator $\IP^2_{BPS} =\IP_{BPS}$ acting on the supersymmetry charge
$Q_{SUSY}$, such that:
\begin{equation}
 \left(\IP_{BPS} \,Q_{SUSY} \right) \, \vert \, \mbox{BPS state} \, >
 \,=  \, 0
 \label{bstato}
\end{equation}
Since the supersymmetry transformation rules of any supersymmetric
field theory are linear in the first derivatives of the fields
eq.\eqn{bstato} is actually a {\it system of first order differential
equations}. This system has to be combined with the second order
field equations of supergravity and the common solutions to both
system of equations is a classical BPS saturated state. That it is
actually an exact state of non--perturbative string theory follows
from supersymmetry representation theory. The classical BPS state is
by definition an element of a {\it short supermultiplet}
and, if supersymmetry is unbroken, it cannot be renormalized to
a {\it long supermultiplet}.
\par
Translating eq. \eqn{bstato} into an explicit first order
differential system requires knowledge of the supersymmetry
transformation rules of supergravity and it is at this level that
solvable Lie algebras can play an important role. In order to grasp the
significance of the above statement let us first rapidly review, as
an example, the algebraic definition of $D=4$, $N=2\nu$ BPS states and then the idea
of the solvable Lie algebra representation of the scalar sector.

The $D=4$ supersymmetry algebra with an even number
$N=2\nu$ of supersymmetry charges  can be written in the following
form:
\begin{eqnarray}
&\left\{ {\bar Q}_{aI \vert \alpha }\, , \,{\bar Q}_{bJ \vert \beta}
\right\}\, = \,  {\rm i} \left( C \, \gamma^\mu \right)_{\alpha \beta} \,
P_\mu \, \delta_{ab} \, \delta_{IJ} \, - \, C_{\alpha \beta} \,
\epsilon_{ab} \, \times \, \ZZ_{IJ}& \nonumber\\
&\left( a,b = 1,2 \qquad ; \qquad I,J=1,\dots, \nu \right)&
\label{susyeven}
\end{eqnarray}
where the SUSY charges ${\bar Q}_{aI}\equiv Q_{aI}^\dagger \gamma_0=
Q^T_{ai} \, C$ are Majorana spinors, $C$ is the charge conjugation
matrix, $P_\mu$ is the 4--momentum operator, $\epsilon_{ab}$ is the
two--dimensional Levi Civita symbol and the symmetric tensor
$\ZZ_{IJ}=\ZZ_{JI}$ is the central charge operator. It
can always be diagonalized $\ZZ_{IJ}=\delta_{IJ} \, Z_J$ and its $\nu$ eigenvalues
$Z_J$ are the central charges.
\par
The Bogomolny bound on the mass of a generalized monopole state:
\begin{equation}
M \, \ge \, \vert \, Z_I \vert \qquad \forall Z_I \, , \,
I=1,\dots,\nu
\label{bogobound}
\end{equation}
is an elementary consequence of the supersymmetry algebra and of the
identification between {\it central charges} and {\it topological
charges}. To see this it is convenient to introduce the following
reduced supercharges:
\begin{equation}
{\bar S}^{\pm}_{aI \vert \alpha }=\frac{1}{2} \,
\left( {\bar Q}_{aI} \gamma _0 \pm \mbox{i} \, \epsilon_{ab} \,  {\bar Q}_{bI}\,
\right)_\alpha
\label{redchar}
\end{equation}
They can be regarded as the result of applying
a projection operator to the supersymmetry
charges:
\begin{eqnarray}
{\bar S}^{\pm}_{aI} &=& {\bar Q}_{bI} \, \IP^\pm_{ba} \nonumber\\
 \IP^\pm_{ba}&=&\frac{1}{2}\, \left({\bf 1}\delta_{ba} \pm \mbox{i} \epsilon_{ba}
 \gamma_0 \right)
 \label{projop}
\end{eqnarray}
Combining eq.\eqn{susyeven} with the definition \eqn{redchar} and
choosing the rest frame where the four momentum is $P_\mu$ =$(M,0,0,0)$, we
obtain the algebra:
\begin{equation}
\left\{ {\bar S}^{\pm}_{aI}  \, , \, {\bar S}^{\pm}_{bJ} \right\} =
\pm \epsilon_{ac}\, C \, \IP^\pm_{cb} \, \left( M \mp Z_I \right)\,
\delta_{IJ}
\label{salgeb}
\end{equation}
By positivity of the operator $\left\{ {S}^{\pm}_{aI}  \, , \, {\bar S}^{\pm}_{bJ} \right\} $
it follows that on a generic state the Bogomolny bound \eqn{bogobound} is
fulfilled. Furthermore it also follows that the states which saturate
the bounds:
\begin{equation}
\left( M\pm Z_I \right) \, \vert \mbox{BPS state,} i\rangle = 0
\label{bpstate1}
\end{equation}
are those which are annihilated by the corresponding reduced supercharges:
\begin{equation}
{\bar S}^{\pm}_{aI}   \, \vert \mbox{BPS state,} i\rangle = 0
\label{susinvbps}
\end{equation}
On one hand eq.\eqn{susinvbps} defines {\sl short multiplet
representations} of the original algebra \eqn{susyeven} in the
following sense: one constructs a linear representation of \eqn{susyeven}
where all states are identically
annihilated by the operators ${\bar S}^{\pm}_{aI}$ for $I=1,\dots,n_{max}$.
If $n_{max}=1$ we have the minimum shortening, if $n_{max}=\nu$ we
have the maximum shortening. On the other hand eq.\eqn{susinvbps}
can be translated into a first order differential equation on the
bosonic fields of supergravity.

Indeed, let us consider
a configuration where all the fermionic fields are zero.
Setting the fermionic SUSY rules appropriate to such a background equal
to zero we find the following Killing spinor equation:
\begin{equation}
0=\delta \mbox{fermions} = \mbox{SUSY rule} \left( \mbox{bosons},\epsilon_{AI} \right)
\label{fermboserule}
\end{equation}
where the SUSY parameter satisfies the following conditions:
\begin{equation}
\begin{array}{rclcl}
\xi^\mu \, \gamma_\mu \,\epsilon_{aI} &=& \mbox{\rm i}\, \varepsilon_{ab}
\,  \epsilon^{bI}   & ; &   I=1,\dots,n_{max}\\
\epsilon_{aI} &=& 0  &;&   I > n_{max} \\
\end{array}
\end{equation}
Here $\xi^\mu$ is a time-like Killing vector for the space-time metric and
$ \epsilon _{aI}, \epsilon^{aI}$ denote the two chiral projections of
a single Majorana spinor:
\begin{equation}
\gamma _5 \, \epsilon _{aI} \, = \, \epsilon _{aI} \quad ; \quad
\gamma _5 \, \epsilon ^{aI} \, = - \epsilon ^{aI}
\label{chiralpro}
\end{equation}
Eq.\eqn{fermboserule} has two features which we want to stress
as main motivations for the developments presented in later sections:
\begin{enumerate}
\item{It requires an efficient parametrization of the scalar field
sector}
\item{It breaks the original $SU(2\nu)$ automorphism of the
supersymmetry algebra to the subgroup $SU(2)\times SU(2\nu-2)\times U(1)$}
\end{enumerate}
The first feature is the reason why the use of the solvable Lie
algebra $Solv$ associated with $U/SU(2\nu)\times H^\prime$ is of great help in this problem.
The second feature is the
reason why the solvable Lie algebra $Solv$ has to be decomposed in
a way appropriate to the decomposition of the isotropy group $H=SU(2\nu)\times H^\prime$
with respect to the subgroup $SU(2)\times SU(2\nu-2)\times U(1) \times H^\prime$.
\par
To explain what is involved in the above statements, a
quick review of the solvable Lie algebra representation will be given in section $5$ (see also ref.
\cite{noialtri,noialtri2}).
\subsection{N=2 decomposition in the $N=8$ theory}
Although our goal is that of developing general methods for the study
of BPS $p$--branes in all higher dimensional supergravities, in the
present paper we concentrate on the maximally extended four--dimensional
theory, namely $N=8$ supergravity.
Hence the relevant $U$--duality group is $E_{7(7)}$ and the
relevant solvable Lie algebra is that associated with the homogeneous
manifold $E_{7(7)}/SU(8)$. Since it is maximally non compact the rank
of $Solv \left( E_{7(7)}/SU(8) \right)$ is seven. Henceforth we introduce the
notation:
\begin{equation}
Solv_7 \, \equiv \,  Solv \left( E_{7(7)}/SU(8) \right)
\label{sol7defi}
\end{equation}
According to the
previous discussion the Killing spinor equation for $N=8$ BPS states
requires that $Solv_7$ should be decomposed  according to the decomposition
of the isotropy subgroup: $SU(8) \longrightarrow SU(2)\times U(6)$. We
show in later sections that the corresponding decomposition of the solvable
Lie algebra is the following one:
\begin{equation}
Solv_7   =  Solv_3 \, \oplus \, Solv_4
\label{7in3p4}
\end{equation}
\begin{equation}
\begin{array}{rclrcl}
Solv_3 & \equiv & Solv \left( SO^\star(12)/U(6) \right) & Solv_4 &
\equiv & Solv \left( E_{6(4)}/SU(2)\times SU(6) \right) \\
\mbox{rank }\, Solv_3 & = & 3 & \mbox{rank }\,Solv_4 &
= & 4 \\
\mbox{dim }\, Solv_3 & = & 30 & \mbox{dim }\,Solv_4 &
= & 40 \\
\end{array}
\label{3and4defi}
\end{equation}
The rank three  Lie algebra $Solv_3$ defined above describes the
thirty dimensional scalar sector of $N=6$ supergravity, while the rank four
solvable Lie algebra $Solv_4$ contains the remaining fourty scalars
belonging to $N=6$ spin $3/2$ multiplets. It should be noted
that, individually, both manifolds $ \exp \left[ Solv_3 \right]$ and
$ \exp \left[ Solv_4 \right]$ have also an $N=2$ interpretation since we have:
\begin{eqnarray}
\exp \left[ Solv_3 \right] & =& \mbox{homogeneous special K\"ahler}
\nonumber \\
\exp \left[ Solv_4 \right] & =& \mbox{homogeneous quaternionic}
\label{pincpal}
\end{eqnarray}
so that the first manifold can describe the interaction of
$15$ vector multiplets, while the second can describe the interaction
of $10$ hypermultiplets. Indeed if we decompose the $N=8$ graviton
multiplet in $N=2$ representations we find:
\begin{equation}
\mbox{N=8} \, \mbox{\bf spin 2}  \,\stackrel{N=2}{\longrightarrow}\,
 \mbox{\bf spin 2} + 6 \times \mbox{\bf spin 3/2} + 15 \times \mbox{\bf vect. mult.}
 +
 10 \times \mbox{\bf hypermult.}
 \label{n8n2decompo}
\end{equation}
Although at the level of linearized representations of supersymmetry
we can just delete the $6$ spin $3/2$ multiplets and obtain a
perfectly viable $N=2$ field content, at the full interaction level
this truncation is not consistent. Indeed, in order to get a consistent
$N=2$ truncation the complete scalar manifold must be the {\sl direct product}
of {\sl a special K\"ahler} manifold with {\sl a quaternionic manifold}. This is
not true in our case since putting together $ \exp \left[ Solv_3 \right]$
with $ \exp \left[ Solv_4 \right]$ we reobtain the $N=8$ scalar
manifold $E_{7(7)}/SU(8)$ which is neither a direct product nor
K\"ahlerian, nor quaternionic. The blame for this can be put on the
decomposition \eqn{7in3p4} which is a direct
sum of vector spaces but not a direct sum of Lie algebras: in other
words we have
\begin{equation}
\left[ Solv_3 \, , \, Solv_4 \right] \, \ne \, 0
\label{nocommut}
\end{equation}
The problem of deriving consistent $N=2$ truncations is most
efficiently addressed in the language of Alekseveeskian solvable
algebras \cite{alex}. $Solv_3$ is a K\"ahler solvable Lie algebra,
while $Solv_4$ is a quaternionic solvable Lie algebra. We must
determine a K\"ahler subalgebra ${\cal K}\,  \subset\, Solv_3$ and a
quaternionic subalgebra ${\cal Q}\, \subset\, Solv_4$ in such a way
that:
\begin{equation}
\left[ {\cal K} \, , \, {\cal Q} \right] \, = \, 0
\label{ycommut}
\end{equation}
Then the truncation to the vector multiplets described by ${\cal K}$ and
the hypermultiplets described by ${\cal Q}$ is consistent at the
interaction level. An obvious solution is to take no vector
multiplets ( ${\cal K}=0 $) and all hypermultiplets ( ${\cal Q}=Solv_4 $) or
viceversa  ( ${\cal K}=Solv_3 $), ( ${\cal Q}=0 $). Less obvious is
what happens if we introduce just one hypermultiplet,
corresponding to the minimal one--dimensional quaternionic algebra.
In later sections we show  that in that case the maximal number of
admitted vector multiplets is $9$. The corresponding K\"ahler
subalgebra is of rank 3 and it is given by:
\begin{equation}
 Solv_3 \, \supset \, {\cal K}_3 \, \equiv \, Solv \left(
 SU(3,3)/SU(3)\times U(3) \right)
 \label{su33ka}
\end{equation}
Note that, as we will discuss in the following, the 18 scalars
parametrizing the manifold $ SU(3,3)/SU(3)\times U(3) $ are all the scalars in
the NS-NS sector of $SO^*(12)$.
A thoroughful discussion of the N=2 truncation problem and of its
solution in terms of solvable Lie algebra decompositions
is discussed in section \ref{secsu33}. At the level
of the present introductory section we want to stress the relation
of the decomposition \eqn{7in3p4} with the
Killing spinor equation for BPS black-holes.

Indeed, as just pointed out, the decomposition \eqn{7in3p4} is implied by
the $SU(2) \times U(6)$ covariance of the Killing spinor equation.
As we show in section (3.2) this equation splits into
various components corresponding to different $SU(2) \times U(6)$
irreducible representations. Introducing the decomposition \eqn{7in3p4}
we will find that the $40$ scalars belonging to $Solv_4$ are constants
independent of the radial variable $r$. Only the $30$ scalars in the
K\"ahler algebra $Solv_3$ can have a radial dependence. In fact their
radial dependence is governed by a first order differential equation
that can be extracted from a suitable component of the Killing spinor
equation. In this way we see that the same solvable Lie algebra
decompositions occurring in the problem of N=2 truncations of N=8
supergravity occur also in the problem of constructing N=8 BPS black holes.
\vskip .5cm
\par
We present now our plan for the next sections.
\par
In section \ref{n8sugra} we discuss the structure of the scalar
sector in $N=8$ supergravity and its supersymmetry transformation
rules. As just stated, our goal is to develop methods for
the analysis of BPS states as classical solutions of supergravity
theories in all dimensions and for all values of $N$. Many results
exist in the literature for the $N=2$ case in four dimensions \cite{feka,n2kal,n2lust},
where the number of complex scalar fields involved just equals the number of
differential equations one obtains from the Killing spinor condition.
Our choice to focus on the $N=8$, $D=4$ case is motivated by the
different group--theoretical structure of the Killing spinor equation
in this case and by the fact that it is the maximally extended
supersymmetric theory.
\par
In section \ref{bhansazzo} we introduce the black--hole ansatz and we
show how using roots and weights of the $E_{7(7)}$ Lie algebra we can
rewrite in a very intrinsic way the Killing spinor equation. We
analyse its components corresponding to irreducible representations
of the isotropy subalgebra $U(1)\times SU(2) \times SU(6)$ and we show
the main result, namely that the $40$ scalars in the $Solv_4$
subalgebra are constants.
\par
In section \ref{cartadila} we exemplify our method by explicitly
solving the simplified model where the only non--zero fields are
the dilatons in the Cartan subalgebra. In this way we retrieve the
known $a$--model solutions of N=8 supergravity.
\par
The following two  sections 5 and \ref{solvodecompo} are concerned with the method and the results of our
computer aided calculations on the embedding of the subalgebras
$ U(1)\times SU(2)\times SU(6) \subset SU(8)$ in $E_{7(7)}$ and with the structure of the
 solvable Lie algebra decomposition  already introduced in eq.\eqn{7in3p4}.
 In particular we study the problem of consistent N=2 truncations using
 Alekseevski formalism.
 These two sections, being rather technical, can be skipped in a
 first reading by the non interested reader.
 We note however  that many of the results there obtained  are used for
 the discussion of the subsequent section. In particular, these results are
 preliminary for the allied
 project of gauging the maximal gaugeable abelian ideal ${\cal G}_{abel}$,
 which we outlined in our earlier paper \cite{noialtri2} and which will be postponed
 to a future publication. Note that such a gauging
 should on one hand produce spontaneous partial breaking of $N=8$
 supersymmetry and on the other hand be interpretable as due to the
 condensation of $N=8$ BPS black-holes.
 \par
 In the final section \ref{concludo} we address the question of the
 most general BPS black-hole. Using the little group of the charge
 vector in its normal form which, following \cite{lastserg}, is
 identified with $SO(4,4)$, we are able to conclude that the only
 relevant scalar fields are those associated with the solvable Lie
 subalgebra:
 \begin{equation}
Solv \left( \frac{SL(2,\IR)^3}{U(1)^3} \right) \, \subset \, Solv_3
\label{rilevanti}
\end{equation}
where the 6 scalars that parametrize the manifold $ \frac{SL(2,\IR)^3}{U(1)^3}$ are all in the
NS-NS sector.

Moreover, with an appropriate
 identification, we show how the calculation of the fixed scalars
 performed by the authors of \cite{STUkallosh} in the N=2 STU model  amounts to
 a solution of the same problem also in the N=8 theory.
 The most general solution can be actually generated by U-duality
 rotations of $E_{7(7)}$.

Therefore, the final result of our whole analysis, summarized in
the conclusions is that
up to U-duality transformations the most general $N=8$ black-hole
is actually an $N=2$ black-hole corresponding however to a very
specific choice of the special K\"ahler manifold, namely $ \frac{SO^*(12)}{U(6)}$
as in eq.\eqn{3and4defi}, \eqn{pincpal}.

\section{N=8 Supergravity and its scalar manifold $E_{7(7)}/SU(8)$ }
\label{n8sugra}
The bosonic Lagrangian of $N=8$ supergravity contains,
besides the metric $28$ vector fields and $70$ scalar fields
spanning the $E_{7(7)}/SU(8)$ coset manifold. This lagrangian falls
into the general type of lagrangians admitting
electric--magnetic duality rotations  considered in
\cite{n2paperone},\cite{mylecture},\cite{ricsertoi}.
For the case where all the scalar fields of the coset manifold have been
switched on the Lagrangian, according to the normalizations of
\cite{n2paperone} has the form:
\begin{equation}
{\cal L} = \sqrt{-g} \, \left( 2\, R[g] + \frac{1}{4}\, \mbox{Im}\,
{\cal N}_{\Lambda\Sigma} \, {\cal F}^{\Lambda\vert \mu \nu}\,{\cal
F}^{\Sigma}_{\mu \nu}  + \frac{1}{4}
\mbox{Re}\, {\cal N}_{\Lambda\Sigma} \, {\cal F}^{\Lambda}_{\mu \nu} \, {\cal
F}^{\Sigma}_{\rho \sigma}\, \epsilon^{\mu \nu\rho\sigma}
+ \frac{\alpha^2}{2}\, g_{ij}(\phi) \, \partial_\mu \phi^i \, \partial^\mu \phi^j \right)
\label{lagrared}
\end{equation}
where the indeces $\Lambda,\Sigma$ enumerate the $28$ vector fields,
$g_{ij}$ is the $E_{7(7)}$ invariant metric on the scalar coset
manifold, $\alpha$ is a real number fixed by supersymmetry
and the period matrix ${\cal N}_{\Lambda\Sigma}$ has the
following general expression holding true for all symplectically embedded
coset manifolds \cite{gaiazumin}:
\begin{equation}
{\cal N}_{\Lambda\Sigma}=h \cdot f^{-1}
\label{gaiazuma}
\end{equation}
The complex $ 28 \times 28 $ matrices $ f,h$ are
defined by the $Usp(56)$ realization $\IL_{Usp} \left(\phi\right)$ of the
coset representative which is related to its $Sp(56,\IR)$ counterpart
$\IL_{Sp}(\phi)$ through a Cayley transformation, as dispayed in the
following formula \cite{amicimiei}:
\begin{eqnarray}
\IL_{Usp} \left(\phi\right) &=& \frac{1}{\sqrt{2}}\left(\matrix{ f+ {\rm i}h & \bar f+ {\rm i}\bar h \cr f- {\rm i}h
& \bar f - {\rm i}\bar h  \cr } \right) \nonumber\\
& \equiv & {\cal C} \, \IL_{Sp}\left(\phi\right) {\cal C}^{-1} \nonumber\\
\IL_{Sp}(\phi) & \equiv & \exp \left[ \phi^i \, T_i \right ] \, = \,
\left(\matrix{ A(\phi) & B(\phi) \cr C(\phi) & D(\phi) \cr } \right) \nonumber\\
{\cal C}& \equiv & \frac{1}{\sqrt{2}} \, \left(\matrix{ \bfone & {\rm i} \, \bfone \cr
\bfone & - \, {\rm i} \, \bfone \cr } \right)
\label{cayleytra}
\end{eqnarray}
In eq. \eqn{cayleytra} we have implicitly utilized the solvable
Lie algebra parametrization of the coset, by assuming that the
matrices $T_i$ ($i=1,\dots ,70$) constitute some basis of the solvable Lie
algebra $Solv_7=Solv\left(E_{7(7)}/SU(8)\right)$.
\par
Obviously, in order to make eq.\eqn{cayleytra} explicit one has to choose a
basis for the ${\bf 56}$ representation of $E_{7(7)}$. In the sequel,
according to our convenience, we utilize
two different bases for a such a representation.
\begin{enumerate}
\item {{\it The Dynkin basis}. In this case,
hereafter referred to as $SpD(56)$, the basis vectors of the real
symplectic representation  are eigenstates
of the Cartan generators with eigenvalue one of the $56$ weight vectors
($\pm {\vec \Lambda } =\{ \Lambda_1 , \dots , \Lambda_7 \}$ pertaining
to the representation:
\begin{eqnarray}
\label{dynkbas}
(W=1,\dots\, 56 )& :&  \vert \, {W} \, \rangle \, = \, \cases
{ \vert {\vec \Lambda} \rangle ~~~~\, : \quad H_i \vert \, {\vec \Lambda} \, \rangle
~~~~ = ~~~\Lambda_i \,
\vert \, {\vec \Lambda} \, \rangle   ~~~\quad (\Lambda= 1,\dots \, 28)
\cr
\vert \, -{\vec \Lambda} \rangle \, :
\quad H_i \vert \, -{\vec \Lambda} \, \rangle = \, -\Lambda_i \,
\vert \, -{\vec \Lambda} \, \rangle   \quad (\Lambda= 1,\dots \, 28)
\cr}\nonumber\\
\vert \, V  \, \rangle & = & f^{\Lambda} \,   \vert {\vec \Lambda} \rangle
\, \,  \oplus \,\,  g_{\Lambda} \,   \vert {- \vec \Lambda} \rangle
\nonumber\\
\mbox{or in matrix notation} && \nonumber\\
\nonumber\\
{\vec V}_{SpD} & = & \left(\matrix{ f^\Lambda \cr g_\Sigma \cr }\right )
\end{eqnarray}
}
\item{{\it The Young basis}. In this case, hereafter referred to as $UspY(56)$,
the basis vectors of the complex pseudounitary
representation correspond to the natural basis of the
${\bf 28}$ + ${\bar {\bf 28}}$ antisymmetric representation
of the maximal compact subgroup $SU(8)$. In other
words, in this  realization of the fundamental $E_{7(7)}$
representation a generic vector is of the following form:
\begin{eqnarray}
\label{youngbas}
\vert {V} \rangle &=& u^{AB} \,
\mbox{$\begin{array}{|c|}
\hline
\stackrel{ }{A}\cr
\hline
\stackrel{}{B}\cr
\hline
\end{array}
$} \,\,    \oplus \, \,   {v}_{AB} \,  \mbox{$\begin{array}{|c|}
\hline
\stackrel{ }{\bar A}\cr
\hline
\stackrel{ }
{\bar B}\cr
\hline
\end{array} $} \quad ; \quad (A,B=1,\dots,8) \nonumber\\
&&\null\nonumber\\
 \mbox{ or in matrix notation}&& \nonumber\\
&&\null\nonumber\\
{\vec V}_{UspY} &=&  \left( \matrix { u^{AB} \cr  v_{AB} \cr }\right)
\end{eqnarray}
}
\end{enumerate}
Although their definitions are respectively  given in terms of the real and the
complex case, via a Cayley transformation  each of the two basis has both a real
symplectic and a complex pseudounitary  realization. Hence
we will  actually deal with four bases:
\begin{enumerate}
\item {The $SpD(56)$--basis}
\item {The $Usp_D(56)$--basis}
\item {The $SpY(56)$--basis}
\item {The $UspY(56)$--basis}
\end{enumerate}
Each of them has distinctive advantages depending on the aspect of
the theory one addresses. In particular the $UspY(56)$ basis is
that originally utilized by de Wit and Nicolai in their construction
of gauged $N=8$ supergravity \cite{dwni}. In the considerations
of the present paper the $SpD(56)$ basis will often offer the best picture
since it is that where the structure of the solvable Lie algebra is
represented in the simplest way. In order to use the best features
of each basis we just need to have full control on the
matrix that shifts fron one to the other. We name such matrix
${\cal S}$ and we write:
\begin{eqnarray}
\label{bfSmat}
\left( \matrix { u^{AB} \cr  v_{AB} \cr }\right) & = & {\cal S} \,
\left( \matrix { f^\Lambda \cr  g_\Sigma \cr }\right) \nonumber\\
&& \null\nonumber\\
 \mbox{where} && \nonumber\\
&& \null\nonumber\\
{\cal S}&=& \left( \matrix{ {\bf S} & {\bf 0} \cr {\bf 0} & {\bf S^\star}
\cr } \right) \, {\cal C} = \frac{1}{\sqrt{2}}\left( \matrix{ {\bf S} & {\rm i}\,{\bf S} \cr
{\bf S^\star} & -{\rm i}\,{\bf S^\star} \cr } \right)\nonumber\\
&& \null \nonumber\\
\mbox{the $ 28 \times 28 $ matrix ${\bf S}$ being unitary} && \nonumber\\
&& \null \nonumber\\
 {\bf S}^\dagger {\bf S} &=& \bfone
\end{eqnarray}
The explicit form of the $U(28)$ matrix ${\bf S}$ is given in
section 5.4. The weights of the $E_{7(7)}$ ${\bf 56}$ representation are
listed in table \ref{e7weight} of appendix B.
\par
\subsection{Supersymmetry transformation rules and central charges}
In order to obtain the $N=8$ BPS saturated Black Holes we cannot
confine ourselves to the bosonic lagrangian, but we also need the
the explicit expression for the supersymmetry transformation rules of the
fermions. Since the $N=8$ theory has no matter multiplets the
fermions are just  the ${\bf 8}$ spin $3/2$ gravitinos and the
${\bf 56}$ spin $1/2$ dilatinos. The two numbers ${\bf 8}$ and ${\bf
56}$ have been written boldfaced since they also single out the
dimensions of the two irreducible $SU(8)$ representations to which
the two kind of fermions are respectively assigned, namely the fundamental and the
three times antisymmetric:
\begin{equation}
 \psi_{\mu\vert A} \, \leftrightarrow \,
 \mbox{$ \begin{array}{|c|}
\hline
\stackrel{ }{A}\cr
\hline
\end{array}$} \, \equiv \,  {\bf 8}\quad ; \quad \chi _{ABC}  \, \leftrightarrow \,
\mbox{$ \begin{array}{|c|}
\hline
\stackrel{ }{A}\cr
\hline
\stackrel{}{B}\cr
\hline
\stackrel{}{C}\cr
\hline
\end{array}
$}   \, \equiv \,  {\bf 56}
\end{equation}
Following the conventions and formalism of
 \cite{amicimiei} and \cite{castdauriafre}
the relevant supersymmetry transformation rules can be written as follows:
\begin{eqnarray}
\delta \psi _{A\mu }&=&\nabla_\mu \epsilon_A - \frac{k}{4} \, c\,
T^{-}_{AB\vert \rho\sigma} \,\gamma^{\rho\sigma} \, \gamma_\mu  \,
\epsilon^B  + \cdots \nonumber\\
\delta \chi_{ABC} &=& a P_{ABCD\vert i } \, \partial_\mu \phi^i
\, \gamma^\mu  \, \epsilon^D + b \,
T^{-}_{[AB \vert \rho\sigma} \,\gamma^{\rho\sigma}
\epsilon_{C ]}  + \cdots
\label{trasforma}
\end{eqnarray}
where $a,b,c$ are numerical coefficients fixed
by superspace  Bianchi identities while, by definition,
$T^{-}_{AB\vert \mu\nu}$ is the antiselfdual part of the
graviphoton field strength
and  $P_{ABCD\vert i }$ is the vielbein of the scalar coset manifold,
completely antisymmetric in $ABCD$ and satisfying the pseudoreality
condition:
\begin{equation}
P_{ABCD}=\frac{1}{4!}\epsilon_{ABCDEFGH}\bar P^{EFGH}.
\end{equation}
What we need is the explicit expression of these objects in terms
of coset representatives. For the vielbein $ P_{ABCD\vert i }$ this
is easily done. Using the $UspY(56)$ basis the left invariant 1--form
has the following form:
\begin{equation}
\IL(\phi)^{-1} \,d \IL(\phi) \, = \, \left( \matrix{
\delta^{[A }_{[C} \, Q^{\phantom{D} B]}_{D]}
& P^{ABEF} \cr P_{CDGH} &   \delta^{[E}_{[G} \, Q^{ F]}_{\phantom{F}H]}\cr }\right)
\label{uspYconnec}
\end{equation}
where the $1$--form $Q^{\phantom{D} B}_{D}$ in the ${\bf 63}$ adjoint
representation of $SU(8)$ is the connection while the $1$--form
$P_{CDGH}$ in the ${\bf 70}$ four times antisymmetric representation
of $SU(8)$ is the vielbein of the coset manifold $E_{7(7)}/SU(8)$.
Later we need to express the same objects in different basis but
their definition is clear from eq.\eqn{uspYconnec}. A little more
care is needed to deal with the graviphoton field strenghts.
To this effect we begin by introducing
the multiplet of electric and magnetic field strengths according
to the standard definitions of \cite{n2paperone},\cite{mylecture}
\cite{ricsertoi}:
\begin{equation}
{\vec V}_{\mu\nu} \equiv \left(\matrix {
F^{\Lambda}_{\mu\nu} \cr G_{\Sigma\vert\mu\nu} \cr}\right)
\label{symvecft}
\end{equation}
where
\begin{eqnarray}
G_{\Sigma\vert\mu\nu} &=& -\mbox{Im}{\cal N}_{\Lambda\Sigma} \,
{\widetilde F}^{\Sigma}_{\mu\nu} -\mbox{Re}{\cal N}_{\Lambda\Sigma} \,
{ F}^{\Sigma}_{\mu\nu}\nonumber\\
{\widetilde F}^{\Sigma}_{\mu\nu}&=&\frac{1}{2}\, \epsilon_{\mu
\nu\rho\sigma} \, F^{\Sigma\vert \rho\sigma}
\end{eqnarray}
The $56$--component field strenght vector ${\vec V}_{\mu\nu}$
transforms in the real symplectic representation of the U--duality
group $E_{7(7)}$. We can also write  a column vector containing
the $ 28 $ components of the graviphoton field strenghts and their
complex conjugate:
\begin{equation}
{\vec T}_{\mu\nu} \equiv \left(\matrix{
T^{\phantom{\mu \nu}\vert AB}_{\mu \nu}  \cr
T_{\mu \nu \vert AB} \cr }\right) \quad T^{\vert AB}_{\mu \nu} =
\left(T_{\mu \nu \vert AB}\right)^\star
\label{gravphotvec}
\end{equation}
in which the upper and lower components  transform in the canonical {\it Young
basis} of $SU(8)$ for the ${\bar {\bf 28}}$ and  ${\bf 28}$
representation respectively.\par
The relation between the graviphoton field strength vectors and the
electric magnetic field strenght vectors involves the coset
representative in the $SpD(56)$ representation and it is the following one:
\begin{equation}
{\vec T}_{\mu \nu} = - {\cal S} \, \IC \, \IL_{SpD}^{-1}(\phi) \, {\vec
V}_{\mu \nu}
\label{T=SCLV}
\end{equation}
The matrix
\begin{equation}
\IC =\left(\matrix { {\bf 0} & \bfone \cr -\bfone & {\bf 0} }\right)
\label{sympinv}
\end{equation}
is the symplectic invariant matrix. Eq.\eqn{T=SCLV} reveal that the
graviphotons transform under the $SU(8)$ compensators associated
with the $E_{7(7)}$ rotations. To show this let $g \in  E_{7(7)}$ be an
element of the U--duality group, $g(\phi)$ the action of $g$ on the
$70$ scalar fields and $\ID (g)$ be the $56$ matrix representing $g$
in the real Dynkin basis. Then by definition of coset representative
we can write:
\begin{equation}
  \ID (g) \, \IL_{SpD}(\phi) = \IL_{SpD}\left(g(\phi)\right) \,
  W_D(g,\phi)  \quad ; \quad  W_D(g,\phi) \, \in \, SU(8) \subset \,
  E_{7(7)}
  \label{compensa}
\end{equation}
where $W_D(g,\phi)$ is the $SU(8)$ compensator in the Dynkin basis.
If we regard the graviphoton composite field as a functional of the
scalars and vector field strenghts, from eq.\eqn{compensa} we
derive:
\begin{eqnarray}
T_{\mu \nu} \left( g(\phi), \ID(g) {\vec V} \right)& =& W^*_Y (g,\phi)
\,  T_{\mu \nu} \left(  \phi ,  {\vec V} \right) \nonumber\\
W_Y(g,\phi) & \equiv & {\cal S}^\star \, W_D(g,\phi) \, {\cal S}^T
\end{eqnarray}
where $W_Y(g,\phi)$ is the $SU(8)$ compensator in the Young basis.
\par
It is appropriate to express the upper and lower components of $\vec{T}$
in terms of the self--dual and antiself--dual parts of the graviphoton
field strenghts, since only the latters enter (\ref{trasforma}) and
therefore the equations for the BPS Black--Hole.\\
These components are defined as follows:
\begin{eqnarray}
T^{+\vert AB}_{ \mu \nu}&=& \frac{1}{2}\left(T^{ \vert AB}_{\mu \nu}+\frac{{\rm i}}{2}
\, \epsilon_{\mu\nu\rho\sigma}g^{\rho\lambda} g^{\sigma\pi}
\, T^{ \vert AB}_{\lambda \pi}\right) \nonumber\\
T^-_{ AB \vert \mu \nu}&=& \frac{1}{2}\left(T_{  AB \vert\mu \nu}-\frac{{\rm i}}{2}
\, \epsilon_{\mu\nu\rho\sigma}g^{\rho\lambda} g^{\sigma\pi}
\, T_{ AB \vert\lambda \pi} \right)
\label{selfdual}
\end{eqnarray}
\par
Indeed the following equalities hold true:
\begin{eqnarray}
 T_{\mu  \nu}^{\phantom{\mu  \nu}\vert AB}&=& T^{+ \vert AB}_{\mu  \nu} \nonumber\\
 T_{ \mu  \nu\vert AB}&=& T^{-}_{ \mu  \nu\vert AB}
\label{symprop}
\end{eqnarray}
In order to understand the above properties \cite{amicimiei}, let us first rewrite equation
(\ref{T=SCLV}) in components:
\begin{eqnarray}
 &&T^{\phantom{\mu  \nu} AB}_{\mu  \nu} =
 \bar{h}^{AB}_{\phantom{AB}\Sigma}F^{\Sigma}_{\mu  \nu}-
 \bar{f}^{ AB \vert\Lambda}G_{\Lambda\vert \mu  \nu}\nonumber \\
 &&T^{-}_{ \mu  \nu\vert AB} = h_{AB \vert \Sigma}F^{\Sigma}_{\mu  \nu}-
 f_{AB}^{\phantom{AB}\Lambda}G_{\Lambda\vert \mu  \nu}
 \label{T=SCLVcomp}
 \end{eqnarray}
having defined the matrices $ h,\, f,\,\bar{h},\,\bar{f}$ in
the following way:
\begin{equation}
 - {\cal S} \, \IC \, \IL_{SpD}^{-1}(\phi) =\left(\matrix{\bar{h} &
 -\bar{f}\cr h & -f }\right)
 \label{LM}
\end{equation}
Next step is to express the self--dual and antiself--dual components
of $G_\Sigma $ (defined in the same way as for $T_{\mu \nu}$ in
(\ref{selfdual}))in terms of the corresponding components of $F^\Sigma$
through the period matrix
\begin{eqnarray}
&& G^{+}_{\Sigma}={\cal N}_{\Sigma\Lambda}F^{+ \Lambda}\nonumber\\
&& G^{-}_{\Sigma}=\bar{{\cal N}}_{\Sigma\Lambda}F^{- \Lambda}
\label{Gpm=NFpm}
\end{eqnarray}
Projecting the two equations (\ref{T=SCLVcomp}) along its self--dual
and antiself--dual parts, and taking into account
(\ref{Gpm=NFpm}), one can deduce the following conditions:
\begin{eqnarray}
 &&T^{- \vert AB}_{\mu  \nu} =0 \nonumber \\
 &&T^{+}_{ \mu  \nu\vert AB} =0
\label{TpTm}
\end{eqnarray}
which imply in turn equations (\ref{symprop}). The symplectic vector
${\vec T}_{\mu \nu}$ of the graviphoton field strenghts
may therefore be rewritten in the following form:
\begin{equation}
{\vec T}_{\mu \nu} \equiv \left(\matrix{T^{+\vert AB}_{\mu \nu}  \cr
T^{-}_{\mu \nu \vert AB} \cr }\right)
\end{equation}
These preliminaries completed we are now ready to consider the
Killing spinor equation and its general implications.
\section{The Black Hole ansatz and the Killing spinor equation}
\label{bhansazzo}
The BPS saturated black holes we are interested in are classical
field configurations with rotational symmetry and time translation
invariance. As expected on general grounds we must allow for the
presence of both electric and magnetic charges. Hence we introduce
the following ansatz for the elementary bosonic fields of the theory
\subsection{The Black Hole ansatz}
We introduce isotropic coordinates:
\begin{eqnarray}
\left\{ x^\mu \right\} &=&\left\{ t , {\vec x} = x^a \right\}  \quad ;
\quad a=1,2,3 \nonumber\\
 r&=&\sqrt{ {\vec x}\, \cdot \, {\vec x} }
 \label{isocoord}
\end{eqnarray}
and we parametrize the metric, the vector fields and the scalar fields as follows:
\begin{eqnarray}
ds^2 &=& \exp \left [ 2U(r)\right ]\,  dt^2 - \exp \left [ -2U(r)\right ] \, d{\vec x}^2
\label{metansaz}\\
F^{-{\vec \Lambda}}_{\mu\nu}&=& \frac{1}{4\pi} \, t^{{\vec \Lambda}}(r) \,
E^{-}_{\mu\nu}\label{vecansaz}\\
\phi^i &=& \phi^i (r)
\label{ansaz}
\end{eqnarray}
where
\begin{equation}
 t^{{\vec \Lambda}}(r) \, \equiv \, 2 \pi \, \left( 2 \, p^{{\vec \Lambda}}+
 \mbox{i}\, q^{{\vec \Lambda}}(r) \right)
\label{tvecvero}
\end{equation}
is a $28$--component  complex vector whose real part is constant,
while the imaginary part is a radial fucntion to be determined. We
will see in a moment the physical interpretation of $p^{{\vec \Lambda}}$
and $q^{{\vec \Lambda}}(r)$.
$E^{-}_{\mu\nu}$ is the unique antiself--dual $2$--form in the background of
the chosen metric and  it reads
as follows \cite{pietrobh1}, \cite{pietrobh2} :
\begin{equation}
E^- \, = \, E^{-}_{\mu\nu} \, dx^\mu  \wedge dx^\nu =
\mbox{i} \frac{e^{2U(r)}}{r^3} \, dt \wedge {\vec x}\cdot
d{\vec x}  +  \frac{1}{2}   \frac{x^a}{r^3} \, dx^b   \wedge
dx^c   \epsilon_{abc}
\label{eaself}
\end{equation}
and it is normalized so that:
  $ \int_{S^2_\infty} \, E^-  =   2 \, \pi $.

Combining eq.\eqn{eaself} with eq.\eqn{vecansaz} and \eqn{tvecvero}
we conclude that
\begin{eqnarray}
 p^{{\vec \Lambda}} &=&\mbox{magnetic charge} \\
 q^{{\vec \Lambda}}(r=\infty) &=&\mbox{electric charge}
 \label{pqinterpret}
\end{eqnarray}
At the same time we can also identify:
\begin{equation}
 q^{{\vec \Lambda}}(r) \, = \, r^2 \, \frac{d C^{{\vec \Lambda}}(r)}{dr} \,
 \exp \left[C^{{\vec \Lambda}}(r) - 2 U \right]
 \label{identqc}
\end{equation}
where $\exp \left[C^{{\vec \Lambda}}(r)\right]$ is a function
parametrizing the electric potential:
\begin{equation}
A^{{\vec \Lambda}}_{elec} \, = \, dt \, \exp \left[C^{{\vec \Lambda}}(r)\right].
\end{equation}
\subsection{The Killing spinor equations}
We can now analyse the Killing spinor equation combining the results
  (\ref{symprop}), (\ref{Gpm=NFpm}) with our ansatz (\ref{ansaz}). This
allows us to rewrite (\ref{T=SCLVcomp}) in the following form:
\begin{eqnarray}
 &&T^{+\phantom{\mu  \nu} AB}_{\mu  \nu} = \frac{1}{4\pi}
 \left(\bar{h}^{AB}_{\phantom{AB}\Sigma}t^{\star\Sigma}-
 \bar{f}^{ AB \vert\Lambda}{\cal N}_{\Lambda\Sigma}t^{\star\Sigma}\right)E^{+}_{\mu \nu}
 \nonumber \\
 &&T^{-}_{ \mu  \nu\vert AB} = \frac{1}{4\pi}
 \left(h_{AB \vert \Sigma}t^{\Sigma}-
 f_{AB}^{\phantom{AB}\Lambda}\bar{{\cal N}}_{\Lambda\Sigma}t^{\Sigma}\right)
 E^{-}_{\mu \nu}
  \label{ThfE}
\end{eqnarray}
where
$ E^+_{\mu\nu}  =\left( E^-_{\mu\nu} \right)^\star $.

Then we can use the general result (obtained in \cite{pietrobh1}, \cite{pietrobh2}):
\begin{eqnarray}
&&E^-_{\mu\nu} \, \gamma^{\mu\nu}  = 2 \,\mbox{i} \frac{e^{2U(r)}}{r^3}  \,
\gamma_a x^a \, \gamma_0 \, \frac{1}{2}\left[ {\bf 1}+\gamma_5 \right] \nonumber \\
&&E^+_{\mu\nu} \, \gamma^{\mu\nu}  =-2 \,\mbox{i} \frac{e^{2U(r)}}{r^3}  \,
\gamma_a x^a \, \gamma_0 \, \frac{1}{2}\left[ {\bf 1}-\gamma_5 \right] \nonumber\\
\label{econtr}
\end{eqnarray}
and contracting both sides of (\ref{ThfE}) with $\gamma^{\mu\nu}$
one finally gets:
\begin{eqnarray}
&&T^{+\phantom{\mu  \nu} AB}_{\mu  \nu}\gamma^{\mu\nu}  = -\frac{{\rm i}}{2\pi}
\frac{e^{2U(r)}}{r^3}  \,
\gamma_a x^a \, \gamma_0 \, \left(\bar{h}^{AB}_{\phantom{AB}\Sigma}t^{\star\Sigma}-
 \bar{f}^{ AB \vert\Lambda}{\cal N}_{\Lambda\Sigma}t^{\star\Sigma}\right)
 \frac{1}{2}\left[ {\bf 1}-\gamma_5 \right]  \nonumber \\
 &&T^{-}_{ \mu  \nu\vert AB} \gamma^{\mu\nu} =\frac{{\rm i}}{2\pi} \frac{e^{2U(r)}}{r^3}
 \, \gamma_a x^a \, \gamma_0 \, \left(h_{AB \vert \Sigma}t^{\Sigma}-
 f_{AB}^{\phantom{AB}\Lambda}\bar{{\cal N}}_{\Lambda\Sigma}t^{\Sigma}\right)
 \frac{1}{2}\left[ {\bf 1}+\gamma_5 \right]
 \label{Tpmgamma}
\end{eqnarray}
At this point we specialize the supersymmetry parameter to be of the
form analogue to the form utilized in \cite{pietrobh1,pietrobh2}:
\begin{equation}
\epsilon_A = e^{f(r)} \xi_A
\label{kilspinor1}
\end{equation}
It is useful
to split the $SU(8)$ index $A=1,\dots ,8$ into an $SU(6)$ index
$X=1,\dots ,6$ and an $SU(2)$ index $a=7,8$. Since we look
 for BPS states belonging to {\sl just once shortened multiplets}
 ({\it i.e.} with  $N=2$ residual supersymmetry)
we require that $ \xi^X=0,\, X=1,\dots ,6$ and furthermore that:
\begin{equation}
\gamma_0\, \xi_a  = -\mbox{i} \, \epsilon_{ab} \, \xi^b
\label{kilspinor2}
\end{equation}
The vanishing of the gravitino transformation rule implies conditions on both functions
$U(r)$ and $f(r)$. The equation for the latter is uninteresting since
it simply fixes the form of the Killing spinor parameter.
The equation for $U$ instead is relevant since it yields the form
of the black hole metric.  It can be written in the following form:
\begin{equation}
\frac{dU}{dr}= - {k}\frac{e^U}{r^2}\left(h_\Sigma t^\Sigma - f^\Lambda \bar\cN_{\Lambda\Sigma}t^\Sigma \right)
\label{gratra}
\end{equation}

Furthermore the $56$ differential equations from the dilatino sector can
be written in the form:
\begin{eqnarray}
&& a\, P_{ABCa\vert i} \, \frac{d\phi^i}{dr} =\frac{b}{2\pi} \frac{e^{U(r)}}{r^2}
\left(h_{\Sigma}t^{\Sigma}-
 f^{\Lambda}\bar{{\cal N}}_{\Lambda\Sigma}t^{\Sigma}\right)_{[AB}
 \delta_{C]}^{b}\epsilon_{ba}\nonumber\\
&&a\, P^{ABCa\vert i} \, \frac{d\phi^i}{dr} =\frac{b}{2\pi} \frac{e^{U(r)}}{r^2}
\left(\bar{h}_{\Sigma}t^{\star\Sigma}-
 \bar{f}^{\Lambda}{\cal N}_{\Lambda\Sigma}t^{\star\Sigma}\right)^{[AB}
 \delta^{C]}_{b}\epsilon^{ba}
\label{BPS56eq1}
\end{eqnarray}

Suppose now that the triplet of indices $(A,\, B,\, C)$ is of the
 type   $(X,\, Y,\,Z)$.
 This corresponds to projecting eq. \eqn{BPS56eq1}  into the
 representation $({\bf 1},{\bf 2}, {\bf 20})$ of  $U(1) \times SU(2)\times SU(6) \subset SU(8)$.
 In this case however the right hand side vanishes identically:
 \begin{equation}
  a\, P_{XYZ a\vert i} \, \frac{d\phi^i}{dr} =\frac{b}{2\pi} \frac{e^{U(r)}}{r^2}
\left(h_{\Sigma}t^{\Sigma}-
 f^{\Lambda}\bar{{\cal N}}_{\Lambda\Sigma}t^{\Sigma}\right)_{[XY}
 \delta_{Z]}^{b}\epsilon_{ba} \equiv 0
 \label{pxyz}
 \end{equation}
 so that we find that the corresponding 40 scalar fields are actually
 constant.

In the case where the triplet of indices $(A,\, B,\, C)$ is $(X,\, Y,\, a )$
the equations may be put in the following matrix form:
\begin{eqnarray}
\left(\matrix{ P^{XY\vert i}\cr P_{XY\vert i}}\right)\,
\frac{d\phi^i}{dr}=\frac{b}{ 3 \, a \,\pi} \frac{e^{U(r)}}{r^2}
\left(\matrix{\bar{h} & -\bar{f}\cr h & -f }\right)_{\vert XY}\,\left(
\matrix{{\bf Re}(\, t\, )\cr {\bf Re}(\, \bar{{\cal N}}t \,)}\right)
\label{pxy}
\end{eqnarray}
The above equations are obtained by projecting the terms on the left and on the right
side of eq. \eqn{BPS56eq1} (transforming respectively in the ${\bf 70}$ and in
the ${\bf 56}$ of $SU(8)$) on the common representation ${\bf
(1, 1,15) \oplus \bar{(1,1,15)}}$ of the subgroup
$U(1) \times SU(2)\times SU(6) \subset SU(8)$).

Finally, when the  triplet  $(A,\, B,\, C)$ takes the values $(X,\, b,\, c )$,
the l.h.s. of eq. \eqn{BPS56eq1}  vanishes   and we are left with the
equation:
\begin{equation}
 0=\frac{e^{U(r)}}{r^2}
\left(h_{\Sigma}t^{\Sigma}-
 f^{\Lambda}\bar{{\cal N}}_{\Lambda\Sigma}t^{\Sigma}\right)_{[Xb}
 \epsilon_{c]a}
 \label{xbc}
\end{equation}
This corresponds to the projection of eq. \eqn{BPS56eq1}  into the
representation $({\bf 1},{\bf 2},{\bf 6}) + \bar{({\bf 1},{\bf 2},{\bf 6})} \subset \bf{28} +\bar{{ \bf{28}}}$.

Let us consider the basis vectors $ \vert \vec\Lambda > \, , \, \vert - \vec\Lambda > \in {\bf 56}= {\bf 28 + \bar 28} $
defined in ref. \eqn{dynkbas} and let us introduce the eigenmatrices $\IK^{\vec\lambda}$ of the $SU(8)$ Cartan generators
${\cal H}_i$ diagonalized on the subspace of non-compact $E_{7(7)}$ generators $\IK $ defined by the
Lie algebra Cartan decomposition $E_{7(7)}=\IK \oplus \IH$ ($\vec \lambda $
being  the weights of the ${\bf 70}$ of $SU(8)$).
It is convenient to use  a real basis for both representations ${\bf 56}$ and ${\bf 70}$, namely:
\begin{eqnarray}
  \vert \vec\Lambda_x > &=& \frac {\vert \vec\Lambda >  + \vert - \vec\Lambda >}{2} \nonumber\\
   \vert \vec\Lambda_y > &=& \frac {\vert \vec\Lambda >  - \vert - \vec\Lambda >}{2{\rm i}} \nonumber\\
   \IK^{\vec \lambda}_x &=& \frac {\IK^{\vec \lambda}  + \IK^{-\vec \lambda}}{2} \nonumber\\
   \IK^{\vec \lambda}_y &=& \frac {\IK^{\vec \lambda}  - \IK^{-\vec \lambda}}{2 {\rm i}}
   \label{realbas}
\end{eqnarray}
such that they satisfy the following relations:
\begin{equation}
\begin{array}{lcrcll}
\mbox{projectors on irrep ${\bf 70}$} &:&
 \left[{\cal H}_i\, ,\, \IK^{\vec \lambda}_{x}\right]\,
 &=& \null & \left({\vec \lambda},
{\vec a}_i\right)\IK^{\vec \lambda}_{y}  \\
\null & \null &
\left[{\cal H}_i\, ,\, \IK^{\vec \lambda}_{y}\right]\, &=&  - & \left({\vec \lambda},
{\vec a}_i\right)\IK^{\vec \lambda}_{x}  \\
\mbox{projectors on irrep ${\bf 28}$ } &:&
{\cal H}_i\vert \,{\vec \Lambda}_{x} \, \rangle\, &=& \null & \left({\vec \Lambda},
{\vec a}_i\right)\vert \,{\vec \Lambda}_{y} \, \rangle  \\
\null & \null &
{\cal H}_i\vert \,{\vec \Lambda}_{y} \, \rangle\, &=&  - & \left({\vec \Lambda},
{\vec a}_i\right)\vert \,{\vec \Lambda}_{x} \, \rangle \\
\end{array}
\label{procione}
\end{equation}

 As it will be explained in section \ref{su8ine7}, the  seven Cartan
 generators ${\cal H}_i$
are given by appropriate linear combinations of the $E_{7(7)}$ step
operators (see eq.s \eqn{su8cartan}).
\par
Using the definitions given above, it is possible
to rewrite equations \eqn{gratra} and \eqn{BPS56eq1} in an
algebraically intrinsic way, which, as we will see, will prove to be useful following.
This purpose can be achieved by expressing the l.h.s. and r.h.s. of
eqs. \eqn{gratra} and \eqn{BPS56eq1}
in terms of suitable projections on the real bases $\IK^{\vec\lambda}_{x,y}$
and $\vert \vec \Lambda_{x,y} >$ respectively.
In particular eqn.  \eqn{pxyz},  \eqn{xbc} become respectively:
 \begin{equation}
\begin{array}{rclcrcl}
\mbox{Tr}\left ( \IK^{{\vec \lambda}}_x \, \IL^{-1} d\IL \right )  & = &  0
& ; & {\vec \lambda} & \in & {\bf (1,2,20)}\subset {\bf 70}\\
0 &=&\langle \, {\vec \Lambda}_x={\vec \lambda}_D \vert \, \IC \, \IL^{-1} (\phi) \,
\vert \, {\vec t}, \phi \, \rangle &; & \quad {\vec \Lambda} & \in &
{\bf (1,2,6)} \subset {\bf 28}  \nonumber\\
0 &=&\langle \, {\vec \Lambda}_y={\vec \lambda}_D \vert \, \IC \, \IL^{-1} (\phi) \,
\vert \, {\vec t}, \phi \, \rangle &; & \quad {\vec \Lambda} & \in &
{\bf (1,2,6)} \subset {\bf 28}+ \bar{\bf 28}
\end{array}
\label{22026eq}
\end{equation}
where we have set:
\begin{equation}
\vert \, {\vec t}, \phi \, \rangle = \left (
\matrix{{\bf Re}(\, t\, )\cr {\bf Re}(\, \bar{{\cal N}}t \,)}\right)
\end{equation}
The real vectors $ \vert \vec\Lambda_x >$ and $ \vert \vec\Lambda_y >$ in eq. \eqn{22026eq} are related
by eq. \eqn{realbas} to $ \vert \vec\Lambda >$ and $ -\vert \vec\Lambda >$,  which are now restricted to
the subrepresentations ${\bf (1,2,6)} $ and $\bar{{\bf (1,2,6)}}$
respectively.
The pair of eq.s \eqn{22026eq} can now be  read as very clear statements. The
first in the pair tells us that $40$ out of the $70$ scalar fields in
the theory must be constants in the radial variable. Comparison with
the results of section \ref{solvodecompo} shows that the fourty
constant fields are those belonging to the solvable subalgebra
$Solv_4 \subset Sol_7$ defined in eq.\eqn{3and4defi}.
\par
Hence those scalars that in an N=2 truncation belong to hypermultiplets are
constant in any BPS black hole solution. \par
The second equation in the pair \eqn{22026eq} can  be read as
a statement on the available charges. Indeed since it must be zero
everywhere, the right hand side of this equation can be evaluated at
infinity where the vector $\vert {\vec t} ,\phi \rangle$ becomes the
$56$-component vector of electric and magnetic charges defined as:
\begin{equation}
\left(\matrix { g^{\vec {\Lambda}} \cr e_{\vec {\Sigma}}\cr } \right)
= \left( \matrix { \int_{S^2_\infty} \, F^{\vec {\Lambda}} \cr
 \int_{S^2_\infty} \, G_{\vec {\Sigma}} \cr }\right)
 \label{gedefi}
\end{equation}
As we will see in the following, the explicit evaluation of  \eqn{pappaequa}
implies that $24$ combinations of  the charges are zero.
This, together  with the fact that the last of eqn. \eqn{22026eq}
yields the vanishing of $24$ more independent combinations, implies that there are only
$8$ surviving charges.
\par
On the other hand, eq. \eqn{pxy} can be rewritten in the following form:
\begin{equation}
\mbox{Tr}\left ( \IK^{{\vec \lambda}} \, \IL^{-1} d\IL \right ) =
\frac{b}{ 3 \, a \,\pi} \frac{e^{U(r)}}{r^2} \, \langle \,
{\vec \Lambda}={\vec \lambda}  \vert_D \, \IC \, \IL^{-1} (\phi) \,
\vert \, {\vec t}, \phi \, \rangle
\label{pappaequa}
\end{equation}
In equation \eqn{pappaequa}  both the weights ${\vec \lambda}$ and ${\vec \Lambda}$ defining the projections in the
l.h.s. and r.h.s. belong to the common representation $ {\bf (1, 1,15) \oplus \bar{(1,1,15)}}$ (
$\lambda \, \in \, {\bf (1, 1,15) \oplus \bar{(1,1,15)}} \,
\subset \, {\bf 70} $,
$\left({\vec \Lambda}\, ,\, -{\vec \Lambda}\right)
\, \in \, {\bf (1, 1,15) \oplus \bar{(1,1,15)}} \,
\subset \, {\bf 28+\bar{28}} $).
Finally, in this intrinsic formalism eq. \eqn{gratra}  takes the
form:
\begin{equation}
\frac{dU}{dr}\,=\,2k \frac{e^U}{r^2} \langle {\vec \Lambda}\vert_D\, \IC \, \IL^{-1} (\phi) \,
\vert \, {\vec t}, \phi \, \rangle \quad; \quad {\vec \Lambda}\in
{\bf (1,1,1)}\subset {\bf 28}
\label{Ueq}
\end{equation}
Eq. (\ref{Ueq}) implies that the projection on $\vert \,{\vec \Lambda}_{x} \,
\rangle$ of the right-hand side equals $\frac{dU}{dr}$ and thus gives
the differential equation for $U$, while the projection on
$\vert \,{\vec \Lambda}_{y} \, \rangle$ equals zero, which means in turn that
the central charge must be real.
\par

In the next section, restricting our attention to a simplified case
where the only non-zero scalar fields are taken in the Cartan subalgebra,
we show how the above implications of the Killing spinor equation can
be made explicit.
\section{A simplified model: BPS black-holes reduced to the Cartan subalgebra}
\label{cartadila}
\par
Just as an illustrative exercise
in the present section we consider the explicit BPS black--hole
solutions where the only scalar fields excited out of zero are those
in the Cartan subalgebra of $E_{7(7)}$.
Having set to zero all the fields except these seven, we will see
that
the Killing spinor equation implies that $52$ of the $56$ electric and
magnetic charges are also zero. So by restricting our attention to
the Cartan subalgebra we retrieve a BPS black hole solution that
depends only on $4$ charges. Furthermore in this solution $4$ of
the $7$ Cartan fields are actually set to constants and only the
remaining $3$ have a non trivial radial dependence. Which is which
is related to the basic solvable Lie algebra decomposition \eqn{7in3p4}:
the $4$ Cartan fields in $Solv_4$ are constants, while the $3$ Cartan fields in
$Solv_3$ are radial dependent. This type of solution reproduces the
so called $a$-model black-holes studied in the literature, but it is
not the most general. However, as we shall argue in the last section, the toy
model presented here misses full generality by little. Indeed
the general solution that depends on $8$, rather than $4$ charges,
involves, besides the $3$ non trivial Cartan fields other $3$
nilpotent fields which correspond to axions of the compactified string
theory.
\par
Let us then begin examining this simplified case.
For its study it is
particularly useful to utilize the Dynkin basis since there the
Cartan generators have a diagonal action on the ${\bf 56}$ representation
and correspondingly on the vector fields. Namely in the Dynkin basis
the matrix ${\cal N}_{\Lambda\Sigma}$ is purely imaginary and
diagonal once the coset representative is restricted to the Cartan
subalgebra.  Indeed if we name $h^i(x)$ ($i=1,\dots\, 7$ ) the
Cartan subalgebra scalar fields, we can write:
\begin{eqnarray}
\label{LSpD}
\IL_{SpD}\left( h \right)& \equiv &\exp \left[ {\vec h} \, \cdot \,
{\vec H} \right ] \, = \, \left(\matrix{ {\bf A}(h) &  {\bf 0} \cr
{\bf 0} & {\bf D}(h) \cr }\right) \nonumber\\
&& \null \nonumber\\
&&\mbox{where}\nonumber\\
&& \null \nonumber\\
 {\bf A}(\phi)^{{\vec \Lambda}}_{\phantom{{\vec \Lambda}}{\vec \Sigma}}&=&
 \delta^{{\vec \Lambda}}_{\phantom{{\vec \Lambda}}{\vec \Sigma}} \,
 \exp\left[ {\vec \Lambda} \, \cdot \, {\vec h} \right]\nonumber\\
 {\bf D}(\phi)_{{\vec \Lambda}}^{\phantom{{\vec \Lambda}}{\vec \Sigma}}&=&
 \delta_{{\vec \Lambda}}^{\phantom{{\vec \Lambda}}{\vec \Sigma}} \,
 \exp\left[-\, {\vec \Lambda} \, \cdot \, {\vec h} \right]\nonumber\\
\end{eqnarray}
so that combining equation \eqn{gaiazuma} with \eqn{cayleytra} and
\eqn{LSpD} we obtain:
\begin{equation}
{\cal N}_{{\vec \Lambda}{\vec \Sigma}} = {\rm i} \,
\left( {\bf A}^{-1} \, {\bf D} \right)_{
{\vec \Lambda}{\vec \Sigma}}
= {\rm i} \, \delta_{{\vec \Lambda}{\vec \Sigma}} \,\exp\left[-\,2 \,
{\vec \Lambda} \, \cdot \, {\vec h} \right]
\label{diagenne}
\end{equation}
Hence in the Dynkin basis the lagrangian \eqn{lagrared} reduced to the
Cartan sector takes the following form:
\begin{equation}
{\cal L} = \sqrt{-g} \, \left( 2\,   R[g] -\frac{1}{4}
\sum_{{\vec \Lambda} \in \Pi^+}
\, \exp\left[-\,2 \,
{\vec \Lambda} \, \cdot \, {\vec h} \right] \, {\cal F}^{{\vec \Lambda}\vert \mu \nu}\,{\cal
F}^{{\vec \Lambda}}_{\mu \nu}
+ \frac{\alpha ^2}{2}\, \sum_{i=1}^{7} \, \partial_\mu h^i \, \partial^\mu h^i \right)
\label{lagranew}
\end{equation}
where by $\Pi^+$ we have denoted the set of positive weights for the
fundamental representation of the U--duality group $E_{7(7)}$ and
$\alpha$ is a real number fixed by supersymmetry already introduced in
eq.\eqn{lagrared}.

\par
Let us examine in detail the constraints imposed by (\ref{pappaequa}),(\ref{22026eq})
and (\ref{Ueq}) on the $28$ complex vectors $t^\Lambda$.  Note that these vectors are
naturally split  in two subsets:
\begin{itemize}
\item{The set: $t_z\equiv \{t^{17},t^{18},t^{23},t^{24}\}$ that
parametrizes $8$ real charges which, through 4 suitable linearly independent combinations,
transform
in the representations ${\bf (1,1,1)+\bar{(1,1,1)}}$
$\oplus{\bf (1,1,15)+\bar{(1,1,15)} }$   and  contribute to the 4 central
charges of the theory.}
\item{the remaining 24 complex vectors $t_\ell$.  Suitable linear combinations of these
vectors transform in the representations ${\bf (1,1,1)+\bar{(1,1,1)}}$ ${\bf  \oplus(1,2,6)
+\bar{(1,2,6)}}$ and are orthogonal to the set $t_z$
}
\end{itemize}

Referring now to the present simplified model we can analyze the
consequences of the projections
 (\ref{Ueq}) and (\ref{pappaequa}).
 They are $16$ complex conditions which
split into:
\begin{enumerate}
\item {a set of 4 equations whose coefficients depend only
on $t_z$ and give rise to 4 real conditions on
the real and immaginary parts of $t_z$ and 4 real differential equations on the
Cartan fields $h_{1,2,7}$ belonging to vector multiplets, namely to
the solvable Lie subalgebra $Solv_3$ (see section \ref{solvodecompo})
}
\item {a set of 12 equations which contribute,
together with the 12 conditions coming from the second
of eq.s (\ref{22026eq}), to set the 24 $t_\ell$ to zero}
\end{enumerate}
In obtaining the above results, we used the fact that all the Cartan fields in $Solv_4$,
(namely ($h_{3,4,5,6}$) which fall into
hypermultiplets when the theory is $N=2$ truncated) are constants, and thus can
be set to zero (modulo duality rotations) as was discussed in the previous section. \\
\\

After the projections have been taken into account we are left with
a reduced set of non vanishing fields that includes only four vectors and
three scalars, namely:
\begin{eqnarray}
\mbox{vector fields}&=&\cases{F^{\Lambda^{17}}_{\mu \nu} \equiv {\cal F}^{17}_{\mu \nu}\cr
F^{\Lambda^{18}}_{\mu \nu} \equiv {\cal F}^{18}_{\mu \nu}\cr
F^{\Lambda^{23}}_{\mu \nu} \equiv {\cal F}^{23}_{\mu \nu}\cr
F^{\Lambda^{24}}_{\mu \nu} \equiv {\cal F}^{24}_{\mu \nu}\cr}\nonumber\\
\mbox{scalar fields}&=&\cases{h_1   \cr
h_2   \cr
h_7   \cr }
\end{eqnarray}
In terms of these fields, using the scalar products displayed in table \ref{scalaprod},
the lagrangian has the following explicit expression:
\begin{eqnarray}
{\cal L}&=& \sqrt{-g} \, \Biggl \{ 2\, R[g] \, + \,
\frac{\alpha ^2}{2}\, \left[ \left(\partial_\mu h_1\right)^2 +
\left(\partial_\mu h_2\right)^2 +
\left(\partial_\mu h_3\right)^2 \right]\nonumber\\
&& -\, \frac{1}{4} \,
\exp \left[ 2\,\sqrt{\frac{2}{3}}\, h_1 \, -\,\frac{2}{\sqrt{3}}\, h_7 \right] \,
\left({\cal F}_{\mu \nu}^{17} \right)^2 \,- \frac{1}{4} \,
\exp\left[ 2\,\sqrt{\frac{2}{3}}\, h_2 \, -\,\frac{2}{\sqrt{3}}\, h_7 \right] \,
\left({\cal F}_{\mu \nu}^{18} \right)^2 \nonumber\\
&& -\, \frac{1}{4} \,
\exp\left[ 2\,\sqrt{\frac{2}{3}}\, h_1 \, +\,\frac{2}{\sqrt{3}}\, h_7 \right] \,
\left({\cal F}_{\mu \nu}^{23} \right)^2 \,- \frac{1}{4} \,
\exp\left[ 2\,\sqrt{\frac{2}{3}}\, h_2 \, +\,\frac{2}{\sqrt{3}}\, h_7 \right] \,
\left({\cal F}_{\mu \nu}^{24} \right)^2  \Biggr \}
\label{reduclagra}
\end{eqnarray}
\par
Introducing an index $\alpha$ that takes the four values $\alpha =17,18,23,24 $ for
the four field strenghts, and moreover four undetermined radial functions to
be fixed by the field equations:
 \begin{equation}
    q^\alpha(r)\, \equiv \, C_\alpha ^\prime e^{C_\alpha-2U} \, r^2
    \label{pseudocharge}
\end{equation}
and four real constants $p^\alpha$,
the ansatz for the vector fields can be parametrized as follows:
\begin{eqnarray}
 \cF^{\alpha }_{el}  &=& - \frac{q^\alpha(r) \, e^{2U(r)} }{r^3} \,
dt \wedge \vec x \cdot d\vec x \, \equiv \, - {C_\alpha ^\prime e^{C_\alpha}   \over r}
dt \wedge \vec x \cdot d\vec x
\nonumber\\
 \cF^{\alpha }_{mag}  &=& {p^\alpha   \over 2r^3}x^a dx^b \wedge d x^c \epsilon_{abc}
\nonumber\\
   \cF^{\alpha } &=&    \cF^{\alpha }_{el} +  \cF^{\alpha }_{mag} \nonumber\\
 \cF^{-\alpha }_{\mu\nu} &=& \frac{1}{4\pi} \, t^{\alpha } E^{-}_{\mu\nu}  \nonumber\\
   t^{\alpha }  &=& 2\pi\left(2p^\alpha  + {\rm i} q^\alpha (r) \right)\equiv
 2\pi (2p^\alpha  + {\rm i}  C^\prime_\alpha  e^{C_\alpha -2U}r^2)
 \end{eqnarray}
\par
The physical interpretation  of the above data is the following:
\begin{equation}
p^\alpha = \mbox{mag. charges} \quad ; \quad q^\alpha(\infty) =
\mbox{elec. charges}
\label{physinter}
\end{equation}
From the effective lagrangian of the reduced system we derive
the following set of Maxwell-Einstein-dilaton field equations, where in addition
to the index $ \alpha $ enumerating the vector fields an index $i$ taking the three values
$i=1,2,7$ for the corresponding three scalar fields has also been introduced:
\begin{eqnarray}
 \mbox{Einstein eq.} &:&
  -2 R_{MN} =T_{MN} = {1\over 2}\alpha^2 \sum_{i} \, \partial_M h_i
\partial_N h_i + S_{MN} \nonumber\\
\Biggl (  S_{MN} & \equiv &-{1 \over 2} \sum _{\alpha }  e^{-2\vec \Lambda_\alpha  \cdot h}
\left[ \cF^\alpha _{M\cdot}  \cF^\alpha _{N\cdot}  -{1 \over 4}
\cF^\alpha _{\cdot\cdot}  \cF^\alpha _{\cdot\cdot}   \eta_{MN}\right]  \Biggr ) \nonumber\\
  \mbox{Maxwell eq.}&:&
   \partial_\mu  \left( \sqrt{-g}   \exp \left[ -2\vec \Lambda_\alpha \cdot h
   \right ] \,
 F^{\alpha \vert \mu \nu} \right) =  0 \nonumber\\
 \mbox{Dilaton eq.s } &:&
    \frac{\alpha^2}{\sqrt{-g}}\, \partial_\mu  \left(\sqrt{-g}\,  g^{\mu \nu} \,
    \partial_\nu h_i(r) \right)
     +\frac{1}{2}\sum_{\alpha} \, \Lambda^\alpha_i \, \exp[-2 \,\Lambda^\alpha \cdot h] \,
      {\cal F}_{\cdot\cdot}^{\alpha} {\cal F}_{\cdot\cdot}^{\alpha} =0  \nonumber\\
 \label{campequa}
 \end{eqnarray}
 In eq.\eqn{campequa} we have denoted by dots the contraction of indices. Furthermore
 we have used the capital latin letters $M,N$ for the flat Lorentz indeces obtained through
 multiplication by the inverse or direct vielbein according to the case. For instance:
 \begin{equation}
\partial_M \equiv V_M^\mu \, \partial_\mu = \cases{\partial_0 = e^{-U}
\, \frac{\partial}{\partial t} \cr
\partial_I = e^{U}
\, \frac{\partial}{\partial x^i} \quad (I=1,2,3)\cr }
\end{equation}
Finally the components of the four relevant weights, restricted to
the three relevant scalar fields are:
\begin{eqnarray}
 \vec\Lambda_{17} &=& \left(-\sqrt{\frac{2}{3} } , 0  , \sqrt{\frac{1}{3} } \right)\nonumber\\
 \vec\Lambda_{18} &=& \left(   0, -\sqrt{\frac{2}{3} } , \sqrt{\frac{1}{3} } \right) \nonumber\\
 \vec\Lambda_{23} &=& \left(-\sqrt{\frac{2}{3} } , 0  , -\sqrt{\frac{1}{3} } \right)\nonumber\\
 \vec\Lambda_{24} &=& \left(   0, -\sqrt{\frac{2}{3} } , -\sqrt{\frac{1}{3} } \right) \nonumber\\
\end{eqnarray}
The flat indexed stress--energy tensor $T_{MN}$ can be evaluated by
direct calculation and we easily obtain:
\begin{eqnarray}
T_{00} &=& S_{00} = -{1\over 4}\sum_{\alpha} \exp\left[{-2\vec \Lambda_\alpha \cdot \vec h}
+4 U \right ] \, \frac{1}{r^4} \,
\left [ \left(q^\alpha(r)\right)^2  +  \left(p^\alpha \right)^2 \right]
\nonumber\\
   T_{\ell m} &=& \left(\delta_{\ell m} -2 {x_\ell x_m \over r^2}
\right)S_{00} + {\alpha^2\over 2} {x_\ell x_m \over r^2}{\partial \vec h
\over \partial r }\cdot {\partial \vec h \over \partial r }
\label{stresstens}
\end{eqnarray}
The next part of the calculation involves the evaluation of
the flat--indexed Ricci tensor for the metric in eq.\eqn{metansaz}.
From the definitions:
\begin{eqnarray}
0 &=& d V^M - \omega^{MN} \, \wedge \, V^N \, \eta_{NR}  \nonumber\\
R^{MN} & = & d\omega^{MN} - \omega^{MR} \, \wedge \, \omega^{SN} \,
\eta_{RS} \equiv R^{MN}_{RS} \,
V^R \, \wedge V^S \nonumber\\
V^R &=& \cases{V^0 = dt e^U\cr
V^I= dx^i e^{-U} \cr }
\label{curvdef}
\end{eqnarray}
we obtain the spin connection:
\begin{equation}
\omega^{0I} = -\frac{x^i}{r} \,  dx^i\, U^\prime\, \exp\left[ 2U \right]
\quad ; \quad \omega^{IJ} =  2 \frac{x^{[i}\,dx^{j]} }{r} \,  U^\prime\,
\label{conspin}
\end{equation}
and the Ricci tensor:
\begin{eqnarray}
R_{00} & = & -\frac{1}{2}\exp[2U] \, \left( U^{\prime\prime} +\frac{2}{r}
U^\prime \right)\nonumber\\
R_{ij} &=& \frac{x^i \, x^j}{r^2} \, \exp[2U] \, \left(U^\prime \right)^2
+ \delta_{ij} R_{00}
\label{rictens}
\end{eqnarray}
Correspondingly the field equations reduce to a set of first order
differential equations for the eight unknown functions:
\begin{equation}
U(r) \quad ; \quad h_i(r) \quad ; \quad q^{\alpha}(r)
\label{unknowns}
\end{equation}
From {\sl Einstein equations} in \eqn{campequa} we get the two
ordinary differential equations:
\begin{eqnarray}
U^{\prime\prime} +\frac{2}{r} U^\prime  &=& S_{00}\, \exp \left[\, -2U\,\right]
\nonumber\\
\left( U^\prime \right)^2 &=& \left(S_{00}  - \frac{\alpha^2}{4} \sum_{i}
\left(h^\prime_i\right)^2\right)\, \exp\left[\, -2U\,\right]
\label{paireinst}
\end{eqnarray}
 from which we can eliminate the contribution of the vector fields
 and obtain an equation involving only the scalar fields and the metric:
 \begin{equation}
 U^{\prime\prime} +\frac{2}{r} U^\prime - \left( U^\prime \right)^2 - \frac{1}{4} \sum_{i}
\left(h^\prime_i\right)^2 = 0
\label{ceccus}
\end{equation}
From {\sl the dilaton equations} in \eqn{campequa} we get the three
ordinary differential equations:
\begin{equation}
h^{\prime\prime}_i +\frac{2}{r} h^\prime_i  =
\frac{1}{\alpha^2} \sum_{\alpha} \, \Lambda^\alpha_i \, \exp \left[
-2\Lambda^\alpha \cdot h +2 U\right] \, \left[ \left(q^\alpha \right)^2 -
\left(p^\alpha \right)^2 \right]\frac{1}{r^4}
\label{dilequa}
\end{equation}
Finally {\sl form the Maxwell equations} in \eqn{campequa} we obtain:
\begin{equation}
0 =  \frac{d}{dr} \, \left (
\exp\left[-2\Lambda_\alpha \cdot h \right] \, q^\alpha (r)
\right )
\label{maxwelequa}
\end{equation}
\subsection{The first order equations from the projection ${\bf (1,1,15)}\oplus
{\bf (\bar 1,\bar 1,\bar {15})}$ }
If we reconsider the general form of eq.s \eqn{pappaequa}, \eqn{Ueq},
we find that out of these 32 equations $24$ are identically satisfied when the fields are restricted to
be non--zero only in the chosen sector.
The remaining $8$ non trivial equations take the following form:
\begin{eqnarray}
\label{nontriv15}
0& =& {{c\,{\frac{\bf e^U}{r^2}}\,
     \left( {e^{{{{\sqrt{2}}\,{\bf h_2} - {\bf h_7}}\over
             {{\sqrt{3}}}}}}\,{\bf p^{18}} +
       {e^{{{{\sqrt{2}}\,{\bf h_1} + {\bf h_7}}\over
             {{\sqrt{3}}}}}}\,{\bf p^{23}} -
       {e^{{{{\sqrt{2}}\,{\bf h_1} - {\bf h_7}}\over
             {{\sqrt{3}}}}}}\,{\bf q^{17}} +
       {e^{{{{\sqrt{2}}\,{\bf h_2} + {\bf h_7}}\over
             {{\sqrt{3}}}}}}\,{\bf q^{24}} \right) }  }\nonumber\\
0&=&{{c\,{\frac{\bf e^U}{r^2}}\, \left( {e^{{{{\sqrt{2}}\,{\bf h_1} -
              {\bf h_7}}\over {{\sqrt{3}}}}}
          }\,{\bf p^{17}} +
       {e^{{{{\sqrt{2}}\,{\bf h_2} + {\bf h_7}}\over {{\sqrt{3}}}}}
          }\,{\bf p^{24}} -
       {e^{{{{\sqrt{2}}\,{\bf h_2} -
              {\bf h_7}}\over {{\sqrt{3}}}}}
          }\,{\bf q^{18}} +
       {e^{{{{\sqrt{2}}\,{\bf h_1} +
              {\bf h_7}}\over {{\sqrt{3}}}}}
          }\,{\bf q^{23}} \right) }  } \nonumber \\
  0&=&{{c\,{\frac{\bf e^U}{r^2}}\,
     \left( {e^{{{{\sqrt{2}}\,{\bf h_2} -
              {\bf h_7}}\over {{\sqrt{3}}}}}
          }\,{\bf p^{18}} -
       {e^{{{{\sqrt{2}}\,{\bf h_1} +
              {\bf h_7}}\over {{\sqrt{3}}}}}
          }\,{\bf p^{23}} -
       {e^{{{{\sqrt{2}}\,{\bf h_1} -
              {\bf h_7}}\over {{\sqrt{3}}}}}
          }\,{\bf q^{17}} -
       {e^{{{{\sqrt{2}}\,{\bf h_2} +
              {\bf h_7}}\over {{\sqrt{3}}}}}
          }\,{\bf q^{24}} \right) }  } \nonumber \\
 0&=&{{c\,{\frac{\bf e^U}{r^2}}\,
     \left( {e^{{{{\sqrt{2}}\,{\bf h_1} -
              {\bf h_7}}\over {{\sqrt{3}}}}}
          }\,{\bf p^{17}} -
       {e^{{{{\sqrt{2}}\,{\bf h_2} +
              {\bf h_7}}\over {{\sqrt{3}}}}}
          }\,{\bf p^{24}} -
       {e^{{{{\sqrt{2}}\,{\bf h_2} -
              {\bf h_7}}\over {{\sqrt{3}}}}}
          }\,{\bf q^{18}} -
       {e^{{{{\sqrt{2}}\,{\bf h_1} +
              {\bf h_7}}\over {{\sqrt{3}}}}}
          }\,{\bf q^{23}} \right) }  } \nonumber \\
{\bf h^\prime_7}&=&
 {{ \frac{c}{2} \,{\frac{\bf e^U}{r^2}}\,\left( {e^{{{{\sqrt{2}}\,{\bf h_2} -
              {\bf h_7}}\over {{\sqrt{3}}}}}}\,{\bf p^{18}} -
       {e^{{{{\sqrt{2}}\,{\bf h_1} +
              {\bf h_7}}\over {{\sqrt{3}}}}}
          }\,{\bf p^{23}} +
       {e^{{{{\sqrt{2}}\,{\bf h_1} - {\bf h_7}}\over {{\sqrt{3}}}}}}\,
       {\bf q^{17}} + {e^{{{{\sqrt{2}}\,{\bf h_2} +
              {\bf h_7}}\over {{\sqrt{3}}}}}
          }\,{\bf q^{24}} \right) }  } \nonumber \\
\left( {\bf h^\prime_1} - {\bf h^\prime_2} \right) &=&
 {{ \frac{c}{{\sqrt{2}}} \,{\frac{\bf e^U}{r^2}}\,\left( {e^{{{{\sqrt{2}}\,{\bf h_1} -
              {\bf h_7}}\over {{\sqrt{3}}}}}
          }\,{\bf p^{17}} +
       {e^{{{{\sqrt{2}}\,{\bf h_2} + {\bf h_7}}\over {{\sqrt{3}}}}}
          }\,{\bf p^{24}} +
       {e^{{{{\sqrt{2}}\,{\bf h_2} -
              {\bf h_7}}\over {{\sqrt{3}}}}}
          }\,{\bf q^{18}} -
       {e^{{{{\sqrt{2}}\,{\bf h_1} +
              {\bf h_7}}\over {{\sqrt{3}}}}}
          }\,{\bf q^{23}} \right) }  } \nonumber \\
\left( {\bf h^\prime_1} + {\bf h^\prime_2} \right) &=&
 {{ \frac{c}{{\sqrt{2}}} \,{\frac{\bf e^U}{r^2}}\,\left( {e^{{{{\sqrt{2}}\,{\bf h_2} -
              {\bf h_7}}\over {{\sqrt{3}}}}}
          }\,{\bf p^{18}} + {e^{{{{\sqrt{2}}\,{\bf h_1} +
              {\bf h_7}}\over {{\sqrt{3}}}}}
          }\,{\bf p^{23}} +
       {e^{{{{\sqrt{2}}\,{\bf h_1} -
              {\bf h_7}}\over {{\sqrt{3}}}}}
          }\,{\bf q^{17}} -
       {e^{{{{\sqrt{2}}\,{\bf h_2} +
              {\bf h_7}}\over {{\sqrt{3}}}}}
          }\,{\bf q^{24}} \right) }  } \nonumber \\
\frac{dU}{dr}  &=&  - \frac{k}{4\sqrt{2}}{{{\frac{\bf e^U}{r^2}}\,\left( {e^{{{{\sqrt{2}}\,{\bf h_1} -
                       {\bf h_7}}\over {{\sqrt{3}}}}}}\,{\bf p^{17}} -
       {e^{{{{\sqrt{2}}\,{\bf h_2} + {\bf h_7}}\over
             {{\sqrt{3}}}}}}\,{\bf p^{24}} +
       {e^{{{{\sqrt{2}}\,{\bf h_2} - {\bf h_7}}\over
             {{\sqrt{3}}}}}}\,{\bf q^{18}} +
       {e^{{{{\sqrt{2}}\,{\bf h_1} + {\bf h_7}}\over
             {{\sqrt{3}}}}}}\,{\bf q^{23}} \right) }  }  \nonumber \\
\end{eqnarray}
where we have defined the coefficient
\begin{equation}
c \equiv \frac{b}{ 18 \times \sqrt{2} \, \pi }
\end{equation}
$b/a, k$ being the relative coefficient between the left and right hand
side of equations \eqn{gratra}, \eqn{BPS56eq1} respectively,
which are completely fixed by the supersymmetry
transformation rules of the $N=8$ theory \eqn{trasforma}.

From the homogeneous equations of the first order system we get:
\begin{eqnarray}
q_{17}(r)\,\exp\left[-{\vec \Lambda}_{17}\cdot {\vec h}\right]\, &=&\,
p_{18}\,\exp\left[-{\vec \Lambda}_{18}\cdot {\vec h}\right]\nonumber\\
q_{18}(r)\, \exp\left[-{\vec \Lambda}_{18}\cdot {\vec h}\right]\, &=&\,
p_{17}\, \exp\left[-{\vec \Lambda}_{17}\cdot {\vec h}\right]\nonumber\\
q_{23}(r)\, \exp\left[-{\vec \Lambda}_{23}\cdot {\vec h}\right]\, &=&\,
-p_{24}\, \exp\left[-{\vec \Lambda}_{24}\cdot {\vec h}\right]\nonumber\\
q_{24}(r)\, \exp\left[-{\vec \Lambda}_{24}\cdot {\vec h}\right]\, &=&\,
-p_{23}\, \exp\left[-{\vec \Lambda}_{23}\cdot {\vec h}\right]
\label{qpeq}
\end{eqnarray}
Then, from the Maxwell equations we get:
\begin{equation}
q_{\alpha}(r)\, =\, A_{\alpha} \, \exp\left[2{\vec \Lambda}_{\alpha}\cdot {\vec h}\right]
\end{equation}
where $A_\alpha$ are integration constants.
By substituting these into the inhomogeneous first order equations one obtains:
\begin{eqnarray}
q_{17}\, &=&\, q_{24}\, =\, p_{18}\, =\, p_{23}\, =\, 0\nonumber\\
h_{1}^{\prime}\, &=&\, -h_{2}^{\prime}\nonumber\\
h_{7}^{\prime}\, &=&\, 0
\end{eqnarray}
Introducing the field:
\begin{equation}
\phi =\sqrt{\frac{2}{3}}h_1-\frac{1}{\sqrt{3}}h_7
\end{equation}
so that:
\begin{eqnarray}
h_1\, &=&\, \sqrt{\frac{3}{2}}\left(\phi +\frac{1}{\sqrt{3}}h_7\right)\nonumber\\
h_2\, &=&\, -\sqrt{\frac{3}{2}}\left(\phi +\frac{1}{\sqrt{3}}h_7-\log{B}\right)
\nonumber\\
\phi^{\prime}\, &=&\, \sqrt{\frac{2}{3}}h_1^{\prime}\, =\, -\sqrt{\frac{2}{3}}h_2^{\prime}
\label{unosolo}
\end{eqnarray}
where $B$ is an arbitrary constant, the only independent first order equations become:
\begin{equation}
\phi^{\prime}\, =\,\frac{{c}}{\sqrt{3}}\frac{{\bf e^U}}{r^2}
\left(p_{17} e^{\phi}\, +\, p_{24} B e^{-\phi}\right)
\label{eqphi}
\end{equation}
\begin{equation}
U^\prime =\,- \frac{k}{2 \sqrt{2}}\frac{{\bf e^U}}{r^2}
\left(p_{17} e^{\phi}\, -\, p_{24} B e^{-\phi}\right)
\label{equ}
\end{equation}
and correspondingly the second order scalar field equations become:
\begin{eqnarray}
\phi^{\prime \prime}+\frac{2}{r}\phi^{\prime}\, &=&\, \frac{1}{3 \alpha^2 \pi^2}
\left(p_{17}^2 e^{2\phi}-p_{24}^2 B^2 e^{-2\phi}\right)\\
h^{\prime \prime}_7+\frac{2}{r}h^{\prime}_7\, &=&\, 0\\
\label{scaleveronesi}
\end{eqnarray}

The system of first and second order differential equations given by eqn.
\eqn{eqphi}, \eqn{equ},  the Einstein
equations  \eqn{paireinst} and the scalar fields equations
\eqn{scaleveronesi} can now be solved and gives:
\begin{eqnarray}
\phi &=& - \frac{1}{2} {\mbox {log}}\left(1+ \frac{b}{r}\right) +
\frac{1}{2} {\mbox {log}}\left(1+ \frac{d}{r}\right)
\label{phi}  \\
U &=& - \frac{1}{2} {\mbox {log}}\left(1+ \frac{b}{r}\right) -
\frac{1}{2} {\mbox {log}}\left(1+ \frac{d}{r}\right)
+  {\mbox {log}}A
\label{u}
\end{eqnarray}
 with:
 \begin{equation}
b= - \frac{1}{\pi \sqrt{2}}p_{17} \quad ; \quad d= - \frac{1}{\pi \sqrt{2}}Bp_{24}
\end{equation}
fixing at the same time the coefficients (which could be
alternatively fixed with supersymmetry techniques):

\begin{eqnarray}
\alpha^2\, &=&\, \frac{4}{3}\nonumber\\
c\, &=&\, -\frac{\sqrt{3}}{2}\nonumber\\
k \, &=& \, \sqrt{2}
\label{susi}
\end{eqnarray}
\par
This concludes our discussion of the simplified model.
\vskip 5mm

We can now identify the simplified $N=8$ model (reduced to the Cartan
subalgebra) that we have studied with a class of black holes well
studied in the literature. These are the black--hole generating
solutions of the heterotic string compactified on a six torus.
As described in \cite{extrortin}, these heterotic black--holes can be
found as solutions of the following truncated action:
\begin{eqnarray}
S^{het} &=& \int \, d^4 x \, \sqrt{-g} \, \, \Biggl \{
2 \, R + 2 \bigl [ (\partial \phi)^2 +(\partial\sigma)^2 +(\partial\rho)^2
\bigr ] \nonumber \\
&& -\frac{1}{4} \, e^{-2\phi} \, \Bigl [ e^{-2(\sigma+\rho) } (F_1)^2
 + e^{-2(\sigma-\rho) } (F_2)^2 \nonumber \\
&& + e^{ 2(\sigma+\rho) } (F_1)^2 + e^{2(\sigma-\rho) } (F_2)^2 \Bigr ]  \Biggr \}
\label{ortinact}
\end{eqnarray}
and were studied in \cite{12ref},\cite{13ref},\cite{14ref}. The
truncated action \eqn{ortinact} is nothing else but our truncated
action \eqn{reduclagra}. The translation vocabulary is given by:
\begin{equation}
\begin{array}{rclcrcl}
h_1 &=& \frac{\sqrt{3}}{2}\, \left( \sigma - \phi\right) & ; &
F_{17} &=& F_{4} \\
h_2 &=& \frac{\sqrt{3}}{2}\, \left( -\sigma - \phi\right) & ; &
F_{18} &=& F_{1} \\
h_7 &=&  \sqrt{3} \, \rho & ; &
F_{23} &=& F_{3} \\
\null &\null &  \null  & \null &
F_{24} &=& F_{2} \\
\end{array}
\label{zingarelli}
\end{equation}
As discussed in \cite{extrortin} the extreme multi black--hole solutions to the
truncated action \eqn{ortinact} depend on four harmonic functions
$H_i({\vec x})$
and for a single black hole solution the four harmonic functions are
simply:
\begin{equation}
H_i = 1+ \frac{\vert k_i \vert }{r}
\end{equation}
which introduces four electromagnetic charges. These are the four
surviving charges $p_{17},p_{23},q_{18},q_{24}$ that we have found in
our BPS saturated solution. It was observed in
\cite{extrortin} that among the general extremal solutions of this
model only a subclass are BPS saturated states, but in the way we
have derived them, namely through the Killing spinor equation, we
have automatically selected the BPS saturated ones.
\par
To make contact with the discussion in \cite{extrortin} let us
introduce the following four harmonic functions:
\begin{equation}
\begin{array}{rclcrcl}
H_{17}(r) &=& 1+\frac{\vert g^{17} \vert }{r} &;& H_{24}(r) &=&
1+\frac{\vert g^{24} \vert }{r} \\
H_{18}(r) &=& 1+\frac{\vert e_{18} \vert }{r} &;& H_{23}(r) &=&
1+\frac{\vert e_{23} \vert }{r} \\
\end{array}
\end{equation}
where $g^{17},g^{24},e_{18},e_{23}$ are four real parameters.
Translating the extremal general solution of the lagrangian \eqn{ortinact}
(see eq.s (30) of \cite{extrortin} )
into our notations we can write it as follows:
\begin{equation}
\begin{array}{rcrl}
h_1(r) &=& -\frac{\sqrt{3}}{4} & \log \,
\left[H_{17}/H_{23}\right]
\\
h_2(r) &=& -\frac{\sqrt{3}}{4} & \log \, \left[ H_{24}/ H_{18}\right] \\
h_7(r) &=& -\frac{\sqrt{3}}{4} & \log \, \left[ H_{18} \, H_{24}/ H_{17} \, H_{23} \right]
\\
U(r) &=& -\frac{1}{4} & \log \left[ H_{17} \, H_{18} \, H_{23} \,
H_{24} \right] \\
q^{18}(r) &=& -e_{18} & H_{18}^{-2}\\
q^{23}(r) &=& -e_{23} & H_{23}^{-2} \\
p^{24}    &=& g^{24} &\null \\
p^{17}    &=& g^{17} &\null \\
\end{array}
\label{idecariche}
\end{equation}
and we see that indeed $e_{18} = -q^{18}(\infty)$, $e_{23}= -q^{23}(\infty)$
are the electric charges, while $g^{24}=p^{24}$, $g^{17}=g^{17}$ are
the magnetic charges for the general extremal black--hole solution.
\par
Comparing now eq.\eqn{zingarelli} with our previous result
\eqn{unosolo} we see that having enforced the Killing spinor
equation, namely the BPS condition, we have the restrictions:
\begin{equation}
 h_1+h_2 = \mbox{const} \quad ; \quad h_7 = \mbox{const}
\end{equation}
which yield:
\begin{equation}
H_{17}^{2} = H_{18}^{2} \quad ; \quad H_{23}^{2} = H_{24}^{2}
\end{equation}
and hence
\begin{equation}
  e_{18} = g^{17} \quad ; \quad e_{23} = g^{24}
\end{equation}
Hence the BPS condition imposes that the electric charges are
pairwise equal to the magnetic charges.

\section{Solvable Lie algebra representation}
Utilizing a well established mathematical
framework \cite{helgason} in all extended supergravities the scalar
coset manifold $U/H$ can be identified with the group manifold of a
normed solvable Lie algebra:
\begin{equation}
U/H \, \sim \, \exp \left[ Solv \right]
\label{solvidea}
\end{equation}
The representation of the supergravity scalar manifold ${\cal M}_{scalar}= U/H$
as the group manifold associated with a {\it  normed solvable Lie algebra}
introduces a one--to--one correspondence between the scalar fields $\phi^I$ of
supergravity and the generators $T_I$ of the solvable Lie algebra $Solv\, (U/H)$.
Indeed the coset representative $\IL (U/H)$ of the homogeneous space $U/H$ is
identified with:
\begin{equation}
\IL(\phi) \, =\, \exp [ \phi^I \, T_I ]
\label{cosrep1}
\end{equation}
where $\{ T_I \}$ is a basis of $Solv\, (U/H)$.
\par
As a consequence of this fact the tangent bundle to the scalar manifold $T{\cal M}_{scalar}$
is identified with the solvable Lie algebra:
\begin{equation}
T{\cal M}_{scalar} \, \sim \,Solv \, (U/H)
\label{cosrep2}
\end{equation}
and any algebraic property of the solvable algebra has a corresponding physical interpretation in
terms of string theory massless field modes.
\par
Furthermore, the local differential geometry of the scalar manifold is described
 in terms of the solvable Lie algebra structure.
Given the euclidean scalar product on $Solv$:
\begin{eqnarray}
  <\, , \, > &:& Solv \otimes Solv \rightarrow \IR
\label{solv1}\\
<X,Y> &=& <Y,X>\label{solv2}
\end{eqnarray}
the covariant derivative with respect to the Levi Civita connection is given by
the Nomizu operator \cite{alex}:
\begin{equation}
\forall X \in Solv : \IL_X : Solv \to Solv
\end{equation}
\begin{eqnarray}
  \forall X,Y,Z \in Solv & : &2 <Z,\IL_X Y> \nonumber\\
&=& <Z,[X,Y]> - <X,[Y,Z]> - <Y,[X,Z]>
\label{nomizu}
\end{eqnarray}
and the Riemann curvature 2--form is given by the commutator of two Nomizu
operators:
\begin{equation}
 <W,\{[\IL_X,\IL_Y]-\IL_{[X,Y]}\}Z> = R^W_{\ Z}(X,Y)
\label{nomizu2}
\end{equation}

In the case of maximally extended supergravities in $D=10-r$ dimensions the scalar
manifold has a universal structure:
\begin{equation}
 { U_D\over H_D}  = {E_{r+1(r+1)} \over H_{r+1}}
\label{maximal1}
\end{equation}
where the Lie algebra of the $U_D$--group $E_{r+1(r+1)} $ is the
maximally non compact real section of the exceptional $E_{r+1}$ series  of the simple complex
Lie Algebras
and $H_{r+1}$ is its maximally compact subalgebra \cite{cre}.
As we noted  in references \cite{noialtri,noialtri2},
the manifolds $E_{r+1(r+1)}/H_{r+1}$
share the distinctive  property of being non--compact homogeneous spaces of maximal rank
$r+1$, so that the associated solvable Lie algebras,
 such that ${E_{r+1(r+1)}}/{H_{r+1}} \, = \, \exp \left [ Solv_{(r+1)} \right ]
$,  have the particularly simple structure:
\begin{equation}
Solv\, \left ( E_{r+1}/H_{r+1} \right )\, = \, {\cal H}_{r+1} \, \oplus_{\alpha \in
\Phi^+(E_{r+1})} \, \IE^\alpha
\label{maxsolv1}
\end{equation}
where $\IE^\alpha \, \subset \, E_{r+1}$ is the 1--dimensional subalgebra associated
with the root $\alpha$
and $\Phi^+(E_{r+1})$ is the positive part of the $E_{r+1}$--root--system.
\par
The generators of the solvable Lie algebra  are in one to one
correspondence with the scalar fields of the theory.
Therefore they can be characterized as NS-NS or R-R
depending on their origin in compactified string theory. To identify them
algebraically it suffices to select the appropriate subgroup $T_{r}\subset E_{r+1}$
of the U-duality group that acts as a group of T--dualities. For instance in
compactifications of TypeIIA superstrings we have $T_{r}=SO(r,r)$. From the
algebraic point of view the generators of the solvable algebra are then of
three possible types:
\begin{enumerate}
\item {Cartan generators }
\item { Roots that belong to the adjoint representation of the
$T_r  \subset E_{r+1(r+1)}$ subalgebra (= the T--duality algebra) }
\item {Roots which are weights of an irreducible representation
 of the T--duality algebra $T_r$}
\end{enumerate}
The scalar fields associated with generators of type 1 and 2 in the above
list are Neveu--Schwarz fields while the fields of type 3 are
Ramond--Ramond fields.
\subsection{${\bf  U(1)\times SU(2) \times SU(6) \subset SU(8) \subset E_{7(7)}}$}
\label{su8ine7}
In order to make formulae \eqn{pappaequa} \eqn{22026eq},\eqn{Ueq} explicit and
in order to derive the solvable Lie algebra decompositions we are
interested in  a preliminary work  based  on standard Lie
algebra techniques.
\par
The  ingredients that we have already tacitly used in the previous sections
and that are needed for a thoroughful discussion of the solavable Lie
algebra splitting in \eqn{7in3p4} are:
\begin{enumerate}
\item{The explicit listing of all the positive roots of the $E_{7(7)}$
Lie algebra}
\item{The explicit listing of all the weight vectors of the fundamental
${\bf 56}$ representation of $E_{7(7)}$}
\item{The explicit construction of the $56 \times 56$ matrices
realizing the 133 generators of $E_{7(7)}$ real Lie algebra in the fundamental
representation}
\item{The canonical Weyl--Cartan decomposition of the $SU(8)$
maximally compact subalgebra of $E_{7(7)}$. This involves the
construction of a Cartan subalgebra of $A_7$ type made out of
$E_{7(7)}$ step operators and the construction of all $A_7$
step operators also in terms of suitable combinations of
$E_{7(7)}$ step operators.}
\item{The determination of the embedding ${\bf SU(2)\times U(6)\subset
SU(8)}\subset\Es$.}
\item{The decomposition of the maximal non compact subspace $\IK\subset\Es$
with respect to ${\bf U(1)} \times {\bf SU(2)\times SU(6)}$: $${\bf 70}\rightarrow {\bf
(1,1,15)}\oplus {\bf\bar{(1,1,15)}}\oplus {\bf (1,2,20)}$$}
\item{Using the ${\bf 56}$ representation of $\Es$ in the {\bf Usp}-basis,
the construction of the subalgebra ${\bf SO^{*}(12)}$ }
\end{enumerate}
The work-plan described in the above points has been completed by
means of a computer programme written in  MATHEMATICA \cite{Wolfram}.
In the present section we just outline the logic of our calculations
and we describe the results that are summarized in various tables
in appendix B. In particular we explain the method to
generate the matrices of the ${\bf 56}$ representation whose explicit form is
the basic tool of our calculations but that are too large and
too little instructive to display on paper.
\par
\subsection{Roots and Weights and the fundamental representation of $E_{7(7)}$}
\label{pesanti}
Let us begin with the construction of the fundamental representation
of the U--duality group. \par
In \cite{noialtri,noialtri2} we showed that the $63$--dimensional
positive part $\Phi^+(E_7)$ of the $E_7$ root space can be decomposed
as follows:
\begin{equation}
\Phi^+(E_7) = \ID^+_1 \oplus \ID^+_2  \oplus \ID^+_3  \oplus \ID^+_4
\oplus \ID^+_5  \oplus \ID^+_6
\label{filtro}
\end{equation}
where $\ID^+_{r}$ are the maximal abelian ideals of the nested
U--duality algebras $\dots \subset E_{r(r)}\subset
E_{r+1(r+1)}\subset \dots$ in dimension $D=10-r$ ($\ID^+_{r}$ being
the  ideal of $E_{r+1(r+1)}$). The dimensions of these abelian
ideals is:
\begin{equation}
\begin{array}{rclcrcl}
\mbox{dim}\,\ID_1 &=& 1 &;& \mbox{dim}\,\ID_2 &=&  3  \\
\mbox{dim}\,\ID_3 &=& 6 &;& \mbox{dim}\,\ID_4 &=&  10  \\
\mbox{dim}\,\ID_5 &=& 16 &;& \mbox{dim}\,\ID_6 &=& 27   \\
\end{array}
\label{didimens}
\end{equation}
The filtration \eqn{filtro} provides a convenient way to
enumerate the $63$ positive roots which in \cite{noialtri2}
were associated in one--to--one way with the massless bosonic fields of
compactified string theory (for instance the TypeIIA theory).
We name the roots as follows:
\begin{equation}
 {\vec \alpha }_{i,j} \, \in \, \ID_i \quad ; \quad \cases { i=1,
 \dots , 6 \cr
 j=1,\dots , \mbox{dim} \, \ID_i \cr }
 \label{filtroname}
\end{equation}
Each positive root can be decomposed along a basis of simple roots
$\alpha_\ell$ (i=1,\dots, 7):
\begin{equation}
{\vec \alpha }_{i,j} = n_{i,j}^\ell \, \alpha_\ell  \, \quad  n_{i,j}^\ell \in \ZZ
\end{equation}
It turns out that as simple roots we can choose:
\begin{equation}
\begin{array}{rclcrclcrcl}
\alpha_1 & = & {\vec \alpha }_{6,2} & ; & \alpha_2 & = & {\vec \alpha }_{5,2}
& ; & \alpha_3 & = & {\vec \alpha }_{4,2} \\
\alpha_4 & = & {\vec \alpha }_{3,2} & ; & \alpha_5 & = & {\vec \alpha }_{2,2}
& ; & \alpha_6 & = & {\vec \alpha }_{2,1} \\
\alpha_7 & = & {\vec \alpha }_{1,1} & \null & \null & \null & \null
& \null & \null & \null & \null \\
\end{array}
\label{simplerut}
\end{equation}
Having fixed this basis, each root is intrinsically identified by its
Dynkin labels, namely by its integer valued components in the basis
\eqn{simplerut}. The listing of all positive roots is given in
table \ref{dideals} were we give their name \eqn{filtroname} according to the abelian
ideal filtration, their Dynkin labels and the correspondence with massless
fileds in a TypeIIA toroidal compactification.
\par
Having identified the roots, the next step for the construction of
real fundamental representation $SpD(56)$  of our
U--duality Lie algebra $E_{7(7)}$ is the knowledge of
the corresponding weight vectors ${\vec W}$.\par
A particularly relevant property of the maximally non--compact
real sections of a simple complex Lie algebra is that all
its irreducible representations are real.  $E_{7(7)}$ is the
maximally non compact real section of the complex Lie algebra $E_7$, hence
all its irreducible representations $\Gamma$  are real.
This implies that if an element of the  weight lattice ${\vec W} \, \in \, \Lambda_w$ is
a weight of a given irreducible representation
${\vec W}\in \Gamma$ then also its negative is a weight of the
same representation: $-{\vec W}\in \Gamma$. Indeed changing sign to
the weights corresponds to complex conjugation.
\par
According to standard Lie algebra lore
every irreducible representation of a simple Lie algebra $\IG$ is identified by a unique
{\it highest} weight ${\vec W}_{max}$. Furthermore all weights can be expressed as
integral non--negative linear  combinations of the {\it simple}
weights ${\vec W}_{\ell}\,(\ell=1,...,r=\mbox{rank}(\IG)) $, whose components
are named the Dynkin labels of the weight.
The simple weights ${\vec W}_{i}$ of $\IG$ are the generators of the
dual lattice to the root lattice and are defined by the condition:
\begin{equation}
\frac{2 ({\vec W}_{i}\, ,\, {\vec \alpha}_{j})}{({\vec \alpha}_{j}\, ,\,
{\vec \alpha}_{j})}=\delta_{ij}
\end{equation}
In the simply laced $E_{7(7)}$ case, the previous equation simplifies as follows
\begin{equation}
({\vec W}_{i}\, ,\, {\vec \alpha}_{j})=\delta_{ij}
\label{simw}
\end{equation}
where ${\vec \alpha}_{j}$ are the the simple roots.
Using eq.\eqn{simplerut}, table \ref{dideals} and the Dynkin diagram of
$E_{7(7)}$ (see fig.\ref{stande7}) from eq.\eqn{simw} we can easily
obtain the explicit expression of the simple weights.
\iffigs
\begin{figure}
\caption{$E_7$ Dynkin diagram}
\label{stande7}
\epsfxsize = 10cm
\epsffile{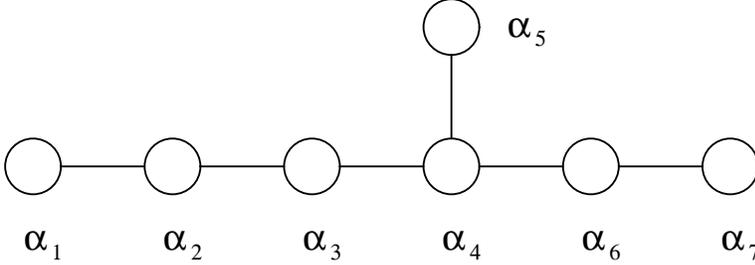}
\vskip -0.1cm
\unitlength=1mm
\end{figure}
\fi
The Dynkin labels of the highest weight of an irreducible
representation $\Gamma$ gives the Dynkin labels of the
representation. Therefore the representation is usually denoted by
$\Gamma[n_1,...,n_{r}]$. All the weights ${\vec W}$ belonging
to the representation $\Gamma$ can be described by $r$ integer non--negative numbers
$q^\ell$ defined by the following equation:
\begin{equation}
{\vec W}_{max}-{\vec W}=\sum_{\ell=1}^{r}q^\ell{\vec \alpha}_{\ell}
\label{qi}
\end{equation}
where $\alpha_\ell$ are the simple roots.
According to this standard formalism the fundamental real representation $SpD(56)$
of $E_{7(7)}$ is $\Gamma[1,0,0,0,0,0,0]$
and the expression of its weights in terms of $q^\ell$ is given in table
\ref{e7weight}, the highest weight being ${\vec W}^{(51)}$.
\par
We can now explain the specific ordering of the weights we have
adopted.
\par
First of all we have separated the $56$ weights in two
groups of $28$ elements so that the first group:
\begin{equation}
{\vec \Lambda}^{(n)}={\vec W}^{(n)} \quad n=1,...,28
\label{elecweight}
\end{equation}
are the weights for  the irreducible {\bf 28} dimensional representation of the
{\sl electric} subgroup {\bf $SL(8,\IR) \subset E_{7(7)}$}.
The remaining group of $28$ weight vectors are the weights for the
transposed representation of the same group that we name ${\bf \bar{28}}$.
\par
Secondly the $28$ weights ${\vec \Lambda}$
have been arranged according to the decomposition with respect to the
T--{\it duality} subalgebra $SO(6,6)\subset E_7(7)$:  the first $16$
correspond to R--R vectors and are the weights of the spinor
representation of $SO(6,6)$ while the last $12$ are associated with N--S fields
and correspond to the weights of the vector representation of $SO(6,6)$.
\par
Eq.\eqn{elecweight} makes explicit the adopted labeling for the
electric gauge fields $A_\mu^{{\vec \Lambda}  }$ and their field
strenghts $F_{\mu\nu}^{{\vec \Lambda}}$ adopted throughout the
previous sections of the paper.
\par
Equipped with the weight vectors we can now proceed to the explicit
construction of the ${\bf SpD(56)}$ representation
of $E_{7(7)}$. In our construction the basis vectors
are the $56$ weights, according to the enumeration of table \ref{e7weight}.
What we need are the $56\times 56$ matrices associated with the  $7$
Cartan generators $H_{{\vec \alpha}_i}$ ($i=1,\dots , 7$) and with
the $126$ step operators $E^{\vec \alpha}$ that are defined by:
\begin{eqnarray}
\left[ SpD_{56}\left( H_{{\vec \alpha}_i} \right)\right]_{nm}&  \equiv &  \langle
{\vec W}^{(n)} \vert \,  H_{ {\vec \alpha}_i } \,\vert {\vec W}^{(m)}
\rangle \nonumber\\
\left[ SpD_{56}\left( E^{ {\vec \alpha} } \right)\right]_{nm}&  \equiv &  \langle
{\vec W}^{(n)} \vert \,  E^{ {\vec \alpha} }  \,\vert {\vec W}^{(m)}
\rangle
\label{sp56defmat}
\end{eqnarray}
Let us begin with the Cartan generators. As a basis of the
Cartan subalgebra we use the generators $H_{\vec \alpha_i}$ defined by the
commutators:
\begin{equation}
\left[ E^{{\vec \alpha }_i}, E^{-{\vec \alpha }_i} \right] \, \equiv
\, H_{\vec \alpha_i}
\label{cartbadefi}
\end{equation}
In the $SpD(56)$ representation the corresponding matrices are
diagonal and of the form:
\begin{equation}
\langle {\vec W}^{(p)} \vert\, H_{\vec \alpha_{i}}\, \vert
{\vec W}^{(q)}\rangle\, =\,\left({\vec W}^{(p)},{\vec \alpha_{i}}\right)
\delta_{p\, q}\quad ; \quad ( p,q\, =\, 1,...,56)
\label{cartane7}
\end{equation}
The scalar products
\begin{equation}
\left({\vec \Lambda}^{(n)} \, \cdot \, {\vec h},
-{\vec \Lambda}^{(m)} \, \cdot \, {\vec h}\right)\, =\, \left({\vec W}^{(p)}
 \, \cdot \,
{\vec h}\right) \quad ; \quad(n,m=1,...,28\, ;\, p=1,...,56)
\end{equation}
appearing in the definition
\ref{LSpD} of the coset representative restricted to the Cartan fields, are
therefore to be understood in the following way:
\begin{equation}
{\vec W}^{(p)} \, \cdot \, {\vec h}\, =\, \sum_{i=1}^{7}\left({\vec W}^{(p)},
{\vec \alpha_{i}}\right)h^i
\end{equation}
The explicit form of these scalar products is given in table
\ref{scalaprod}  \par
Next we construct the matrices associated with the step operators. Here the first
observation is that it suffices to consider the positive roots. Indeed because
of the reality of the representation, the matrix associated with the
negative of a root is just the transposed of that associated with the
root itself:
\begin{equation}
E^{-\alpha} = \left[ E^\alpha \right]^T \, \leftrightarrow \, \langle
{\vec W}^{(n)} \vert \,  E^{- {\vec \alpha} }  \,\vert {\vec W}^{(m)} \rangle \, = \,
\langle {\vec W}^{(m)} \vert \,  E^{ {\vec \alpha} }  \,\vert {\vec W}^{(n)} \rangle
\label{transpopro}
\end{equation}
The method we have followed to obtain the matrices for all the
positive roots is that of constructing first the $56\times 56$
matrices for the step operators $E^{\vec \alpha_{\ell}}\, (\ell=1,...,7)$
associated with the simple roots and then generating all the others
through their commutators. The construction rules for the $SpD(56)$
representation of  the six operators $E^{\alpha_{\ell}} \, (\ell\neq 5)$
are:
\begin{equation}
\ell \, \neq \, 5 \quad \,
\Biggl \{ \matrix {\langle {\vec W}^{(n)} \vert\, E^{\vec \alpha_{\ell}}\, \vert
{\vec W}^{(m)}\rangle & = & \delta_{{\vec W}^{(n)},
{\vec W}^{(m)}+{\vec \alpha}_\ell} &;& n,m=1,\dots, 28 \cr
\langle {\vec W}^{(n+28)} \vert\, E^{\vec \alpha_{\ell}}\, \vert
{\vec W}^{(m+28)}\rangle & = & -\delta_{{\vec W}^{(n+28)},
{\vec W}^{(m+28)}+{\vec \alpha}_\ell} & ; & n,m=1,\dots, 28 \cr }
\label{repineq5}
\end{equation}
The six simple roots ${\vec \alpha_{\ell}} $ with $ \ell \neq 5$
belong also to the the Dynkin diagram of the electric subgroup {\bf SL(8,\IR)}
(see fig.\ref{compsu8}).  Thus their shift
operators have a block diagonal action on the {\bf 28} and ${\bf \bar{28}}$
subspaces of the $SpD(56)$ representation that are irreducible
under the electirc subgroup. Indeed from eq.\eqn{repineq5} we conclude
that:
\begin{equation}
\ell \, \neq \, 5 \quad \,SpD_{56}\left(E^{{\vec \alpha}_\ell} \right)=
\left(\matrix { A[{{\vec \alpha}_\ell}]
& {\bf 0} \cr {\bf 0} & - A^T[{{\vec \alpha}_\ell}] \cr} \right)
\label{spdno5}
\end{equation}
the $28 \times 28$ block  $A[{{\vec \alpha}_\ell}]$ being defined
by the first line of eq.\eqn{repineq5}.\par
On the contrary the operator $E^{\vec \alpha_{5}}$, corresponding to the only
root of the $E_7$ Dynkin diagram that is not also part of the $A_7$
diagram is represented by a matrix whose non--vanishing $28\times 28$ blocks
are off--diagonal. We have
\begin{equation}
SpD_{56}\left(E^{{\vec \alpha}_5} \right)=\left(\matrix { {\bf 0} & B[{{\vec \alpha}_5}]
\cr  C[{{\vec \alpha}_5}] & {\bf 0} \cr} \right)
\label{spdyes5}
\end{equation}
where both $B[{{\vec \alpha}_5}]=B^T[{{\vec \alpha}_5}]$ and
$C[{{\vec \alpha}_5}]=C^T[{{\vec \alpha}_5}]$ are symmetric $28
\times 28$ matrices. More explicitly the  matrix
 $SpD_{56}\left(E^{{\vec \alpha}_5} \right)$
is given by:
\begin{eqnarray}
&& \langle {\vec W}^{(n)} \vert\, E^{\vec \alpha_{5}}\, \vert
{\vec W}^{(m+28)}\rangle \, =\,  \langle {\vec W}^{(m)} \vert\,
E^{\vec \alpha_{5}}\, \vert {\vec W}^{(n+28)}\rangle \nonumber \\
&& \langle {\vec W}^{(n+28)} \vert\, E^{\vec \alpha_{5}}\, \vert
{\vec W}^{(m)}\rangle \, =\,  \langle {\vec W}^{(m+28)} \vert\,
E^{\vec \alpha_{5}}\, \vert {\vec W}^{(n)}\rangle
\label{sim5}
\end{eqnarray}
with
\begin{equation}
\begin{array}{rcrcrcrcrcr}
  \langle {\vec W}^{(7)} \vert\, E^{\vec \alpha_{5}}\, \vert
{\vec W}^{(44)}\rangle & =& -1 & \null &
  \langle {\vec W}^{(8)} \vert\, E^{\vec \alpha_{5}}\, \vert
{\vec W}^{(42)}\rangle & = & 1  & \null &
   \langle {\vec W}^{(9)} \vert\, E^{\vec \alpha_{5}}\, \vert
{\vec W}^{(43)}\rangle & = & -1  \\
   \langle {\vec W}^{(14)} \vert\, E^{\vec \alpha_{5}}\, \vert
{\vec W}^{(36)}\rangle & = & 1 & \null &
   \langle {\vec W}^{(15)} \vert\, E^{\vec \alpha_{5}}\, \vert
{\vec W}^{(37)}\rangle & = & -1 & \null &
 \langle {\vec W}^{(16)} \vert\, E^{\vec \alpha_{5}}\, \vert
{\vec W}^{(35)}\rangle & = & -1  \\
   \langle {\vec W}^{(29)} \vert\, E^{\vec \alpha_{5}}\, \vert
{\vec W}^{(6)}\rangle  & = & -1 & \null &
   \langle {\vec W}^{(34)} \vert\, E^{\vec \alpha_{5}}\, \vert
{\vec W}^{(1)}\rangle  & = & -1 & \null &
  \langle {\vec W}^{(49)} \vert\, E^{\vec \alpha_{5}}\, \vert
{\vec W}^{(28)}\rangle & = & 1 \\
  \langle {\vec W}^{(50)} \vert\, E^{\vec \alpha_{5}}\, \vert
{\vec W}^{(27)}\rangle & = & -1 & \null &
   \langle {\vec W}^{(55)} \vert\, E^{\vec \alpha_{5}}\, \vert
{\vec W}^{(22)}\rangle & = &  -1 & \null &
   \langle {\vec W}^{(56)} \vert\, E^{\vec \alpha_{5}}\, \vert
{\vec W}^{(21)}\rangle & =& 1 \\
\end{array}
\label{sim5bis}
\end{equation}
In this way we have completed the construction of the $E^{{\vec \alpha}_\ell}$
operators associated with simple roots. For the matrices associated
with higher roots we just proceed iteratively in the following way.
As usual we organize the roots by height :
\begin{equation}
{\vec \alpha}=n^\ell \, {\vec \alpha}_\ell \quad \rightarrow
\quad \mbox{ht}\,{\vec \alpha} \, = \, \sum_{\ell=1}^{7} n^\ell
\label{altezza}
\end{equation}
and for the roots $\alpha_i + \alpha_j$ of height $\mbox{ht}=2$ we
set:
\begin{equation}
SpD_{56} \left( E^{ \alpha _i + \alpha _j} \right) \equiv \left[
SpD_{56}\left(E^{\alpha _i} \right) \, , \,
SpD_{56}\left(E^{\alpha _i} \right) \right] \quad ; \quad i<j
\label{alto2}
\end{equation}
Next for the roots of $\mbox{ht}=3$ that can be written as $\alpha_i
+ \beta $ where $\alpha_i$ is simple and $\mbox{ht}\, \beta\, =\, 2$
we write:
\begin{equation}
SpD_{56} \left( E^{ \alpha _i + \beta} \right) \equiv \left[
SpD_{56}\left(E^{\alpha _i} \right) \, , \,
SpD_{56}\left(E^{\beta} \right) \right]
\label{alto3}
\end{equation}
Obtained the matrices for the roots of $\mbox{ht}=3$ one proceeds in
a similar way for those of the next height and so on up to exhaustion
of all the $63$ positive roots.
\par
This concludes our description of the algorithm by means of which our
computer programme constructed all the $70$ matrices spanning the
solvable Lie algebra $Solv_7$ in the $SpD(56)$ representation. Taking
into account property \eqn{transpopro} once the representation of the
solvable Lie algebra is given, also the remaining $63$ operators
corresponding to negative roots are also given.
\par
The next point in our programme is the organization of the
maximal compact subalgebra $SU(8)$ in a canonical Cartan Weyl basis.
This is instrumental for a decomposition of the full algebra and of
the solvable Lie algebra in particular into irreducible
representations of the subgroup $U(1)\times SU(2) \times SU(6)$.
\subsection{Cartan Weyl decomposition of the maximal compact subalgebra $SU(8)$}
The Lie algebra $\IG$ of $\Es$ is written, according to the Cartan decomposition,
in the form:
\begin{equation}
\IG=\IH\oplus \IK
\label{chicco}
\end{equation}
where $ \IH$ denotes its maximal {\sl compact} subalgebra (i.e. the Lie algebra of
${\bf SU(8)}$) and $\IK$ its maximal {\sl non-compact} subspace.
Starting from the knowledge of the $\Es$ generators in the symplectic
$SpD(56)$ representation, for brevity now denoted  $H_{\alpha_{i}}\quad E^{\alpha}$,
the generators in $\IH$ and in $\IK$ are obtained from the following identifications:
\begin{eqnarray}
\IH&=&\{ E^{\alpha}-E^{-\alpha}\}=\{ E^{\alpha}-(E^{\alpha})^T\}\\
\IK&=&\{H_{\alpha_{i}};\quad E^{\alpha}+E^{-\alpha}\}=\{
H_{\alpha_{i}};\quad E^{\alpha}+(E^{\alpha})^T \}
\label{compnocomp}
\end{eqnarray}
In eq. \eqn{compnocomp} what is actually meant is that both $\IH$ and $\IK$
are the vector spaces generated by the linear combinations with
{\sl real} coefficients of the specified generators.
\par
In order to find out the generators belonging to
the subalgebra ${\bf U(1) \times SU(2)\times SU(6)}$  within $\IH$
the generators of $\IH$ have to be rearranged according to the canonical form of
the ${\bf SU(8)}$ algebra. This was achieved by first fixing seven commuting
matrices in $\IH$ to be the Cartan generators of ${\bf SU(8)}$ and then diagonalizing
with a computer programme their adjoint action over $\IH$. Their eigenmatrices were identified
with the shift operators of ${\bf SU(8)}$. In the sequel we will use
the following notation: $a$ will denote a generic root of $\IH$ of the form
$a=\pm\epsilon_i \pm \epsilon_j$, $E^a$ the corresponding shift operator,
${\cal H}_{a_i}$ the Cartan generator associated with the simple root $a_i$ and
$B^{\alpha_{i,j}}$ the compact combination $E^{\alpha_{i,j}}-E^{-\alpha_{i,j}}$ where
$ \alpha_{i,j} $ is the $j^{th}$ positive root in the $i^{th}$ abelian
ideal $\ID_i \quad i=1,...,6 $ of $ \Es $, according to the enumeration
of table \ref{dideals}. \par
A basis ${\cal H}_i$ of Cartan operators  was chosen as follows:
\begin{eqnarray}
{\cal H}_1 &=&E^{{\vec \alpha}_{2,1} } - E^{-{\vec \alpha}_{2,1}}\nonumber \\
{\cal H}_2 &=&E^{{\vec \alpha}_{2,2} } - E^{-{\vec \alpha}_{2,2}}\nonumber \\
{\cal H}_3 &=&E^{{\vec \alpha}_{4,1} } - E^{-{\vec \alpha}_{4,1}}\nonumber \\
{\cal H}_4 &=&E^{{\vec \alpha}_{4,2} } - E^{-{\vec \alpha}_{4,2}}\nonumber \\
{\cal H}_5 &=&E^{{\vec \alpha}_{6,1} } - E^{-{\vec \alpha}_{6,1}}\nonumber \\
{\cal H}_6 &=&E^{{\vec \alpha}_{6,2} } - E^{-{\vec \alpha}_{6,2}}\nonumber \\
{\cal H}_7 &=&E^{{\vec \alpha}_{6,11} } - E^{-{\vec \alpha}_{6,11}}
\label{su8cartan}
\end{eqnarray}
The reason for this choice is that the seven roots:
\begin{equation}
\begin{array}{ccccc}
 \{1,2,2,2,1,1,0\}& \leftrightarrow & {\vec \alpha}_{6,1} & = &
 \epsilon _1 + \epsilon _2  \\
 \{1,0,0,0,0,0,0\}& \leftrightarrow & {\vec \alpha}_{6,2} & = &
 \epsilon _1 - \epsilon _2  \\
\{0,0,1,2,1,1,0\}& \leftrightarrow & {\vec \alpha}_{4,1} & = &
 \epsilon _3 + \epsilon _4  \\
\{0,0,1,0,0,0,0\}& \leftrightarrow & {\vec \alpha}_{4,1} & = &
 \epsilon _3 - \epsilon _4  \\
 \{0,0,0,0,1,0,0\}& \leftrightarrow & {\vec \alpha}_{2,1} & = &
 \epsilon _5 + \epsilon _6  \\
\{0,0,0,1,0,0,0\}& \leftrightarrow & {\vec \alpha}_{2,2} & = &
 \epsilon _5 -\epsilon _6  \\
\{1,2,3,4,2,3,2\}& \leftrightarrow & {\vec \alpha}_{6,11} & = &
 \sqrt{2} \, \epsilon_7 \\
\end{array}
\label{spiegcart}
\end{equation}
are all orthogonal among themselves as it is evident by the last
column of eq.\eqn{spiegcart} where $\epsilon_i$ denote the unit
vectors in a Euclidean $7$--dimensional space.
\par
The roots ${\vec a}$ were obtained by arranging into a vector the seven
eigenvalues associated with each ${\cal H}_i$ for a fixed
eigenmatrix $E^{\vec a}$. Following a very well known procedure,
the {\sl positive} roots were computed
as those vectors ${\vec a}$ that a positive projection along an arbitrarely fixed direction
(not parallel to any of them) and among them the simple roots ${\vec a}_i$ were
identified with the undecomposable ones \cite{Humphrey}. Finally the Cartan
generator corresponding to a generic root ${\vec a}$ was worked out using
the following expression:
\begin{equation}
{\cal H}_{a}=a^j {\cal H}_{j}
\end{equation}
Once the generators of the ${\bf SU(8)}$ algebra were written in the
canonical form, the ${\bf U(1)\times}$ \hfill \break ${\bf SU(2)\times U(6)}$
subalgebra could be
easily extracted. Choosing ${\vec a}_1$ as the root of ${\bf SU(2)}$ and
${\vec a}_i \quad i=3,...,7 $ as the simple roots of ${\bf SU(6) }$, the ${\bf U(1)}$
generator was found to be the following combination of Cartan generators:
\begin{equation}
{\cal H}_{U(1)}= -3{\cal H}_{a_1}-6{\cal H}_{a_2}-5{\cal H}_{a_3}-4{\cal H}_{a_4}
-3{\cal H}_{a_5}-2{\cal H}_{a_6}-{\cal H}_{a_7}
\end{equation}
A suitable combination of the shift operators $E^{\vec a}$ allowed to define
the proper real compact form of the generators of ${\bf SU(8)}$,
denoted by $X^{a},\quad Y^{a}$. By definition, these latter fulfill the following
commutation rules:
\begin{eqnarray}
\left [{\cal H}_{i},X^{a}\right]&=&a^{i} Y^{a} \\
\left[{\cal H}_{i},Y^{a}\right]&=&-a^{i} X^{a} \\
\left[X^{a},Y^{a}\right]&=&a^{i} {\cal H}_{i}
\end{eqnarray}
In tables \ref{su8rutte}, \ref{su8rutte2}, \ref{su8rutte3}
the explicit expressions of the generators
$X^a$ and $Y^a$ are dislpayed as linear combinations of the step
operators $B^{i,j}$. These latter are labeled
according to the labeling of the $E_{7(7)}$ roots  as given in   table \ref{dideals}
where they are classified by the abelian ideal  filtration.
The labeling of the $SU(8)$ positive roots is the standard one
according to their height.
Calling ${\vec a}_1,\dots ,{\vec a}_7$ the simple roots, the full
set of the $28$ positive roots is the following one:
\begin{equation}
{\vec a}_{i,i+i,\dots,j-1,j} \, = \,{\vec a}_{i} +
 \,{\vec a}_{i+1} + \dots +{\vec a}_{j-1}+{\vec a}_{j}
 \quad ; \quad \forall 1\le i <j\le 7
 \label{allroots}
\end{equation}
We stress that the non--compact $A_7$ subalgebra $SL(8,\IR)$ of $E_{7(7)}$ is
regularly embedded, so that it shares the same Cartan subalgebra
and its roots are vectors in the same $7$--dimensional space as
the roots of $E_{7(7)}$. On the other hand the compact
$A_7$ subalgebra of $SU(8)$ is irregularly embedded  and its
Cartan subalgebra has actually intersection zero with the Cartan
subalgebra of $E_{7(7)}$. Hence the $SU(8)$ roots are vectors in
a $7$--dimensional totally different from the space where the
$E_{7(7)}$ roots live. Infact the Cartan generators of $SU(8)$
have been written as linear combinations of the step operators
$E_{7(7)}$. The difference is emphasized in fig.\ref{compsu8}
\iffigs
\begin{figure}
\caption{$A_7$ subalgebras of $E_{7(7)}$}
\label{compsu8}
\epsfxsize = 10cm
\epsffile{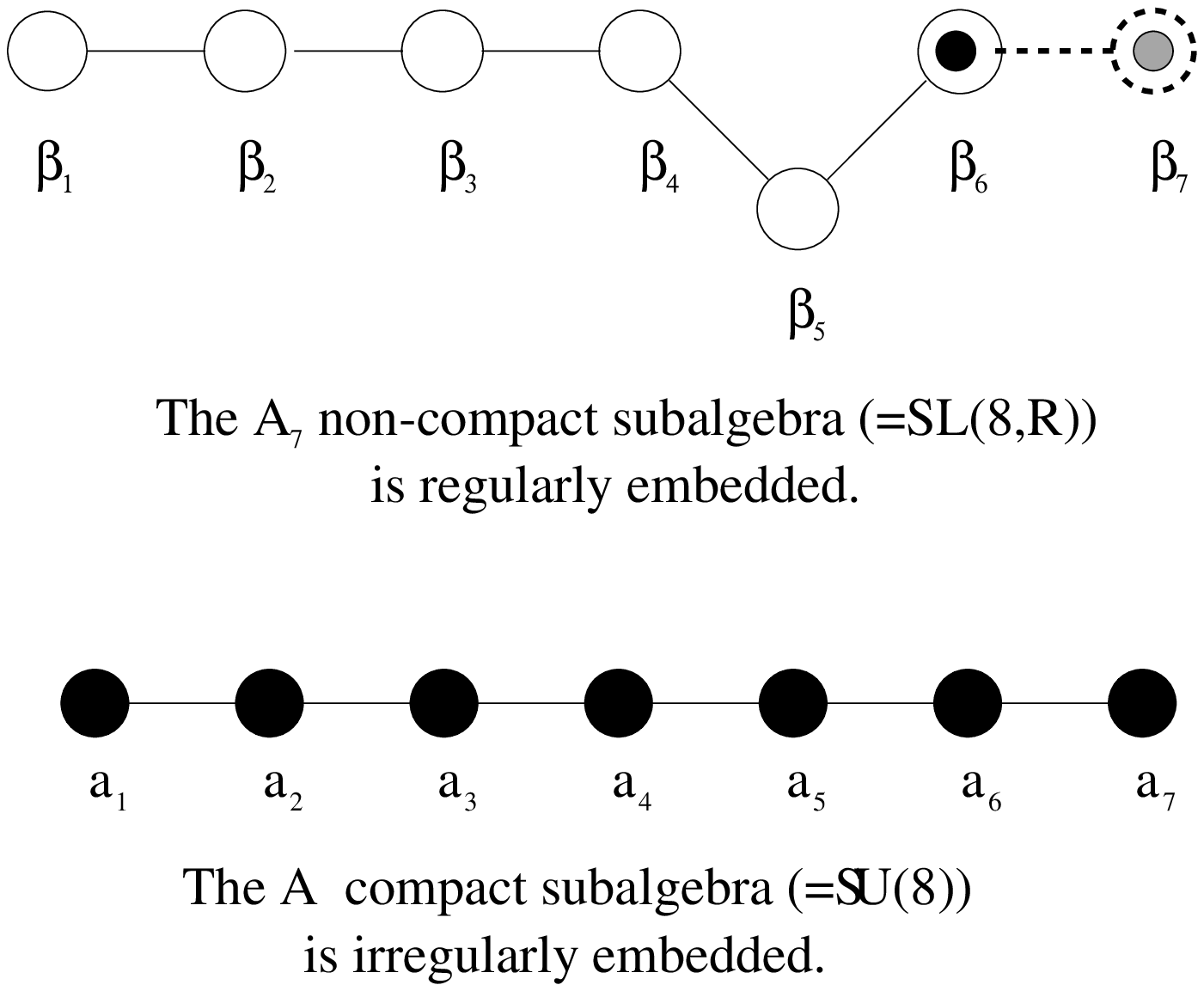}
\vskip -0.1cm
\unitlength=1mm
\end{figure}
\fi
\subsection{The ${\bf UspY(56)}$ basis for fundamental representation of
$E_{7(7)}$}
As outlined in the preceeding subsection, the generators of ${\bf U(1) \times SU(2)\times SU(6)}\subset SU(8)$
were found
in terms of the matrices $B^{\alpha_{ij}}= E^{\alpha_{ij}}-E^{-\alpha_{ij}} $
belonging to the real symplectic representation ${ \bf 56}$ of $E_{7(7)}$ ($SpD(56)$).
By the simultaneous diagonalization of ${\cal H}_{U(1)}$ and
the Casimir operator of ${\bf SU(2)}$.
it was then possible to  decompose
the  $SpD(56)$ with respect to ${\bf U(1)\times SU(2)\times SU(6)}$ ( i.e.
${ \bf 56}\rightarrow [(1,1,1)\oplus (1,1,15)\oplus (1,2,6)]\oplus
\bar{[...]}$). The eigenvector basis of this
decomposition provided the unitary symplectic representation ${ \bf UspY(56)}$
in which the first diagonal block has the standard form for the Young basis:
\begin{eqnarray}
T^{AB}_{CD}&=&\frac{1}{2} \delta^{[A}_{[C}q^{\phantom{D}B]}_{D]}\\
q^{\phantom{D}B}_{D}& \in &{\bf SU(8)} \quad A,...,D=1,...,8
\label{Yourepre}
\end{eqnarray}
the $8 \times 8$ matrix $q^{\phantom{D}B}_{D}$ being the fundamental
octet representation of the corresponding $SU(8)$
generator.
\par
Such a procedure amounts to the determination of the matrix ${\bf S}$
introduced in eq.\eqn{bfSmat}.
The explicit form of ${\bf S}$ is given below:
\begin{eqnarray}
& {\bf S}= & \nonumber
\end{eqnarray}
{\tiny
\begin{eqnarray}
&\frac{1}{\sqrt{2}}\left(\matrix{ 0 & 0 & 0 & 0 & 0 & 0 & 0 & 0 & 0 & 0 & 0 & 0 & 0 & 0 & 0 & 0 &
  {i\over {{\sqrt{2}}}} & {1\over {{\sqrt{2}}}} & 0 & 0 & 0 & 0 &
  {1\over {{\sqrt{2}}}} & {{-i}\over {{\sqrt{2}}}} & 0 & 0 & 0 & 0 \cr 0 & 0
   & 0 & 0 & 0 & 0 & 0 & 0 & 0 & 0 & 0 & 0 & 0 & 0 & 0 & 0 & -i & 0 & 0 & 0 &
  0 & 0 & 0 & -i & 0 & 0 & 0 & 0 \cr 0 & 0 & 0 & 0 & 0 & 0 & 0 & 0 & 0 & 0 & 0
   & 0 & 0 & 0 & 0 & 0 & -{1\over {{\sqrt{3}}}} & 0 & 0 & 0 & 0 & 0 &
  {{-2\,i}\over {{\sqrt{3}}}} & {1\over {{\sqrt{3}}}} & 0 & 0 & 0 & 0 \cr 0 &
  0 & 0 & 0 & 0 & 0 & 0 & 0 & 0 & 0 & 0 & 0 & 0 & 0 & 0 & 0 &
  -{1\over {{\sqrt{6}}}} & -i\,{\sqrt{{3\over 2}}} & 0 & 0 & 0 & 0 &
  {i\over {{\sqrt{6}}}} & {1\over {{\sqrt{6}}}} & 0 & 0 & 0 & 0 \cr 1 & 0 & 0
   & 0 & 0 & 0 & -1 & 0 & 0 & 0 & 0 & 0 & 0 & 0 & 0 & 0 & 0 & 0 & 0 & 0 & 0 &
  0 & 0 & 0 & 0 & 0 & 0 & 0 \cr 0 & 1 & i & 0 & 0 & 0 & 0 & 0 & 0 & 0 & 0 & 0
   & 0 & 0 & 0 & 0 & 0 & 0 & 0 & 0 & 0 & 0 & 0 & 0 & 0 & 0 & 0 & 0 \cr 0 & 0
   & 0 & 1 & 0 & 0 & 0 & 0 & 0 & 0 & 0 & 0 & 1 & 0 & 0 & 0 & 0 & 0 & 0 & 0 & 0
   & 0 & 0 & 0 & 0 & 0 & 0 & 0 \cr 0 & 0 & 0 & 0 & 1 & 0 & 0 & 0 & 0 & 0 & 0
   & -1 & 0 & 0 & 0 & 0 & 0 & 0 & 0 & 0 & 0 & 0 & 0 & 0 & 0 & 0 & 0 & 0 \cr 0
   & 0 & 0 & 0 & 0 & 1 & 0 & 0 & 0 & 0 & 0 & 0 & 0 & 0 & 0 & 1 & 0 & 0 & 0 & 0
   & 0 & 0 & 0 & 0 & 0 & 0 & 0 & 0 \cr 0 & 0 & 0 & 0 & 0 & 0 & 0 & 1 & 0 & 0
   & 0 & 0 & 0 & 0 & i & 0 & 0 & 0 & 0 & 0 & 0 & 0 & 0 & 0 & 0 & 0 & 0 & 0
   \cr 0 & 0 & 0 & 0 & 0 & 0 & 0 & 0 & 1 & 0 & 0 & 0 & 0 & i & 0 & 0 & 0 & 0
   & 0 & 0 & 0 & 0 & 0 & 0 & 0 & 0 & 0 & 0 \cr 0 & 0 & 0 & 0 & 0 & 0 & 0 & 0
   & 0 & 1 & i & 0 & 0 & 0 & 0 & 0 & 0 & 0 & 0 & 0 & 0 & 0 & 0 & 0 & 0 & 0 & 0
   & 0 \cr 0 & 0 & 0 & 0 & 0 & 0 & 0 & 0 & 0 & 0 & 0 & 0 & 0 & 0 & 0 & 0 & 0
   & 0 & 1 & 0 & 0 & 0 & 0 & 0 & i & 0 & 0 & 0 \cr 0 & 0 & 0 & 0 & 0 & 0 & 0
   & 0 & 0 & 0 & 0 & 0 & 0 & 0 & 0 & 0 & 0 & 0 & 0 & 1 & 0 & 0 & 0 & 0 & 0 & i
   & 0 & 0 \cr 0 & 0 & 0 & 0 & 0 & 0 & 0 & 0 & 0 & 0 & 0 & 0 & 0 & 0 & 0 & 0
   & 0 & 0 & 0 & 0 & 1 & 0 & 0 & 0 & 0 & 0 & i & 0 \cr 0 & 0 & 0 & 0 & 0 & 0
   & 0 & 0 & 0 & 0 & 0 & 0 & 0 & 0 & 0 & 0 & 0 & 0 & 0 & 0 & 0 & 1 & 0 & 0 & 0
   & 0 & 0 & i \cr 1 & 0 & 0 & 0 & 0 & 0 & 1 & 0 & 0 & 0 & 0 & 0 & 0 & 0 & 0
   & 0 & 0 & 0 & 0 & 0 & 0 & 0 & 0 & 0 & 0 & 0 & 0 & 0 \cr 0 & 1 & -i & 0 & 0
   & 0 & 0 & 0 & 0 & 0 & 0 & 0 & 0 & 0 & 0 & 0 & 0 & 0 & 0 & 0 & 0 & 0 & 0 & 0
   & 0 & 0 & 0 & 0 \cr 0 & 0 & 0 & 1 & 0 & 0 & 0 & 0 & 0 & 0 & 0 & 0 & -1 & 0
   & 0 & 0 & 0 & 0 & 0 & 0 & 0 & 0 & 0 & 0 & 0 & 0 & 0 & 0 \cr 0 & 0 & 0 & 0
   & 1 & 0 & 0 & 0 & 0 & 0 & 0 & 1 & 0 & 0 & 0 & 0 & 0 & 0 & 0 & 0 & 0 & 0 & 0
   & 0 & 0 & 0 & 0 & 0 \cr 0 & 0 & 0 & 0 & 0 & 1 & 0 & 0 & 0 & 0 & 0 & 0 & 0
   & 0 & 0 & -1 & 0 & 0 & 0 & 0 & 0 & 0 & 0 & 0 & 0 & 0 & 0 & 0 \cr 0 & 0 & 0
   & 0 & 0 & 0 & 0 & 1 & 0 & 0 & 0 & 0 & 0 & 0 & -i & 0 & 0 & 0 & 0 & 0 & 0 &
  0 & 0 & 0 & 0 & 0 & 0 & 0 \cr 0 & 0 & 0 & 0 & 0 & 0 & 0 & 0 & 1 & 0 & 0 & 0
   & 0 & -i & 0 & 0 & 0 & 0 & 0 & 0 & 0 & 0 & 0 & 0 & 0 & 0 & 0 & 0 \cr 0 & 0
   & 0 & 0 & 0 & 0 & 0 & 0 & 0 & 1 & -i & 0 & 0 & 0 & 0 & 0 & 0 & 0 & 0 & 0 &
  0 & 0 & 0 & 0 & 0 & 0 & 0 & 0 \cr 0 & 0 & 0 & 0 & 0 & 0 & 0 & 0 & 0 & 0 & 0
   & 0 & 0 & 0 & 0 & 0 & 0 & 0 & 1 & 0 & 0 & 0 & 0 & 0 & -i & 0 & 0 & 0 \cr 0
   & 0 & 0 & 0 & 0 & 0 & 0 & 0 & 0 & 0 & 0 & 0 & 0 & 0 & 0 & 0 & 0 & 0 & 0 & 1
   & 0 & 0 & 0 & 0 & 0 & -i & 0 & 0 \cr 0 & 0 & 0 & 0 & 0 & 0 & 0 & 0 & 0 & 0
   & 0 & 0 & 0 & 0 & 0 & 0 & 0 & 0 & 0 & 0 & 1 & 0 & 0 & 0 & 0 & 0 & -i & 0
   \cr 0 & 0 & 0 & 0 & 0 & 0 & 0 & 0 & 0 & 0 & 0 & 0 & 0 & 0 & 0 & 0 & 0 & 0
   & 0 & 0 & 0 & 1 & 0 & 0 & 0 & 0 & 0 & -i \cr  }\right ) &
\end{eqnarray}
}
\subsection{Weights of the compact subalgebra $SU(8)$ }
Having gained control over the embedding of the subgroup ${\bf U(1)\times SU(2)
\times SU(6)}$, let us now come back to the fundamental representation
of $\Es$ and consider the further decomposition of its
${\bf 28}$ and ${\bf \bar{28}}$ components, irreducible with respect to
${\bf SU(8)}$, when we reduce this latter to its subgroup  ${\bf U(1)\times SU(2)
\times SU(6)}$.
In the unitary symplectic basis (either $UspD(56)$ or $UspY(56)$)
the general form of an $E_{7(7)}$ Lie algebra matrix is
 \begin{equation}
{\cal S} = \left(\matrix{ T & V \cr V^* & T^*} \right)
\label{genformal}
\end{equation}
where $T$ and $V$ are $28 \times 28$ matrices respectively
antihermitean and symmetric:
\begin{equation}
T= -T^\dagger \quad ; \quad V=V^T
\label{vtpro}
\end{equation}
The subgalgebra $SU(8)$ is represented by matrices where $V=0$.
Hence the subspaces corresponding to the first and second blocks of
$28$ rows are ${\bf 28}$ and ${\bf \bar{28}}$ irreducible
representations, respectively. Under the subgroup
${\bf U(1)\times SU(2) \times SU(6)}$ each blocks decomposes as
follows:
\begin{eqnarray}
\null & \null & \null \nonumber\\
{\bf 28} & \rightarrow & {\bf (1,1,1)\oplus (1,1,15)\oplus (1,2,6)}\nonumber \\
\null & \null & \null \nonumber\\
 {\bf \bar{28}}& \rightarrow & {\bf \bar{(1,1,1)}\oplus \bar{(1,1,15)}\oplus
\bar{(1,2,6)}} \nonumber\\
\null & \null & \null
\label{28decomp}
\end{eqnarray}
This decomposition corresponds to a rearrangement of the
${\vec \Lambda}^{(n)}={\vec W}^{(n)}$ (and therefore $-{\vec \Lambda}^{(n)}=
{\vec W}^{(n+28)}$) in
a new sequence of weights ${\vec \Lambda}^{\prime( n)}$
($-{\vec \Lambda}^{\prime( n)}$), defined in the following way:
\begin{eqnarray}
{\vec \Lambda}^{\prime( n^{\prime})}\, &=&\, {\vec \Lambda}^{(n)}\nonumber \\
n=1,...,28\leftrightarrow n^{\prime}&=&\cases{[7] \, \leftarrow \,  {\bf (1,1,1)}\cr
[5,20,26,16,13,28,1,22,4,19,25,21,6,27,12]  \, \leftarrow \,  {\bf (1,1,15)}\cr
 [9,14,2,17,23,3,24,18,8,15,10,11] \, \leftarrow \,  {\bf (1,2,6)}\cr }
\label{lambdaprime}
\end{eqnarray}
Denoting by ${\vec \Lambda}_i\,(i=1,...,7)$ the simple weights of
 ({\bf SU(8)}), defined by an equation analogous to \ref{simw},
it turns out that:
${\bf 28}=\Gamma[0,0,0,0,0,1,0]$ and ${\bf \bar{28}}=\Gamma[0,1,0,0,0,0,0]$.
Using the labeling \eqn{qi}, the weights ${\vec \Lambda}^{\prime(n)}$ for the {\bf 28}
representation of $SU(8)$ and the weights $-{\vec \Lambda}^{\prime(n)}$ for the
${\bf \bar{28}}$ representation, ordered according to
 the decomposition \ref{28decomp}, have the  form listed in table
 \ref{28weights} and \ref{28bweights}.
\par
In fig.\ref{su2u6} we show the structure of the $SU(8)$ Lie algebra
 elements in the ${\bf UspY(56)}$
 basis for the fundamental representation of $E_{7(7)}$.
\iffigs
\begin{figure}
\caption{$SU(2)\otimes U(6)$  and $SU(8)$  generators}
\label{su2u6}
\epsfxsize = 10cm
\epsffile{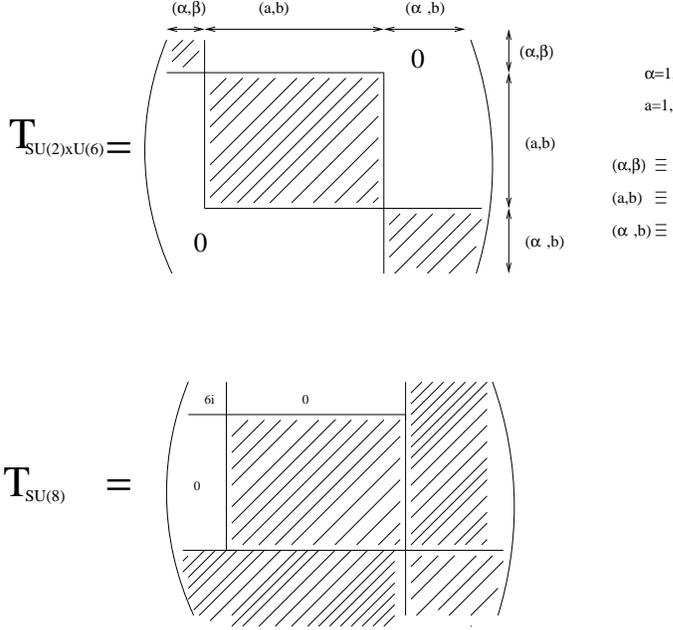}
\vskip -0.1cm
\unitlength=1mm
\end{figure}
\fi
\section{Solvable Lie algebra decompositions}
\label{solvodecompo}
In the present section, the construction of the $ SO^{\star}(12)$
and $ E_{6(4)}$ subalgebras of $E_{7(7)}$ will be discussed in detail.
The starting point of this analysis is eq. (\ref{7in3p4}).
This equality does not  uniquely define the embedding of  $ SO^{\star}(12)$
and $ E_{6(4)}$ into $E_{7(7)}$.
This embedding is  determined
by the requirement that the effective $N=2$ theories corresponding to a truncation of
the $N=8$ scalar manifold to either ${\cal M}_3\sim \exp Solv_3$
or ${\cal M}_4 \sim \exp Solv_4$, be obtained from $D=10 $ type IIA theory through
a compactification on a suitable Calabi--Yau manifold. This condition amounts
to imposing that the fields parametrizing $ Solv_3$ and $Solv_4$ should split
into R--R and N--S  in the following way:
\begin{eqnarray}
Solv_3\, &:& \quad {\bf 30}\rightarrow {\bf 18}\, (N-S) \, +\, {\bf 12} \,
(R-R)\nonumber \\
Solv_4\, &:& \quad {\bf 40}\rightarrow {\bf 20}\, (N-S) \, +\, {\bf 20} \,
(R-R)
\label{rrnssplit}
\end{eqnarray}
The above equations should correspond, according to a procedure defined in
(\cite{noialtri2}) and (\cite{witten}), to the decomposition of $ Solv_3$ and
$Solv_4$ with respect to the solvable algebra of the ST--duality group $O(6,6)
\,\otimes\,  SL(2,\IR)$, which is parametrized
by the whole set of N--S fields in  $D=4$ maximally extended supergravity.
The decomposition is the following one
\footnote{ For notational brevity in this section we use
$Solv \, G \equiv \, Solv \, G/H$, $H$ being the maximal compact subgroup
of $G$}:
\begin{eqnarray}
Solv\left( O(6,6)\,\otimes\, SL(2,\IR)\right)\,&=&\, Solv\left(SU(3,3)_1\right)
\, \oplus\, Solv\left(SU(3,3)_2\right)\, \oplus\,  Solv\left(SL(2,\IR)\right)
\nonumber \\
Solv_3\, &=&\, Solv\left(SU(3,3)_1\right)\, \oplus \, {\cal W}_{12}\nonumber \\
Solv_4\, &=&\,  Solv\left(SL(2,\IR)\right)\, \oplus \, Solv\left(SU(3,3)_2\right)\,
\oplus \, {\cal W}_{20}
\label{nsrrdecom}
\end{eqnarray}
where ${\cal W}_{12}$ and ${\cal W}_{20}$ consist of nilpotent generators
in $ Solv_3$ and
 $Solv_4$ respectively, describing R--R fields in the {\bf 12} and {\bf 20}
irreducible representations of $SU(3,3)$.
\subsection{Structure of $SO^\star(12)$ and $E_{6(4)}$ subalgebras of $E_{7(7)}
$ and some consistent $N=2$ truncations.}
\label{secsu33}
The subalgebras $SO^\star(12)$ and $E_{6(4)}$ in
$E_{7(7)}$ were explicitly constructed starting from their maximal compact subalgebras,
namely $U(6)$ and $ SU(2)\otimes SU(6)\subset \IH$, respectively.
The construction of the algebra $U(1)\otimes SU(2)\otimes SU(6)\subset SU(8)$
was discussed in section $5.2$. As mentioned in earlier sections, by diagonalizing the adjoint
action of $U(1)$ on the $70$--dimensional vector space
$\IK$, (see e.\eqn{chicco}) we could decompose it into
irreducible representations of $U(1)\otimes SU(2)\otimes SU(6)\subset SU(8)$,
namely:
\begin{equation}
\IK\, =\, \IK_{(1,1,15)}\, \oplus\, \IK_{{\bar{(1,1,15)}}}\, \oplus\,
\IK_{(1,2,20)}
\label{ncdec}
\end{equation}
The algebras $SO^\star(12)$ and $E_{6(4)}$ were then constructed as follows:
\begin{eqnarray}
SO^\star(12)\, &=&\, \IK_{(1,1,15)}\, \oplus\, \IK_{{\bar {(1,1,15)}}}\,\oplus
\, U(1)\, \oplus\, SU(6)\nonumber\\
E_{6(4)}\, &=&\,\IK_{(1,2,20)}\, \oplus\, SU(2)\, \oplus\, SU(6)
\label{so12const}
\end{eqnarray}
Unfortunately this construction does not define an embedding $Solv_3\, , \,
Solv_4 \hookrightarrow Solv_7$  fulfilling the requirements (\ref{nsrrdecom}).
However this is not a serious problem. Indeed it suffices to write a
new conjugate solvable Lie algebra $Solv_{7}^{\prime} \, =\,
U^{-1}\, Solv_7 \, U\\ \,\,\, \left( U\in SU(8)/U(1)\otimes SU(2)\otimes SU(6)
\right)$ (recall that $Solv_7$ is not stable with respect to the action of  $SU(8)$)
such that the new embedding $Solv_3\, , \, Solv_4 \hookrightarrow
Solv_{7}^{\prime}
$ fulfills (\ref{nsrrdecom}). We could easily determine
such a matrix $U$. The unitary transformation $U$ depends of course
on the particular embedding $U(1)\otimes SU(2)\otimes SU(6) \hookrightarrow
SU(8)$ chosen to define $SO^\star(12)$ and $E_{6(4)}$. Therefore, in order
to achieve an interpretation of the generators of $Solv_3\, , \, Solv_4$
in terms of $N=8$ fields, the positive roots defining
the two solvable algebras must be viewed as roots of $Solv_7^\prime$ whose
Dynkin diagram consists of the following new simple roots:
\begin{eqnarray}
\tilde{\alpha}_1\, &=&\, {\vec \alpha}_5\quad\tilde{\alpha}_2\, =\,
 {\vec \alpha}_{3,6}\quad \tilde{\alpha}_3\, =\,{\vec \alpha}_{3}\nonumber \\
\tilde{\alpha}_4\, &=&\,{\vec \alpha}_2\quad\tilde{\alpha}_5\, =\,
 {\vec \alpha}_{1}\quad \tilde{\alpha}_6\, =\,{\vec \alpha}_{4,1}\nonumber \\
\tilde{\alpha}_7\, &=&\,-{\vec \alpha}_{6,27}
\label{newdynk}
\end{eqnarray}
Since $Solv_3$ and $Solv_4$  respectively define a special K\"ahler and
a quaternionic manifold, it is useful to describe them  in the Alekseevski's
formalism \cite{alex}.
The algebraic structure of $Solv_3$ and $Solv_4$ can be described in the following way:
\begin{eqnarray}
Solv_3\, :\quad\quad\quad\quad\quad\quad\quad\quad & &\nonumber\\
Solv_3\, &=&\, F_1\, \oplus\, F_2\, \oplus\, F_3\, \oplus\, {\bf X}\, \oplus\,
{\bf Y}\, \oplus\, {\bf Z}\nonumber\\
F_i\, &=&\, \{{\rm h}_i\, ,\,{\rm g}_i\}\quad i=1,2,3\nonumber\\
{\bf X}\, &=&\, {\bf X}^+\, \oplus\, {\bf X}^-\, =\, {\bf X}_{NS}\, \oplus\,
{\bf X}_{RR}\nonumber\\
{\bf Y}\, &=&\, {\bf Y}^+\, \oplus\, {\bf Y}^-\, =\, {\bf Y}_{NS}\, \oplus\,
{\bf Y}_{RR}\nonumber\\
{\bf Z}\, &=&\, {\bf Z}^+\, \oplus\, {\bf Z}^-\, =\, {\bf Z}_{NS}\, \oplus\,
{\bf Z}_{RR}\nonumber\\
Solv\left(SU(3,3)_1\right)\, &=&\, F_1\, \oplus\, F_2\, \oplus\, F_3\, \oplus\,
{\bf X}_{NS}\, \oplus\,
{\bf Y}_{NS}\, \oplus\, {\bf Z}_{NS}\nonumber\\
Solv\left(SL(2,\IR)^3\right)\, &=&\, F_1\, \oplus\, F_2\, \oplus\, F_3\nonumber
\\
{\cal W}_{12}\, &=&\, {\bf X}_{RR}\, \oplus\,{\bf Y}_{RR}\, \oplus\,{\bf Z}_{RR}
\nonumber\\
dim\left( F_i\right)\, &=&\, 2\, ;\quad dim\left( {\bf X}_{NS/RR}\right)\, =\,
dim\left( {\bf X}^{\pm}\right)\, =\,4\nonumber \\
 dim\left( {\bf Y}_{NS/RR}\right)\, &=&\,dim\left( {\bf Y}^{\pm}\right)\, =\,
dim\left( {\bf Z}_{NS/RR}\right)\, =\,dim\left( {\bf Z}^{\pm}\right)\,
=\,4\nonumber\\
\left[{\rm h}_i\, ,\,{\rm g}_i\right]\, &=&\, {\rm g}_i\quad i=1,2,3\nonumber\\
\left[F_i\, ,\,F_j\right]\, &=&\, 0 \quad i\neq j\nonumber\\
\left[{\rm h}_3\, ,\,{\bf Y}^{\pm}\right]\, &=&\,\pm\frac{1}{2}{\bf Y}^{\pm}
\nonumber\\
\left[{\rm h}_3\, ,\,{\bf X}^{\pm}\right]\, &=&\,\pm\frac{1}{2}{\bf X}^{\pm}
\nonumber\\
\left[{\rm h}_2\, ,\,{\bf Z}^{\pm}\right]\, &=&\,\pm\frac{1}{2}{\bf Z}^{\pm}
\nonumber\\
\left[{\rm g}_3\, ,\,{\bf Y}^{+}\right]\, &=&\,\left[{\rm g}_2\, ,\,{\bf Z}^{+}
\right]\, =\,\left[{\rm g}_3\, ,\,{\bf X}^{+}\right]\, =\, 0\nonumber\\
 \left[{\rm g}_3\, ,\,{\bf Y}^{-}\right]\, &=&\,{\bf Y}^+\, ;\,\,
\left[{\rm g}_2\, ,\,{\bf Z}^{-}\right]\, =\,{\bf Z}^+\, ;\,\,
\left[{\rm g}_3\, ,\,{\bf X}^{-}\right]\, =\,{\bf X}^+\nonumber\\
\left[F_1\, ,\,{\bf X}\right]\, &=&\,\left[F_2\, ,\,{\bf Y}\right]\, =\,
\left[F_3\, ,\,{\bf Z}\right]\, =\, 0\nonumber\\
\left[{\bf X}^-\, ,\,{\bf Z}^{-}\right]\, &=&\, {\bf Y}^-
\label{Alekal}
\end{eqnarray}
\begin{eqnarray}
Solv_4\, :\quad\quad\quad\quad\quad\quad\quad\quad\quad\quad\quad\quad\quad
 & &\nonumber\\
Solv_4 & =& F_0\, \oplus  \, F_1^{\prime}\,\oplus \, F_2^{\prime}
\,\oplus \, F_2^{\prime}\,\nonumber\\
 & &\oplus \, {\bf X}_{NS}^{\prime}\,\oplus \,
{\bf Y}_{NS}^{\prime}\,\oplus \, {\bf Z}_{NS}^{\prime}\,\oplus \,
{\cal W}_{20}\nonumber\\
Solv\left(SL(2,\IR)\right)\,\oplus \,Solv\left(SU(3,3)_2\right)\, &=&\,
\left[F_0\right]\,\oplus\, \Biggl [ F_1^{\prime}\,\oplus \, F_2^{\prime}
\,\oplus \, F_2^{\prime} \nonumber \\
 & & \,\oplus \, {\bf X}_{NS}^{\prime}\,\oplus \,\Biggr ]\nonumber\\
F_0\, &=&\, \{{\rm h}_0\, ,\,{\rm g}_0\}\quad \left[{\rm h}_0\, ,\,
{\rm g}_0\right]\, =\, {\rm g}_0\nonumber\\
F_i^\prime\, &=&\, \{{\rm h}_i^\prime\, ,\,{\rm g}_i^\prime\}
\quad i=1,2,3\nonumber\\
\left[F_0\, , \, Solv\left(SU(3,3)_2\right)\right]\,&=& \, 0\, ;\quad
\left[{\rm h}_0\, ,\,{\cal W}_{20}\right]\, =\, \frac{1}{2}{\cal W}_{20}\nonumber\\
\left[{\rm g}_0\, ,\,{\cal W}_{20}\right]\, &=&\, \left[{\rm g}_0\, , \,
 Solv\left(SU(3,3)_2\right)\right]\,=\, 0\nonumber\\
\left[Solv\left(SL(2,\IR)\right)\,\oplus \,Solv\left(SU(3,3)_2\right)\, ,\,
{\cal W}_{20}\right]\, &=&\, {\cal W}_{20}
\label{Alekquat}
\end{eqnarray}
The operators ${\rm h}_i\quad i=1,2,3$ are the Cartan generators of
$SO^\star(12)$ and ${\rm g}_i$ the corresponding axions which together
with
${\rm h}_i$ complete the solvable algebra $Solv\left(SL(2,\IR)^3\right)$
For reasons that will be apparent in the next sections we name:
\begin{equation}
 Solv\left(SL(2,\IR)^3\right) = \mbox{the STU algebra}
\end{equation}
In order to achieve a characterization of all the $Solv\left(SO^\star(12)\right)$
 generators
in terms of fields, the next step is to write down the explicit expression of
the $Solv\left(SO^\star(12)\right)$ generators in terms of  roots of
$Solv_7^\prime$, whose field interpretation can be read directly from Table
\ref{dideals}. We have:
\begin{eqnarray}
{\rm h}_1\, &=& \, \frac{1}{2} H_{{\vec \alpha}_{6,1}}\quad {\rm g}_1\, = \,
E^{{\vec \alpha}_{6,1}}\nonumber\\
{\rm h}_2\, &=& \, \frac{1}{2} H_{{\vec \alpha}_{4,1}}\quad {\rm g}_2\, = \,
E^{{\vec \alpha}_{4,1}}\nonumber\\
{\rm h}_3\, &=& \, \frac{1}{2} H_{{\vec \alpha}_{2,2}}\quad {\rm g}_3\, = \,
E^{{\vec \alpha}_{2,2}}\nonumber\\
{\bf X}_{NS}^+\, &=&\, \left( \begin{array}{c}
E^{{\vec \alpha}_{4,3}}+E^{{\vec \alpha}_{3,4}}\\
E^{{\vec \alpha}_{3,1}}-E^{{\vec \alpha}_{4,6}}\end{array}\right)\,\,\,
{\bf X}_{NS}^-\, =\, \left( \begin{array}{c}
E^{{\vec \alpha}_{4,5}}+E^{{\vec \alpha}_{3,2}}\\
E^{{\vec \alpha}_{3,3}}-E^{{\vec \alpha}_{4,4}}\end{array}\right)\nonumber\\
{\bf X}_{RR}^+\, &=&\, \left( \begin{array}{c}
E^{{\vec \alpha}_{6,21}}+E^{-{\vec \alpha}_{6,17}}\\
E^{{\vec \alpha}_{5,16}}-E^{-{\vec \alpha}_{5,12}}\end{array}\right)\,\,\,
{\bf X}_{RR}^-\, =\, \left( \begin{array}{c}
E^{{\vec \alpha}_{6,20}}+E^{-{\vec \alpha}_{6,16}}\\
E^{{\vec \alpha}_{5,15}}-E^{-{\vec \alpha}_{5,11}}\end{array}\right)\nonumber\\
{\bf Y}_{NS}^+\, &=&\, \left( \begin{array}{c}
E^{{\vec \alpha}_{6,10}}+E^{{\vec \alpha}_{5,5}}\\
E^{{\vec \alpha}_{5,8}}-E^{{\vec \alpha}_{6,7}}\end{array}\right)\,\,\,
{\bf Y}_{NS}^-\, =\, \left( \begin{array}{c}
E^{{\vec \alpha}_{6,8}}+E^{{\vec \alpha}_{5,7}}\\
E^{{\vec \alpha}_{5,6}}-E^{{\vec \alpha}_{6,9}}\end{array}\right)\nonumber\\
{\bf Y}_{RR}^+\, &=&\, \left( \begin{array}{c}
E^{{\vec \alpha}_{6,24}}+E^{-{\vec \alpha}_{3,6}}\\
E^{{\vec \alpha}_{6,26}}-E^{-{\vec \alpha}_{4,9}}\end{array}\right)\,\,\,
{\bf Y}_{RR}^-\, =\, \left( \begin{array}{c}
E^{{\vec \alpha}_{6,23}}+E^{-{\vec \alpha}_{3,5}}\\
E^{{\vec \alpha}_{6,25}}-E^{-{\vec \alpha}_{4,8}}\end{array}\right)\nonumber\\
{\bf Z}_{NS}^+\, &=&\, \left( \begin{array}{c}
E^{{\vec \alpha}_{6,5}}+E^{{\vec \alpha}_{5,1}}\\
E^{{\vec \alpha}_{5,3}}-E^{{\vec \alpha}_{6,3}}\end{array}\right)\,\,\,
{\bf Z}_{NS}^-\, =\, \left( \begin{array}{c}
E^{{\vec \alpha}_{6,4}}-E^{{\vec \alpha}_{5,4}}\\
E^{{\vec \alpha}_{5,2}}+E^{{\vec \alpha}_{6,6}}\end{array}\right)\nonumber\\
{\bf Z}_{RR}^+\, &=&\, \left( \begin{array}{c}
E^{{\vec \alpha}_{6,12}}-E^{-{\vec \alpha}_{2,3}}\\
E^{{\vec \alpha}_{6,27}}+E^{-{\vec \alpha}_{1,1}}\end{array}\right)\,\,\,
{\bf Z}_{RR}^-\, =\, \left( \begin{array}{c}
E^{{\vec \alpha}_{6,22}}+E^{-{\vec \alpha}_{4,10}}\\
E^{{\vec \alpha}_{6,13}}+E^{-{\vec \alpha}_{4,7}}\end{array}\right)
\label{so12struc}
\end{eqnarray}
One can finally check  that the axions associated with the STU--algebra, i.e.
with the generators ${\rm g}_i$ are $B_{5,6}\, ,\, B_{7,8}\, ,\, g_{9,10}$.
Futhermore it is worthwhile noticing that the bidimensional subalgebra
$F_0$ of $Solv_4$ is the solvable algebra of the S--duality group $SL(2,\IR)$
of the $N=8$ theory, and therefore is parametrized by the following fields:
\begin{eqnarray}
\phi \,(dilaton)\,&\leftrightarrow& {\rm h}_0\nonumber\\
B_{\mu \nu} \, \,\,\, &\leftrightarrow &{\rm g}_0
\label{Skey}
\end{eqnarray}
Some consistent  $N=2 $ truncations of the $N=8$ theory can be described
in terms of their scalar content in the following way:
\begin{eqnarray}
{\cal M}_{N=8}&\sim& Solv_7^\prime\,\rightarrow \, {\cal M}_{N=2}\equiv
{\cal M}_{vec}\, \otimes\, {\cal M}_{quat}\nonumber\\
{\cal M}_{vec}\, &\sim&\, Solv_3 \,\,\,\,\,\,\,\,\,\,\,\,\,\,\,\,\,\,\,\,\,\,\,
\,\,\,\,\quad {\cal M}_{quat}\, \sim\, \bfone
\nonumber\\
{\cal M}_{vec}\, &\sim&\, Solv\left(SU(3,3)_1\right) \quad {\cal M}_{quat}\, \sim\,
Solv\left(SU(2,1)\right)\nonumber\\
{\cal M}_{vec}\, &\sim&\, Solv\left(Sl(2,\IR)^3\right) \quad {\cal M}_{quat}\, \sim\,
Solv\left(SO(4,6)\right)\nonumber\\
{\cal M}_{vec}\, &\sim&\, \bfone \,\,\,\,\,\,\,\,\,\,\,\,\,\,\,\,\,\quad\qquad
\,\,\,\,\quad {\cal M}_{quat}\, \sim\, Solv\left(
E_{6(4)}\right)
\end{eqnarray}
\section{The general solution and conclusions}
\label{concludo}
In this paper we have considered two separate but closely related issues:
\begin{enumerate}
\item{The $N=2$ decomposition of the $N=8$ solvable Lie algebra
$Solv_7 \equiv Solv \left(E_{7(7)}/SU(8)\right)$}
\item{The system of first and second order equations characterizing
$BPS$ black--holes in the $N=8$ theory}
\end{enumerate}
With respect to issue (1) our treatment has been exhaustive and
we have shown how the decomposition \eqn{7in3p4},\eqn{3and4defi}
corresponds to the splitting of the $N=8$ scalar fields into vector
multiplet scalars and hypermultiplet scalars. We have also shown how
the alekseevskian analysis of the decomposed solvable Lie algebra
$Solv_7$ is the key to determine the consistent $N=2$ truncations of
the $N=8$ theory at the interaction level. In addition the algebraic results
on the embedding of the $U(1)\times SU(2) \times SU(6)$ Lie algebra
into $E_{7(7)}$ and the solvable counterparts of this embedding are
instrumental for the completion of the programme already outlined in
our previous paper \cite{noialtri2}, namely the gauging of the
maximal gaugeable abelian ideal  ${\cal G}_{abel} \subset Solv_7$ which turns out to be of
dimension $7$. This gauging is postponed to a future publication, but
the algebraic results presented in this paper are an
essential step forward in this direction.
\par
With respect to issue (2) we made a general group--theoretical
analysis of the Killing vector equations and we proved that the
hypermultiplet scalars corresponding to the solvable Lie subalgebra
$Solv_4 \subset Solv_7$ are constant in the most general solution.
Next we analysed a simplified model where the only non--zero fields
are those in the Cartan subalgebra $H \, \subset \, Solv_7$ and we
showed how the algebraically decomposed Killing spinor equations work
in an explicit way. In particular by means of this construction we
retrieved the $N=8$ embedding of the $a$--model black-hole solutions known
in the literature \cite{gensugrabh}. It remains to be seen how general
the presented solutions are, modulo U--duality rotations. That they are not
fully general is evident from the fact that by restricting the non--zero
fields to be in the Cartan subalgebra we obtain constraints on the
electric and magnetic charges such that the solution is parametrized
by only four charges: two electric $(q_{18},q_{23})$ and two magnetic
$p_{17},p_{24}$.  We are therefore lead to consider the question
\par
{ \it How many more scalar fields besides those associated with the
Cartan subalgebra have to be set non zero in order to generate the
most general solution modulo U--duality rotations?}
\par
An answer can be given in terms of solvable Lie algebra once again.
The argument is the following.
\par
Let
\begin{equation}
{\vec Q} \equiv \left( \matrix { g^{\vec {\Lambda}} \cr e_{\vec {\Sigma}}\cr } \right)
\label{chavecto}
\end{equation}
be the vector of electric and magnetic charges (see eq.\eqn{gedefi})
that transforms in the ${\bf 56}$ dimensional real representation
of the U duality group $E_{7(7)}$.
Through the Cayley matrix we can convert it to the ${\bf Usp(56)}$ basis namely to:
\begin{equation}
\left(\matrix { t^{{\vec {\Lambda}}_1}=
g^{{\vec {\Lambda}}_1}+ {\rm i} \, e_{{\vec {\Lambda}}_1}\, \cr
{\bar t}_{{\vec {\Lambda}}_1}=
g^{{\vec {\Lambda}}_1}- {\rm i} \, e_{{\vec {\Lambda}}_1}\,\cr
} \right)
\label{uspqvec}
\end{equation}
Acting on ${\vec Q}$ by means of suitable $Solv \left(E_{7(7)}\right)$
transformations, we can reduce it to the following {\it normal} form:
\begin{equation}
{\vec Q}\rightarrow {\vec Q}^N \equiv \left(\matrix { t^0_{(1,1,1)}\cr t^1_{(1,1,15)}
\cr t^2_{(1,1,15)}\cr t^3_{(1,1,15)}\cr 0\cr
\dots \cr 0 \cr  \cr {\bar t}^0_{{\bar (1,1,1)}}\cr {\bar t}^1_{{\bar (1,1,15)}
} \cr
{\bar t}^2_{{\bar (1,1,15)}}\cr {\bar t}^3_{{\bar (1,1,15)}}\cr 0 \cr
\dots \cr 0 \cr  } \right)
\label{qnormalf}
\end{equation}
Consequently also the central charge ${\vec Z}\equiv \left( Z^{AB}\, , \,
Z_{CD}\right)$, which depends on ${\vec Q}$ through the coset representative
in a symplectic--invariant way, will be brought to the {\it normal} form
\begin{equation}
{\vec Z}\rightarrow {\vec Z}^N \equiv \left(\matrix { z^0_{(1,1,1)}\cr z^1_{(1,1,15)}
\cr z^2_{(1,1,15)}\cr z^3_{(1,1,15)}\cr 0\cr
\dots \cr 0 \cr  \cr {\bar z}^0_{{\bar (1,1,1)}}\cr {\bar z}^1_{{\bar (1,1,15)}
} \cr
{\bar z}^2_{{\bar (1,1,15)}}\cr {\bar z}^3_{{\bar (1,1,15)}}\cr 0 \cr
\dots \cr 0 \cr  } \right)
\label{znormalf}
\end{equation}
through a suitable $SU(8)$ transformation. It was shown in
\cite{lastserg}, \cite{savoy} that ${\vec Q}^N$ is invariant with respect to
the action of an $O(4,4)$ subgroup of $ E_{7(7)}$ and its {\it normalizer}
is an $SL(2,\IR)^3 \subset E_{7(7)}$ commuting with it.
 Indeed it turns out that
the eight real parameters in ${\vec Q}^N$ are singlets with respect to
$O(4,4)$ and in a ${\bf (2,2,2)}$ irreducible representation of
$SL(2,\IR)^3$  as it is shown in the following
decomposition of the ${\bf 56}$ with respect to $O(4,4)\otimes SL(2,\IR)^3$:
\begin{equation}
{\bf 56}\rightarrow {\bf (8_v,2,1,1)\, \oplus \, (8_s,1,2,1)\,\oplus \,
(8_{s^\prime},1,1,2)}\, \oplus\, {\bf (1,2,2,2)}
\label{56normaldec}
\end{equation}
The corresponding subgroup of $SU(8)$ leaving ${\vec Z}^N$ invariant
is therefore $SU(2)^4$ which is the maximal compact subgroup of $O(4,4)$.

Note that $SL(2,\IR)^3$ contains a $U(1)^3$ which is in $SU(8)$ and
which can be further used to classify the general normal frame
black-holes by five real parameters, namely four complex numbers with
the same phase.
This corresponds to write the 56 dimensional generic vector in terms
of the five normal frame parameters plus 51 ``angles'' which
parametrize the 51 dimensional compact space $\frac{SU(8)}{SU(2)^4}$, where $ SU(2)^4 $
is the maximal compact subgroup of the stability group
$O(4,4)$ \cite{cvet}.

Consider now the scalar ``geodesic'' potential (see eq.\eqn{T=SCLV}):
\begin{eqnarray}
V(\phi) & \equiv & {\bar Z}^{AB}(\phi) \, Z_{AB}(\phi) \nonumber\\
&=& {\vec Q}^T \, \left[ \IL^{-1}\left(\phi\right) \right]^T \,
\IL^{-1}\left(\phi\right) \, {\vec Q}
\end{eqnarray}
whose minimization determines the fixed values of the scalar fields
at the horizon of the black--hole.
Because of its invariance properties  the scalar potential $V(\phi)$ depends on
  ${\vec Z}$ and therefore on ${\vec Q}$ only  through their normal forms. Since the
 fixed scalars at the horizon of the Black--Hole are obtained minimizing
$V(\phi)$, it can be inferred that the most general solution of this kind will depend
(modulo duality transformations) only on those scalar fields associated with the
{\it normalizer} of the normal form ${\vec Q}^N$. Indeed the dependence of $V(\phi)$ on
a scalar field is achieved by acting on ${\vec Q}$ in the expression
of $V(\phi)$ by means of the transformations in $Solv_7$ associated with that field.
Since at any point of the scalar manifold $V(\phi)$ can be made to depend only on
${\vec Q}^N$, its minimum will be defined only by those scalars that correspond
to trasformations acting on the non--vanishing components of the normal form
({\it normalizer} of ${\vec Q}^N$). Indeed
 all the other isometries were used to rotate  ${\vec Q}$ to the normal form ${\vec Q}^N$.
Among those scalars which are not determined by the fixed point conditions
there are the {\it flat direction fields} namely
those on which the scalar potential does not  depend  at all:
\begin{equation}
  \mbox{flat direction field } \, q_f \quad \leftrightarrow \quad
  \frac{\partial}{\partial q_f} \, V(\phi) =0
\end{equation}
Some of these fields parametrize $Solv\left(O(4,4)\right)$  since
they are associated with isometries leaving ${\vec Q}^N$ invariant, and the remaining ones
are  obtained from the latter by means of duality transformations.
In order to identify the scalars which are {\it flat} directions of $V(\phi)$, let
us consider the way in which  $Solv\left(O(4,4)\right)$ is embedded into $Solv_7$,
referring to the description of $Solv_4$ given in eqs. (\ref{Alekquat}):
\begin{eqnarray}
Solv\left(O(4,4)\right)&\subset& Solv_4\nonumber\\
Solv\left(O(4,4)\right)\, &=&\,  F_0\,\oplus \, F_1^{\prime}\,\oplus \,
F_2^{\prime}\,\oplus \, F_3^{\prime}\, \oplus \, {\cal W}_{8}
\label{o44}
\end{eqnarray}
where the R--R part ${\cal W}_{8}$ of $Solv\left(O(4,4)\right)$ is the quaternionic image
of $ F_0\,\oplus \, \\F_1^{\prime}\,\oplus \, F_2^{\prime}\,\oplus \, F_3^{\prime}$
in ${\cal W}_{20}$. Therefore $Solv\left(O(4,4)\right)$
is parametrized by the $4$ {\it hypermultiplets} containing the Cartan fields of
$Solv\left(E_{6(4)}\right)$. One finds that the other flat directions are all
the remaining parameters of $Solv_4$, that is all the hyperscalars.
\par
Alternatively we can observe that since the hypermultiplet
scalars are flat directions of the potential, then we can use the solvable Lie algebra
$ Solv_4$    to set them to zero at the horizon.
Since we know from the Killing spinor equations that these
$40$ scalars  are  constants
it follows that we can safely set them to zero and forget
about their existence (modulo U--duality transformations).
Hence the non zero scalars required for a general
solution have to be looked for among the vector multiplet scalars
that is in the solvable Lie algebra $Solv_3$. In other words
the most general $N=8$ black--hole (up to U--duality rotations) is
given by the most general $N=2$ black--hole based on the $15$--dimensional
special K\"ahler manifold:
\begin{equation}
{\cal SK }_{15}  \, \equiv \, \exp \left[ Solv_3 \right] \, =
\frac{SO^\star(12)}{U(1) \times SU(6)}
\label{mgeneral}
\end{equation}
Having determined the little group of the normal form enables us
to decide which among the above $30$ scalars have to be
kept alive in order to generate the most general BPS black--hole
solution (modulo U--duality).
\par
We argue as follows. The {\it normalizer} of the normal form
is contained in the largest subgroup
of $E_{7(7)}$ commuting with $O(4,4)$.
Indeed, a necessary condition for a group $G^N$ to be the {\it normalizer}
of ${\vec Q}^N$ is to commute with the {\it little group} $G^L=O(4,4)$ of
${\vec Q}^N$:
\begin{eqnarray}
{\vec Q}^{\prime N}\, &=&\, G^N\cdot {\vec Q}^N\quad {\vec Q}^N\, =\,
 G^L\cdot {\vec Q}^{N}\nonumber\\
{\vec Q}^{\prime N}\, &=&\, G^L\cdot {\vec Q}^{\prime N}\Rightarrow
\left[G^N\, ,\, G^L\right]\, =\, 0
\label{gngl}
\end{eqnarray}
As previously mentioned, it was proven that $G^N\, =\, SL(2,\IR)^{3} \subset SO^{\star}(12)$ whose solvable
algebra is defined by the last
 of eqs. (\ref{Alekal}). Moreover $G^N$ coincides with the largest subgroup of $Solv_7$
 commuting with $G^L$.\\ The duality transformations associated with
 the $SL(2,\IR)^{3}$ isometries act only on the eight non vanishing components
of ${\vec Q}^N$ and therefore belong to ${\bf Sp(8)}$.\par
{\it In conclusion the most general $N=8$ black--hole solution is  described
by the 6 scalars
parametrizing  $Solv\left(SL(2,\IR)^{3}\right)$, which are the only ones involved in the
fixed point conditions at the horizon.}
\par
 Another way of seeing this is to
notice that all the other $64$ scalars are either the $16$ parameters of
$Solv\left(O(4,4)\right)$
which are flat directions of $V\left(\phi\right)$, or coefficients of the
$48=56-8$ transformations needed to rotate ${\vec Q}$ into ${\vec Q}^N$ that is to set $48$
components of ${\vec Q}$ to zero as shown in eq. (\ref{qnormalf}).
\par
Let us then reduce our attention to
the Cartan vector multiplet sector, namely to the 6 vectors
corresponding to the solvable Lie algebra $Solv \left ( SL(2,\IR)
\right )$.
\subsection{$SL(2,\IR)^3$ and the fixed scalars at the horizon}
\label{fixascala}
In this paper we have elaborated
the group--theoretical rules of this game and in section \ref{cartadila}
we have worked out the simplified
example where the only non--zero fields are in the Cartan subalgebra.
From the  viewpoint of string toroidal compactifications this means that we have
just introduced the dilaton and the $6$ radii $R_i$ of the torus $T^6$.
An item that so far was clearly missing  are the $3$ commuting axions $B_{5,6}$,
$B_{7,8}$ and $g_{9,10}$. As already pointed out, by looking at table \ref{dideals}
we realize
that they correspond to the roots $ \alpha_{6,1},\alpha_{4,1},
\alpha_{2,2}$. So, as it is evident from eq.\eqn{so12struc} the nilpotent
generators associated with these fields are the $g_1, g_2, g_3$
partners of the Cartan generators $h_1,h_2,h_3$ completing the three
2--dimensional {\it key algebras} $F_1, F_2, F_3$ in the
Alekseveeskian decomposition of the K\"ahler algebra
$Sol_3=Solv \left( SO^\star(12) \right)$
(see eq.\eqn{Alekal}):
\begin{equation}
Solv_3=F_1 \oplus F_2 \oplus F_3 \oplus {\bf X} \oplus {\bf Y} \oplus
{\bf Z}
\label{keyalg}
\end{equation}
This triplet of key algebras is nothing else but the Solvable Lie
algebra of $ \left[ SL(2,\IR)/U(1) \right]^3$ defined above as the
normalizer of the little group of the normal form ${\vec Q}^N$:
\begin{equation}
F_1 \oplus F_2 \oplus F_3 = Solv \left( SL(2,\IR)\otimes SL(2,\IR)
\otimes SL(2,\IR) \right)
\label{guarunpo}
\end{equation}
The above considerations have reduced the quest for the most
general $N=8$ black--hole to the solution of the model containing only the
$6$ scalar fields associated with the triplet of key algebras \eqn{guarunpo}.
This model is nothing else but the model of $STU$ N=2 black-holes
studied in \cite{STUkallosh}. Hence we can utilize the results of
that paper and insert them in the general set up we have derived.
In particular we can utilize the determination of the fixed values of the
scalars at the horizon in terms of the charges given in
\cite{STUkallosh}. To make a complete connection between the results
of that paper and our framework we just need to derive the relation
between the fields of the solvable Lie algebra parametrization and
the standard $S,T,U$ complex fields utilized as coordinates of the
special K\"ahler manifold:
\begin{equation}
 {\cal ST}[2,2]\,  \equiv  \, \frac{SU(1,1)}{U(1)} \, \otimes \,
 \frac{SO(2,2)}{SO(2) \times SO(2)}
 \label{st22}
\end{equation}
To this effect we consider the embedding
of the Lie algebra $SL(2,\IR)_1 \times SL(2,\IR)_2 \times SL(2,\IR)_3$
into $Sp(8,\IR)$   such  that the fundamental
${\bf 8}$--dimensional representation of $Sp(8,\IR)$ is irreducible
under the three subgroups and is
\begin{equation}
   {\bf 8}= \left({\bf 2},{\bf 2},{\bf 2} \right)
   \label{222}
\end{equation}
The motivation of this embedding is that the $SO(4,4)$ singlets
in the decomposition \eqn{56normaldec} transform under $SL(2,\IR)^3$
as the representation mentioned in eq.\eqn{222}. Therefore the
requested embedding corresponds to the action of the {\it key
algebras} $ F_1 \oplus F_2 \oplus F_3$ on the non vanishing
components of the charge vector in its normal form.
We obtain the desired result from the standard embedding of
$SL(2,\IR) \, \times \, SO(2,n)$ in $Sp \left( 2\times (2+n) ,\IR \right)$:
\begin{eqnarray}
{\bf A} \, \in \,SO(2,n) & \hookrightarrow &
\left( \matrix { {\bf A} & {\bf 0} \cr {\bf 0} & -{\bf A}^T \cr } \right) \nonumber\\
\left( \matrix { a & b \cr c & d \cr } \right)\,  \in \, SL(2,\IR)
& \hookrightarrow & \left( \matrix { a \, \bfone & b\, \eta  \cr c \, \eta  & d \, \bfone \cr }
\right)
\label{standaimbed}
\end{eqnarray}
used to derive the Calabi Vesentini parametrization of the Special
K\"ahler manifold:
\begin{equation}
{\cal ST}[2,n] \, \equiv \,  \frac{SU(1,1)}{U(1)}\times
\frac{SO(2,n)}{SO(2)\times SO(n)}
\end{equation}
It suffices to set $n=2$ and to use the accidental isomorphism:
\begin{equation}
SO(2,2) \, \sim \, SL(2,\IR) \, \times  \, SL(2,\IR)
\label{accisoph}
\end{equation}
Correspondingly we can write an explicit realization of the
$SL(2,\IR)^3$ Lie algebra:
\begin{equation}
\begin{array}{rcrl}
\left[ L^{(i)}_0 \,  , \, L^{(i)}_\pm \right ] & = &  \pm \,  L^{(i)}_\pm  & i=1,2,3
 \\
\left[ L^{(i)}_+ \,  , \, L^{(i)}_- \right ] & = &  2\,  L^{(i)}_0  & i=1,2,3
\nonumber\\
 \left[ L^{(i)}_A \,  , \, L^{(j)}_B \right ] &=& 0 & i \neq j   \\
 \end{array}
 \label{treLie}
\end{equation}
by means of $8 \times  8$ symplectic matrices satisfying:
\begin{equation}
\left[ L^{(i)}_A \right]^T \,  \IC  + \IC \, L^{(i)}_A = 0
\end{equation}
where
\begin{equation}
  \IC = \left( \matrix{ {\bf 0}_{4 \times 4} & \bfone_{4 \times 4}
  \cr  - \bfone_{4 \times 4} &  {\bf 0}_{4 \times 4} } \right)
\label{symp8}
\end{equation}

Given this structure of the algebra, we can easily construct the
coset representatives by writing:
\begin{eqnarray}
\IL^{(i)}\left(h_i,a_i \right) & \equiv & \exp[ 2 \, h_i \, L_0^{(i)} ] \,
\exp[ a_i \, e^{-h_i} \, L_+^{(i)}] \nonumber \\
 & = & \left( Cosh[h_i] \bfone + Sinh [h_i] \, L_0^{(i)} \right)\, \left(\bfone +
 a_i \, e^{-h_i} \, L_+^{(i)}\right) \nonumber\\
\end{eqnarray}
which follows from the identitities:
\begin{eqnarray}
 L_0^{(i)}\, L_0^{(i)} &=& \frac{1}{4} \, \bfone \\
 L_+^{(i)}\, L_+^{(i)} &=& {\bf 0}\\
\end{eqnarray}
The explicit form of the matrices $L_a^{(i)}$ and $\IL^{(i)}$ is given in appendix A.

We are now ready to construct the central charges and their modulus
square whose minimization with respect to the fields yields the
values of the fixed scalars.
\par
Let us introduce the charge vector:
\begin{equation}
{\vec Q} = \left(\matrix{g^1 \cr g^2 \cr g^3 \cr g^4 \cr e_1 \cr e_2 \cr e_3
\cr e_4 }\right)
\end{equation}
Following our general formulae we can write the central charge vector
as follows:
\begin{equation}
{\vec Z} \, = \, {\cal S}\, \IC \, \prod_{i=1}^{3} \,
\IL^{(i)}(-h_1,-a_i) \, {\vec Q}
\label{centvec}
\end{equation}
where ${\cal S}$ is some unitary matrix and $\IC$ is the symplectic metric.
\par
At this point it is immediate to write down the potential,
whose minimization with respect to the scalar fields
yields the values of the fixed scalars at the horizon.
\par
We have:
\begin{equation}
V({\vec Q}, h, a) \equiv {\vec Z}^\dagger \, {\vec Z} \, = \,
{\vec Q} \, \prod_{i=1}^{3} \, M^{(i)} \left(h_i,a_i\right) \, {\vec
Q}
\label{potential}
\end{equation}
where:
\begin{equation}
 M^{(i)} \left(h_i,a_i \right) \, \equiv \,
 \left[ \IL^{(i)}\left(-h_i,-a_i\right) \right ]^T \, \IL^{(i)}\left(-h_i,-a_i \right)
 \label{mll}
\end{equation}
Rather then working out the derivatives of this potential and
equating them to zero, we can just use the results of paper
\cite{STUkallosh}. It suffices to write the correspondence between our
solvable Lie algebra fields and the $3$ complex scalar fields
$S,T,U$ used in the $N=2$ standard parametrization of the theory.
This correspondence is:
\begin{eqnarray}
T &=&\, a_1 \, + \mbox{\rm i} \, \exp[2 h_1]
\nonumber\\
U & =& \,a_2\,+ \mbox{\rm i} \, \exp[2 h_2]
\nonumber\\
S &=& \,a_3\, + \mbox{\rm i} \, \exp[2 h_3]
\label{pippobaudo}
\end{eqnarray}
and it is established with the following argument. The symplectic
section of special geometry $X^\Lambda$ is defined, in terms of the
$SO(2,2)$ coset representative $L^\Lambda_{\phantom{\Lambda}\Sigma}(\phi)$,
by the formula ( see eq.(C.1) of \cite{n2paperone}):
\begin{equation}
\frac{1}{\sqrt{X^\Lambda \, X^\Sigma}} \, X^\Lambda = \frac{1}{\sqrt{2}}
\, \left( L^\Lambda_{\phantom{\Lambda}1} + \mbox{\rm i}\,L^\Lambda_{\phantom{\Lambda}2}
\right )
\label{C1equa}
\end{equation}
Using for $L^\Lambda_{\phantom{\Lambda}\Sigma}(\phi)$ the upper $4
\times 4$ block of the product $\IL^{(1)}\left(h_1,a_1\right) \,
\IL^{(2)}\left(h_2,a_2\right) $ and using for the symplectic section $X^\Lambda$
that given in eq.(58) of \cite{STUkallosh} we obtain the first two lines
of eq.\eqn{pippobaudo}. The last line of the same equation is
obtained by identifying  the $SU(1,1)$ matrix:
\begin{equation}
{\cal C} \, \left(\matrix{ e^{h_3} & a_3 \cr 0 & e^{-h3} \cr }\right)  \, {\cal C}^{-1}
\end{equation}
where ${\cal C}$ is the $2$--dimensional Cayley matrix with the
matrix $M(S)$ defined in eq.(3.30) of \cite{n2paperone}.
\par
Given this identification of the fields, the fixed values at the horizon
are given by eq.(37) of \cite{STUkallosh}.
\par
We can therefore conclude that we have determined the fixed values of
the scalar fields at the horizon in a general $N=8$ BPS saturated
black--hole.

\section*{Acknowledgements}
We acknowledge useful discussions with Igor Pesando during the first
stage of this work.


\vfill
\eject

\section*{Appendix A}
The explicit expression for the generators of  $SL(2,\IR)^3$ of section 7.1 is:
\begin{eqnarray}
L_{0 }^{(1)}&=&
\left(
\matrix{ 0 & 0 & 0 & {1\over 2} & 0 & 0 & 0 & 0
   \cr 0 & 0 & -{1\over 2} & 0 & 0 & 0 & 0 & 0
   \cr 0 & -{1\over 2} & 0 & 0 & 0 & 0 & 0 & 0
   \cr {1\over 2} & 0 & 0 & 0 & 0 & 0 & 0 & 0
   \cr 0 & 0 & 0 & 0 & 0 & 0 & 0 & -{1\over 2}
   \cr 0 & 0 & 0 & 0 & 0 & 0 & {1\over 2} & 0
   \cr 0 & 0 & 0 & 0 & 0 & {1\over 2} & 0 & 0
   \cr 0 & 0 & 0 & 0 & -{1\over 2} & 0 & 0 & 0
   \cr  } \right) \nonumber\\ \null & \null & \null \nonumber\\
L_{+ }^{(1)}&=& \left(
\matrix{ 0 & -{1\over 2} & -{1\over 2} & 0 & 0 &
  0 & 0 & 0 \cr {1\over 2} & 0 & 0 & -{1\over 2}
   & 0 & 0 & 0 & 0 \cr -{1\over 2} & 0 & 0 &
  {1\over 2} & 0 & 0 & 0 & 0 \cr 0 & -{1\over 2}
   & -{1\over 2} & 0 & 0 & 0 & 0 & 0 \cr 0 & 0 &
  0 & 0 & 0 & -{1\over 2} & {1\over 2} & 0 \cr 0
   & 0 & 0 & 0 & {1\over 2} & 0 & 0 & {1\over 2}
   \cr 0 & 0 & 0 & 0 & {1\over 2} & 0 & 0 &
  {1\over 2} \cr 0 & 0 & 0 & 0 & 0 & {1\over 2}
   & -{1\over 2} & 0 \cr  } \right)\nonumber\\ \null & \null & \null \nonumber\\
L_{- }^{(1)}&=& \left(
\matrix{ 0 & {1\over 2} & -{1\over 2} & 0 & 0 & 0
   & 0 & 0 \cr -{1\over 2} & 0 & 0 & -{1\over 2}
   & 0 & 0 & 0 & 0 \cr -{1\over 2} & 0 & 0 &
  -{1\over 2} & 0 & 0 & 0 & 0 \cr 0 & -{1\over 2}
   & {1\over 2} & 0 & 0 & 0 & 0 & 0 \cr 0 & 0 & 0
   & 0 & 0 & {1\over 2} & {1\over 2} & 0 \cr 0 &
  0 & 0 & 0 & -{1\over 2} & 0 & 0 & {1\over 2}
   \cr 0 & 0 & 0 & 0 & {1\over 2} & 0 & 0 &
  -{1\over 2} \cr 0 & 0 & 0 & 0 & 0 & {1\over 2}
   & {1\over 2} & 0 \cr  }\right)\nonumber\\ \null & \null & \null \nonumber\\
L_{0 }^{(2)}&=& \left(
 \matrix{ 0 & 0 & 0 & -{1\over 2} & 0 & 0 & 0 & 0
    \cr 0 & 0 & -{1\over 2} & 0 & 0 & 0 & 0 & 0
    \cr 0 & -{1\over 2} & 0 & 0 & 0 & 0 & 0 & 0
    \cr -{1\over 2} & 0 & 0 & 0 & 0 & 0 & 0 & 0
    \cr 0 & 0 & 0 & 0 & 0 & 0 & 0 & {1\over 2}
\   \cr 0 & 0 & 0 & 0 & 0 & 0 & {1\over 2} & 0
    \cr 0 & 0 & 0 & 0 & 0 & {1\over 2} & 0 & 0
    \cr 0 & 0 & 0 & 0 & {1\over 2} & 0 & 0 & 0
\   \cr  }  \right)\nonumber\\ \null & \null & \null \nonumber\\
L_{+ }^{(2)}&=& \left(
\matrix{ 0 & -{1\over 2} & -{1\over 2} & 0 & 0 &
  0 & 0 & 0 \cr {1\over 2} & 0 & 0 & {1\over 2}
   & 0 & 0 & 0 & 0 \cr -{1\over 2} & 0 & 0 &
  -{1\over 2} & 0 & 0 & 0 & 0 \cr 0 & {1\over 2}
   & {1\over 2} & 0 & 0 & 0 & 0 & 0 \cr 0 & 0 & 0
   & 0 & 0 & -{1\over 2} & {1\over 2} & 0 \cr 0
   & 0 & 0 & 0 & {1\over 2} & 0 & 0 & -{1\over 2}
   \cr 0 & 0 & 0 & 0 & {1\over 2} & 0 & 0 &
  -{1\over 2} \cr 0 & 0 & 0 & 0 & 0 & -{1\over 2}
   & {1\over 2} & 0 \cr  } \right) \nonumber\\ \null & \null & \null \nonumber\\
L_{- }^{(2)}&=& \left(
\matrix{ 0 & {1\over 2} & -{1\over 2} & 0 & 0 & 0
   & 0 & 0 \cr -{1\over 2} & 0 & 0 & {1\over 2}
   & 0 & 0 & 0 & 0 \cr -{1\over 2} & 0 & 0 &
  {1\over 2} & 0 & 0 & 0 & 0 \cr 0 & {1\over 2}
   & -{1\over 2} & 0 & 0 & 0 & 0 & 0 \cr 0 & 0 &
  0 & 0 & 0 & {1\over 2} & {1\over 2} & 0 \cr 0
   & 0 & 0 & 0 & -{1\over 2} & 0 & 0 &
  -{1\over 2} \cr 0 & 0 & 0 & 0 & {1\over 2} & 0
   & 0 & {1\over 2} \cr 0 & 0 & 0 & 0 & 0 &
  -{1\over 2} & -{1\over 2} & 0 \cr  } \right)\nonumber\\ \null & \null & \null \nonumber\\
L_{0 }^{(3)}&=& \left(
\matrix{ {1\over 2} & 0 & 0 & 0 & 0 & 0 & 0 & 0
   \cr 0 & {1\over 2} & 0 & 0 & 0 & 0 & 0 & 0
   \cr 0 & 0 & {1\over 2} & 0 & 0 & 0 & 0 & 0
   \cr 0 & 0 & 0 & {1\over 2} & 0 & 0 & 0 & 0
   \cr 0 & 0 & 0 & 0 & -{1\over 2} & 0 & 0 & 0
   \cr 0 & 0 & 0 & 0 & 0 & -{1\over 2} & 0 & 0
   \cr 0 & 0 & 0 & 0 & 0 & 0 & -{1\over 2} & 0
   \cr 0 & 0 & 0 & 0 & 0 & 0 & 0 & -{1\over 2}
   \cr  } \right)\nonumber\\ \null & \null & \null \nonumber\\
L_{+ }^{(3)}&=& \left(
\matrix{ 0 & 0 & 0 & 0 & 1 & 0 & 0 & 0 \cr 0 & 0
   & 0 & 0 & 0 & 1 & 0 & 0 \cr 0 & 0 & 0 & 0 & 0
   & 0 & -1 & 0 \cr 0 & 0 & 0 & 0 & 0 & 0 & 0 &
  -1 \cr 0 & 0 & 0 & 0 & 0 & 0 & 0 & 0 \cr 0 & 0
   & 0 & 0 & 0 & 0 & 0 & 0 \cr 0 & 0 & 0 & 0 & 0
   & 0 & 0 & 0 \cr 0 & 0 & 0 & 0 & 0 & 0 & 0 & 0
   \cr  } \right)\nonumber\\ \null & \null & \null \nonumber\\
L_{- }^{(3)}&=& \left(
\matrix{ 0 & 0 & 0 & 0 & 0 & 0 & 0 & 0 \cr 0 & 0
   & 0 & 0 & 0 & 0 & 0 & 0 \cr 0 & 0 & 0 & 0 & 0
   & 0 & 0 & 0 \cr 0 & 0 & 0 & 0 & 0 & 0 & 0 & 0
   \cr 1 & 0 & 0 & 0 & 0 & 0 & 0 & 0 \cr 0 & 1 &
  0 & 0 & 0 & 0 & 0 & 0 \cr 0 & 0 & -1 & 0 & 0 &
  0 & 0 & 0 \cr 0 & 0 & 0 & -1 & 0 & 0 & 0 & 0
   \cr  } \right) \nonumber\\ \null & \null & \null \nonumber\\
\end{eqnarray}

Furthermore, the explicit expression for the coset representatives
of  $\frac{SL(2,\IR)^3}{U(1)^3}$ in the same section is:
 \begin{eqnarray}
&\IL^{(1)}\left(h_1,a_1\right)=& \nonumber\\
&\null& \nonumber\\
& \left(
\matrix{ \cosh h_1 & -a_1 & -a_1 & \sinh h_1 &
  0 & 0 & 0 & 0 \cr a_1 & \cosh h_1 & -\sinh h_1 &
  -a_1 & 0 & 0 & 0 & 0 \cr -a_1 & -\sinh h_1 &
  \cosh h_1 & a_1 & 0 & 0 & 0 & 0 \cr \sinh h_1 &
  -a_1 & -a_1 & \cosh h_1 & 0 & 0 & 0 & 0 \cr 0 & 0 & 0 &
  0 & \cosh h_1 & -a_1 & a_1 & -\sinh h_1 \cr 0
   & 0 & 0 & 0 & a_1 & \cosh h_1 & \sinh h_1 &
  a_1 \cr 0 & 0 & 0 & 0 & a_1 & \sinh h_1 &
  \cosh h_1 & a_1 \cr 0 & 0 & 0 & 0 & -\sinh h_1 &
  a_1 & -a_1 & \cosh h_1 \cr  }  \right)&\nonumber\\
  &\null & \nonumber\\
 &\IL^{(2)}\left(h_2,a_2\right)=& \nonumber\\
 & \left(
 \matrix{ \cosh   h_2 & - a_2 & - a_2 & -\sinh  h_2
    & 0 & 0 & 0 & 0 \cr  a_2 & \cosh   h_2 & -\sinh  h_2
    &  a_2 & 0 & 0 & 0 & 0 \cr - a_2 & -\sinh  h_2 &
   \cosh   h_2 & - a_2 & 0 & 0 & 0 & 0 \cr -\sinh  h_2 &
    a_2 &  a_2 & \cosh   h_2 & 0 & 0 & 0 & 0 \cr 0 & 0 & 0 & 0
    & \cosh   h_2 & - a_2 &  a_2 & \sinh  h_2 \cr 0
    & 0 & 0 & 0 &  a_2 & \cosh   h_2 & \sinh  h_2 &
   - a_2 \cr 0 & 0 & 0 & 0 &  a_2 & \sinh  h_2 &
   \cosh   h_2 & - a_2 \cr 0 & 0 & 0 & 0 & \sinh  h_2 &
   - a_2 &  a_2 & \cosh   h_2 \cr  } \right)& \nonumber\\
   & \null & \nonumber \\
  &\IL^{(3)}\left(h_3,a_3\right)=& \nonumber\\
  & \left(
  \matrix{ 1\,{e^{h_3}} & 0 & 0 & 0 & 2\,a_3 & 0 & 0 & 0 \cr 0 &
    1\,{e^{h_3}} & 0 & 0 & 0 &  \,a_3 & 0 & 0 \cr 0 & 0 &
    1\,{e^{h_3}} & 0 & 0 & 0 & - \,a_3 & 0 \cr 0 & 0 & 0 &
    1\,{e^{h_3}} & 0 & 0 & 0 & - \,a_3 \cr 0 & 0 & 0 & 0 &
    {1\over {{e^{h_3}}}} & 0 & 0 & 0 \cr 0 & 0 & 0 & 0 & 0 &
    {1\over {{e^{h_3}}}} & 0 & 0 \cr 0 & 0 & 0 & 0 & 0 & 0 &
    {1\over {{e^{h_3}}}} & 0 \cr 0 & 0 & 0 & 0 & 0 & 0 & 0 &
    {1\over {{e^{h_3}}}} \cr  }\right)& \nonumber\\
\end{eqnarray}


\par
\vfill
\eject


\section*{Appendix B}
In this appendix we give several tables concerning various results
obtained by computer-aided computations about roots and weights  and
their relations to the physical fields of the solvable Lie algebra of
$E_{7(7)}/SU(8)$.

\begin{table}[ht]
\caption{{\bf
The abelian ideals $ \ID^{+}_{r}$ and the roots of $E_{7(7)}$:}}
\label{dideals}
\begin{center}
\begin{tabular}{||lcl|c|lcl||}
\hline
\hline
  Type IIA & Root & Dynkin & \null  &Type IIA & Root & Dynkin \\
  field    & name & labels & \null  &field    & name & labels \\
\hline
\null & \null & \null & $\ID^+_1$ & \null & \null & \null \\
$ A_{10}$ &  ${\vec \alpha}_{1,1}$ & $\{ 0,0,0,0,0,0,1\}$ & \null &
\null & \null & \null \\
\hline
\null & \null & \null & $\ID^+_2$ & \null & \null & \null \\
$B_{9,10}$ & ${\vec \alpha}_{2,1}$ &$\{ 0,0,0,0,0,1,0\}$ & \null &
$g_{9,10}$ & ${\vec \alpha}_{2,2}$& $\{ 0,0,0,0,1,0,0\}$ \\
$A_9$ & ${\vec \alpha}_{2,3}$ & $\{ 0,0,0,0,0,1,1\}$ & \null &
\null & \null & \null \\
\hline
\null & \null & \null & $\ID^+_3$ & \null & \null & \null \\
$B_{8,9}$ & ${\vec \alpha}_{3,1}$ & $\{ 0,0,0,1,1,1,0\}$ & \null &
$g_{8,9}$ & ${\vec \alpha}_{3,2}$ & $\{ 0,0,0,1,0,0,0\}$ \\
$B_{8,10} $ & ${\vec \alpha}_{3,3} $ & $\{ 0,0,0,1,0,1,0\} $ & \null &
$g_{8,10}$ & ${\vec \alpha}_{3,4} $ & $\{ 0,0,0,1,1,0,0\} $    \\
$A_8 $ & ${\vec \alpha}_{3,5} $ & $\{ 0,0,0,1,1,1,1\} $    & \null &
$A_{8,9,10} $ & ${\vec \alpha}_{3,6} $ & $\{ 0,0,0,1,0,1,1\} $ \\
\hline
\null & \null & \null & $\ID^+_4$ & \null & \null & \null \\
$B_{7,8} $ & ${\vec \alpha}_{4,1} $ & $ \{ 0,0,1,2,1,1,0\}  $ & \null &
$g_{7,8} $ & ${\vec \alpha}_{4,2} $ & $ \{ 0,0,1,0,0,0,0\}  $ \\
$B_{7,9} $ & ${\vec \alpha}_{4,3} $ & $ \{ 0,0,1,1,1,1,0\}  $ & \null &
$g_{7,9} $ & ${\vec \alpha}_{4,4} $ & $ \{ 0,0,1,1,0,0,0\}  $ \\
$B_{7,10} $ & ${\vec \alpha}_{4,5} $ & $ \{ 0,0,1,1,0,1,0\}  $ & \null &
$g_{7,10} $ & ${\vec \alpha}_{4,6} $ & $ \{ 0,0,1,1,1,0,0\}  $ \\
$A_{7,9,10}$ & ${\vec \alpha}_{4,7} $ & $ \{ 0,0,1,2,1,1,1\}  $ & \null &
$A_{7,8,10}$ & ${\vec \alpha}_{4,8} $ & $ \{ 0,0,1,1,1,1,1\}  $ \\
 $A_{7,8,9}$ & ${\vec \alpha}_{4,9} $ & $ \{ 0,0,1,1,0,1,1\}  $ & \null &
$A_{7} $ & ${\vec \alpha}_{4,10} $ & $ \{ 0,0,1,2,1,2,1\} $ \\
\hline
\null & \null & \null & $\ID^+_5$ & \null & \null & \null \\
$ B_{6,7}$ & ${\vec \alpha}_{5,1} $ & $ \{ 0,1,2,2,1,1,0\}  $ & \null &
$ g_{6,7}$ & ${\vec \alpha}_{5,2} $ & $ \{ 0,1,0,0,0,0,0\}  $ \\
$ B_{6,8}$ & ${\vec \alpha}_{5,3} $ & $ \{ 0,1,1,2,1,1,0\}  $ & \null &
$ g_{6,8}$ & ${\vec \alpha}_{5,4} $ & $ \{ 0,1,1,0,0,0,0\}  $ \\
$ B_{6,9}$ & ${\vec \alpha}_{5,5} $ & $ \{ 0,1,1,1,1,1,0\}  $ & \null &
$ g_{6,9}$ & ${\vec \alpha}_{5,6} $ & $ \{ 0,1,1,1,0,0,0\}  $ \\
$ B_{6,10}$ & ${\vec \alpha}_{5,7} $ & $ \{ 0,1,1,1,0,1,0\}  $ & \null &
$ g_{6,10}$ & ${\vec \alpha}_{5,8} $ & $ \{ 0,1,1,1,1,0,0\}  $ \\
$ A_{6,8,9}$ & ${\vec \alpha}_{5,9} $ & $ \{ 0,1,2,2,1,1,1\}  $ & \null &
$ A_{6,7,9}$ & ${\vec \alpha}_{5,10} $ & $ \{ 0,1,1,2,1,1,1\}  $ \\
$ A_{6,7,8}$ & ${\vec \alpha}_{5,11} $ & $ \{ 0,1,1,1,1,1,1\}  $ & \null &
$ A_{\mu\nu\rho}$ & ${\vec \alpha}_{5,12} $ & $ \{ 0,1,1,1,0,1,1\}  $ \\
$ A_{6,7,10}$ & ${\vec \alpha}_{5,13} $ & $ \{ 0,1,1,2,1,2,1\}  $ & \null &
$ A_{6,8,10}$ & ${\vec \alpha}_{5,14} $ & $ \{ 0,1,2,2,1,2,1\}  $ \\
$ A_{6,9,10}$ & ${\vec \alpha}_{5,15} $ & $ \{ 0,1,2,3,1,2,1\}  $ & \null &
$ A_6 $ & ${\vec \alpha}_{5,16} $ & $ \{ 0,1,2,3,2,2,1\}$ \\
\hline
\null & \null & \null & $\ID^+_6$ & \null & \null & \null \\
$ B_{5,6}$ & ${\vec \alpha}_{6,1} $ & $ \{ 1,2,2,2,1,1,0\}  $ & \null &
$ g_{5,6}$ & ${\vec \alpha}_{6,2} $ & $ \{ 1,0,0,0,0,0,0\}  $ \\
$ B_{5,7}$ & ${\vec \alpha}_{6,3} $ & $ \{ 1,1,2,2,1,1,0\}  $ & \null &
$ g_{5,7}$ & ${\vec \alpha}_{6,4} $ & $ \{ 1,1,0,0,0,0,0\}  $ \\
$ B_{5,8}$ & ${\vec \alpha}_{6,5} $ & $ \{ 1,1,1,2,1,1,0\}  $ & \null &
$ g_{5,8}$ & ${\vec \alpha}_{6,6} $ & $ \{ 1,1,1,0,0,0,0\}  $ \\
$ B_{5,9}$ & ${\vec \alpha}_{6,7} $ & $ \{ 1,1,1,1,1,1,0\}  $ & \null &
$ g_{5,9}$ & ${\vec \alpha}_{6,8} $ & $ \{ 1,1,1,1,0,0,0\}  $ \\
$ B_{5,10}$ & ${\vec \alpha}_{6,9} $ & $ \{ 1,1,1,1,0,1,0\}  $ & \null &
$ g_{5,10}$ & ${\vec \alpha}_{6,10} $ & $ \{ 1,1,1,1,1,0,0\}  $ \\
$ B_{\mu\nu}$ & ${\vec \alpha}_{6,11} $ & $ \{ 1,2,3,4,2,3,2\}  $ & \null &
$ A_5 $ & ${\vec \alpha}_{6,12} $ & $ \{ 1,2,3,4,2,3,1\}  $ \\
$ A_{\mu\nu 6}$ & ${\vec \alpha}_{6,13} $ & $ \{ 1,2,2,2,1,1,1\}  $ & \null &
$ A_{\mu\nu 7} $ & ${\vec \alpha}_{6,14} $ & $ \{ 1,1,2,2,1,1,1\}  $ \\
$ A_{\mu\nu 8} $ & ${\vec \alpha}_{6,15} $ & $ \{ 1,1,1,2,1,1,1\}  $ & \null &
$ A_{\mu\nu 9} $ & ${\vec \alpha}_{6,16} $ & $ \{ 1,1,1,1,1,1,1\}  $ \\
$ A_{\mu\nu 10} $ & ${\vec \alpha}_{6,17} $ & $ \{ 1,1,1,1,0,1,1\}  $ & \null &
$ A_{5,6,7} $ & $ {\vec \alpha}_{6,18} $ & $ \{ 1,1,1,2,1,2,1\}  $ \\
$ A_{5,6,8} $ & ${\vec \alpha}_{6,19} $ & $ \{ 1,1,2,2,1,2,1\}  $ & \null &
$ A_{5,6,9} $ & ${\vec \alpha}_{6,20} $ & $ \{ 1,1,2,3,1,2,1\}  $ \\
$ A_{5,6,10} $ & ${\vec \alpha}_{6,21} $ & $ \{ 1,1,2,3,2,2,1\}  $ & \null &
$ A_{5,7,8} $ & ${\vec \alpha}_{6,22} $ & $ \{ 1,2,2,2,1,2,1\}  $ \\
$ A_{5,7,9} $ & ${\vec \alpha}_{6,23} $ & $ \{ 1,2,2,3,1,2,1\}  $ & \null &
$ A_{5,7,10} $ & ${\vec \alpha}_{6,24} $ & $ \{ 1,2,2,3,2,2,1\}  $ \\
$ A_{5,8,9} $ & ${\vec \alpha}_{6,25} $ & $ \{ 1,2,3,3,1,2,1\}  $ & \null &
$ A_{5,8,10} $ & ${\vec \alpha}_{6,26} $ & $ \{ 1,2,3,3,2,2,1\}  $ \\
$ A_{5,9,10} $ & ${\vec \alpha}_{6,27} $ & $ \{ 1,2,3,4,2,2,1\}  $ & \null &
\null & \null & \null \\
\hline
\hline
\end{tabular}
\end{center}
\end{table}
\begin{table}[ht]\caption{{\bf
Weights of the ${\bf 56}$ representation of $E_{7(7)} $:}}
\label{e7weight}
\begin{center}
\begin{tabular}{||cl|c|cl||}
\hline
\hline
    Weight & $q^\ell$  & \null & Weight & $q^\ell$ \\
      name & vector  & \null & name & vector \\
\hline
\null & \null & \null & \null & \null \\
$ {\vec W}^{(1)} \, =\, $ &$  \{ 2,3,4,5,3,3,1\} $  & \null &
$ {\vec W}^{(2)} \, =\, $ &$  \{ 2,2,2,2,1,1,1\} $   \\
$ {\vec W}^{(3)} \, =\, $ &$  \{ 1,2,2,2,1,1,1\} $  & \null &
$ {\vec W}^{(4)} \, =\, $ &$  \{ 1,1,2,2,1,1,1\} $    \\
$ {\vec W}^{(5)} \, =\, $ &$  \{ 1,1,1,2,1,1,1\} $  & \null &
$ {\vec W}^{(6)} \, =\, $ &$  \{ 1,1,1,1,1,1,1\} $    \\
$ {\vec W}^{(7)} \, =\, $ &$  \{ 2,3,3,3,1,2,1\} $  & \null &
$ {\vec W}^{(8)} \, =\, $ &$  \{ 2,2,3,3,1,2,1\} $    \\
$ {\vec W}^{(9)} \, =\, $ &$  \{ 2,2,2,3,1,2,1\} $  & \null &
$ {\vec W}^{(10)} \, =\, $ &$  \{ 2,2,2,2,1,2,1\} $   \\
$ {\vec W}^{(11)} \, =\, $ &$  \{ 1,2,2,2,1,2,1\} $  & \null &
$ {\vec W}^{(12)} \, =\, $ &$  \{ 1,1,2,2,1,2,1\} $    \\
$ {\vec W}^{(13)} \, =\, $ &$  \{ 1,1,1,2,1,2,1\} $  & \null &
$ {\vec W}^{(14)} \, =\, $ &$  \{ 1,2,2,3,1,2,1\} $    \\
$ {\vec W}^{(15)} \, =\, $ &$  \{ 1,2,3,3,1,2,1\} $  & \null &
$ {\vec W}^{(16)} \, =\, $ &$  \{ 1,1,2,3,1,2,1\} $    \\
$ {\vec W}^{(17)} \, =\, $ &$  \{ 2,2,2,2,1,1,0\} $  & \null &
$ {\vec W}^{(18)} \, =\, $ &$  \{ 1,2,2,2,1,1,0\} $    \\
$ {\vec W}^{(19)} \, =\, $ &$  \{ 1,1,2,2,1,1,0\} $  & \null &
$ {\vec W}^{(20)} \, =\, $ &$  \{ 1,1,1,2,1,1,0\} $    \\
$ {\vec W}^{(21)} \, =\, $ &$  \{ 1,1,1,1,1,1,0\} $  & \null &
$ {\vec W}^{(22)} \, =\, $ &$  \{ 1,1,1,1,1,0,0\} $    \\
$ {\vec W}^{(23)} \, =\, $ &$  \{ 3,4,5,6,3,4,2\} $  & \null &
$ {\vec W}^{(24)} \, =\, $ &$  \{ 2,4,5,6,3,4,2\} $    \\
$ {\vec W}^{(25)} \, =\, $ &$  \{ 2,3,5,6,3,4,2\} $  & \null &
$ {\vec W}^{(26)} \, =\, $ &$  \{ 2,3,4,6,3,4,2\} $    \\
$ {\vec W}^{(27)} \, =\, $ &$  \{ 2,3,4,5,3,4,2\} $  & \null &
$ {\vec W}^{(28)} \, =\, $ &$  \{ 2,3,4,5,3,3,2\} $    \\
$ {\vec W}^{(29)} \, =\, $ &$  \{ 1,1,1,1,0,1,1\} $  & \null &
$ {\vec W}^{(30)} \, =\, $ &$  \{ 1,2,3,4,2,3,1\} $    \\
$ {\vec W}^{(31)} \, =\, $ &$  \{ 2,2,3,4,2,3,1\} $  & \null &
$ {\vec W}^{(32)} \, =\, $ &$  \{ 2,3,3,4,2,3,1\} $    \\
$ {\vec W}^{(33)} \, =\, $ &$  \{ 2,3,4,4,2,3,1\} $  & \null &
$ {\vec W}^{(34)} \, =\, $ &$  \{ 2,3,4,5,2,3,1\} $    \\
$ {\vec W}^{(35)} \, =\, $ &$  \{ 1,1,2,3,2,2,1\} $  & \null &
$ {\vec W}^{(36)} \, =\, $ &$  \{ 1,2,2,3,2,2,1\} $   \\
$ {\vec W}^{(37)} \, =\, $ &$  \{ 1,2,3,3,2,2,1\} $  & \null &
$ {\vec W}^{(38)} \, =\, $ &$  \{ 1,2,3,4,2,2,1\} $    \\
$ {\vec W}^{(39)} \, =\, $ &$  \{ 2,2,3,4,2,2,1\} $  & \null &
$ {\vec W}^{(40)} \, =\, $ &$  \{ 2,3,3,4,2,2,1\} $   \\
$ {\vec W}^{(41)} \, =\, $ &$  \{ 2,3,4,4,2,2,1\} $  & \null &
$ {\vec W}^{(42)} \, =\, $ &$  \{ 2,2,3,3,2,2,1\} $    \\
$ {\vec W}^{(43)} \, =\, $ &$  \{ 2,2,2,3,2,2,1\} $  & \null &
$ {\vec W}^{(44)} \, =\, $ &$  \{ 2,3,3,3,2,2,1\} $    \\
$ {\vec W}^{(45)} \, =\, $ &$  \{ 1,2,3,4,2,3,2\} $  & \null &
$ {\vec W}^{(46)} \, =\, $ &$  \{ 2,2,3,4,2,3,2\} $    \\
$ {\vec W}^{(47)} \, =\, $ &$  \{ 2,3,3,4,2,3,2\} $  & \null &
$ {\vec W}^{(48)} \, =\, $ &$  \{ 2,3,4,4,2,3,2\} $    \\
$ {\vec W}^{(49)} \, =\, $ &$  \{ 2,3,4,5,2,3,2\} $  & \null &
$ {\vec W}^{(50)} \, =\, $ &$  \{ 2,3,4,5,2,4,2\} $    \\
$ {\vec W}^{(51)} \, =\, $ &$  \{ 0,0,0,0,0,0,0\} $  & \null &
$ {\vec W}^{(52)} \, =\, $ &$  \{ 1,0,0,0,0,0,0\} $    \\
$ {\vec W}^{(53)} \, =\, $ &$  \{ 1,1,0,0,0,0,0\} $  & \null &
$ {\vec W}^{(54)} \, =\, $ &$  \{ 1,1,1,0,0,0,0\} $   \\
$ {\vec W}^{(55)} \, =\, $ &$  \{ 1,1,1,1,0,0,0\} $  & \null &
$ {\vec W}^{(56)} \, =\, $ &$  \{ 1,1,1,1,0,1,0\} $   \\
\hline
\end{tabular}
\end{center}
\end{table}
\begin{table}[ht]\caption{{\bf
Scalar products of weights and Cartan dilatons}:}
\label{scalaprod}
\begin{center}
\begin{tabular}{||rcl|rcl||}
\hline
\hline
\null & \null & \null & \null & \null & \null \\
$ {\vec \Lambda}^{(1)} \, \cdot \,   {\vec h} $ &=& $ {{-h_1  - h_2  - h_3  - h_4  - h_5  +
     h_6 }\over {{\sqrt{6}}}} $ &
$ {\vec \Lambda}^{(2)} \, \cdot \,   {\vec h} $ &=& $ {{-h_1  + h_2  + h_3  + h_4  + h_5  +
     h_6 }\over {{\sqrt{6}}}} $ \\
$ {\vec \Lambda}^{(3)} \, \cdot \,
  {\vec h} $ &=& $ {{h_1  - h_2  + h_3  + h_4  + h_5  + h_6 }\over {{\sqrt{6}}}} $ &
$ {\vec \Lambda}^{(4)} \, \cdot \,
  {\vec h} $ &=& $ {{h_1  + h_2  - h_3  + h_4  + h_5  + h_6 }\over {{\sqrt{6}}}} $ \\
$ {\vec \Lambda}^{(5)} \, \cdot \,
  {\vec h} $ &=& $ {{h_1  + h_2  + h_3  - h_4  + h_5  + h_6 }\over {{\sqrt{6}}}} $ &
$ {\vec \Lambda}^{(6)} \, \cdot \,
  {\vec h} $ &=& $ {{h_1  + h_2  + h_3  + h_4  - h_5  + h_6 }\over {{\sqrt{6}}}}  $\\
$ {\vec \Lambda}^{(7)} \, \cdot \,
  {\vec h} $ &=& $ {{-h_1  - h_2  + h_3  + h_4  + h_5  -
     h_6 }\over {{\sqrt{6}}}} $ &
$ {\vec \Lambda}^{(8)} \, \cdot \,   {\vec h} $ &=& $ {{-h_1  + h_2  - h_3  + h_4  + h_5  -
     h_6 }\over {{\sqrt{6}}}} $ \\
$ {\vec \Lambda}^{(9)} \, \cdot \,   {\vec h} $ &=& $ {{-h_1  + h_2  + h_3  - h_4  + h_5  -
     h_6 }\over {{\sqrt{6}}}} $ &
$ {\vec \Lambda}^{(10)} \, \cdot \,   {\vec h} $ &=& $ {{-h_1  + h_2  + h_3  + h_4  - h_5  -
     h_6 }\over {{\sqrt{6}}}} $ \\
$ {\vec \Lambda}^{(11)} \, \cdot \,   {\vec h} $ &=& $ {{h_1  - h_2  + h_3  + h_4  - h_5  -
     h_6 }\over {{\sqrt{6}}}} $ &
$ {\vec \Lambda}^{(12)} \, \cdot \,   {\vec h} $ &=& $ {{h_1  + h_2  - h_3  + h_4  - h_5  -
     h_6 }\over {{\sqrt{6}}}} $ \\
$ {\vec \Lambda}^{(13)} \, \cdot \,   {\vec h} $ &=& $ {{h_1  + h_2  + h_3  - h_4  - h_5  -
     h_6 }\over {{\sqrt{6}}}} $ &
$ {\vec \Lambda}^{(14)} \, \cdot \,   {\vec h} $ &=& $ {{h_1  - h_2  + h_3  - h_4  + h_5  -
     h_6 }\over {{\sqrt{6}}}} $ \\
$ {\vec \Lambda}^{(15)} \, \cdot \,   {\vec h} $ &=& $ {{h_1  - h_2  - h_3  + h_4  + h_5  -
     h_6 }\over {{\sqrt{6}}}} $ &
$ {\vec \Lambda}^{(16)} \, \cdot \,   {\vec h} $ &=& $ {{h_1  + h_2  - h_3  - h_4  + h_5  -
     h_6 }\over {{\sqrt{6}}}} $ \\
$ {\vec \Lambda}^{(17)} \, \cdot \,   {\vec h} $ &=& $ {{-\left( {\sqrt{2}}\,h_1  \right )  +
     h_7 }\over {{\sqrt{3}}}} $ &
$ {\vec \Lambda}^{(18)} \, \cdot \,   {\vec h} $ &=& $ {{-\left( {\sqrt{2}}\,h_2  \right )  +
     h_7 }\over {{\sqrt{3}}}} $ \\
$ {\vec \Lambda}^{(19)} \, \cdot \,   {\vec h} $ &=& $ {{-\left( {\sqrt{2}}\,h_3  \right )  +
     h_7 }\over {{\sqrt{3}}}} $ &
$ {\vec \Lambda}^{(20)} \, \cdot \,   {\vec h} $ &=& $ {{-\left( {\sqrt{2}}\,h_4  \right )  +
     h_7 }\over {{\sqrt{3}}}} $ \\
$ {\vec \Lambda}^{(21)} \, \cdot \,   {\vec h} $ &=& $ {{-\left( {\sqrt{2}}\,h_5  \right )  +
     h_7 }\over {{\sqrt{3}}}} $ &
$ {\vec \Lambda}^{(22)} \, \cdot \,   {\vec h} $ &=& $ {{{\sqrt{2}}\,h_6  +
h_7 }\over {{\sqrt{3}}}}$
  \\
$ {\vec \Lambda}^{(23)} \, \cdot \,   {\vec h} $ &=& $ -{{{\sqrt{2}}\,h_1  +
h_7 }\over {{\sqrt{3}}}} $
&
$ {\vec \Lambda}^{(24)} \, \cdot \,   {\vec h} $ &=& $ -{{{\sqrt{2}}\,h_2  + h_7
}\over {{\sqrt{3}}}} $
  \\
$ {\vec \Lambda}^{(25)} \, \cdot \,   {\vec h} $ &=& $ -{{{\sqrt{2}}\,h_3  + h_7 }
\over {{\sqrt{3}}}} $
&
$ {\vec \Lambda}^{(26)} \, \cdot \,   {\vec h} $ &=& $ -{{{\sqrt{2}}\,h_4  + h_7 }
\over {{\sqrt{3}}}} $
  \\
$ {\vec \Lambda}^{(27)} \, \cdot \,   {\vec h} $ &=& $ -{{{\sqrt{2}}\,h_5  + h_7 }
\over {{\sqrt{3}}}} $
&
$ {\vec \Lambda}^{(28)} \, \cdot \,   {\vec h} $ &=& $ {{{\sqrt{2}}\,h_6  - h_7 }
\over {{\sqrt{3}}}} $ \\
\null & \null & \null & \null & \null & \null \\
\hline
\hline
\end{tabular}
\end{center}
\end{table}
\begin{table}[ht]\caption{{\bf
The step operators of the $SU(8)$ subalgebra of $E_{7(7)}$:}}
\label{su8rutte}
\begin{center}
\begin{tabular}{||ccl|l||}
\hline
\hline
  $\#$ & Root & Root & SU(8) step operator in\\
       & name & vector & terms of $E_{7(7)}$ step oper. \\
\hline
1 & ${\vec a}_{1}$ & $ \left \{{-1, -1, 1, 1, 0, 0, 0} \right \} $
& $\cases { \null \cr
X^{a_{1}} =
2 \left ( B^{\alpha_{3,3} } + B^{\alpha_{3,4} }
- B^{\alpha_{4,3} } + B^{\alpha_{4,4} } \right )
\cr
Y^{a_{1}} =
2 \left ( B^{\alpha_{3,1} }
- B^{\alpha_{3,2} } + B^{\alpha_{4,5} } + B^{\alpha_{4,6} } \right )
\cr \null \cr}$ \\ \hline
 2& ${\vec a}_{2}$ & $
\left \{{0, 0, -1, -1, 1, 1, 0} \right \}
$
& $\cases { \null \cr
X^{a_{2}} =
2 \left ( B^{\alpha_{5,3} } +
B^{\alpha_{5,4} } - B^{\alpha_{6,3} } + B^{\alpha_{6,4} } \right )
\cr
Y^{a_{2}} =
2 \left ( B^{\alpha_{5,1} } - B^{\alpha_{5,2} }
+ B^{\alpha_{6,5} } + B^{\alpha_{6,6} } \right )
\cr \null \cr}$ \\ \hline
3& ${\vec a}_{3}$ & $
\left \{{1, 1, 1, 1, 0, 0, 0} \right \}
$
& $\cases { \null \cr
X^{a_{3}} =
2 \left ( B^{\alpha_{3,3} }
+ B^{\alpha_{3,4} } + B^{\alpha_{4,3} } - B^{\alpha_{4,4} } \right )
\cr
Y^{a_{3}} =
2 \left ( -B^{\alpha_{3,1} }
+ B^{\alpha_{3,2} } + B^{\alpha_{4,5} } + B^{\alpha_{4,6} } \right )
\cr \null \cr}$ \\ \hline
4& ${\vec a}_{4}$ & $
\left \{{-1, 0, -1, 0, -1, 0, -1} \right \}
$
& $\cases { \null \cr
X^{a_{4}} =
2 \left ( B^{\alpha_{2,3} }
+ B^{\alpha_{4,7} } - B^{\alpha_{6,12} } + B^{\alpha_{6,13} } \right )
\cr
Y^{a_{4}} =
2 \left ( B^{\alpha_{1,1} } + B^{\alpha_{4,10} } +
B^{\alpha_{6,22} } + B^{\alpha_{6,27} } \right )
\cr \null \cr}$ \\ \hline
5& ${\vec a}_{5}$ & $
\left \{{1, -1, 0, 0, 1, -1, 0} \right \}
$
& $\cases { \null \cr
X^{a_{5}} =
2 \left ( B^{\alpha_{5,5} } +
B^{\alpha_{5,6} } - B^{\alpha_{6,9} } + B^{\alpha_{6,10} } \right )
\cr
Y^{a_{5}} =
-2 \left ( B^{\alpha_{5,7} }
- B^{\alpha_{5,8} } + B^{\alpha_{6,7} } + B^{\alpha_{6,8} } \right )
\cr \null \cr}$ \\ \hline
6 & ${\vec a}_{6}$ & $
\left \{{0, 0, 1, -1, -1, 1, 0} \right \}
$
& $\cases { \null \cr
X^{a_{6}} =
2 \left ( -B^{\alpha_{5,1} } -
B^{\alpha_{5,2} } - B^{\alpha_{6,5} } + B^{\alpha_{6,6} } \right )
\cr
Y^{a_{6}} =
-2 \left ( -B^{\alpha_{5,3} }
+ B^{\alpha_{5,4} } + B^{\alpha_{6,3} } + B^{\alpha_{6,4} } \right )
\cr \null \cr}$ \\ \hline
7& ${\vec a}_{7}$ & $
\left \{{-1, 1, 0, 0, 1, -1, 0} \right \}
$
& $\cases { \null \cr
X^{a_{7}} =
2 \left ( B^{\alpha_{5,5} }
+ B^{\alpha_{5,6} } + B^{\alpha_{6,9} } - B^{\alpha_{6,10} } \right )
\cr
Y^{a_{7}} =
-2 \left ( -B^{\alpha_{5,7} }
+ B^{\alpha_{5,8} } + B^{\alpha_{6,7} } + B^{\alpha_{6,8} } \right )
\cr \null \cr}$ \\ \hline
8 & ${\vec a}_{12}$ & $
\left \{{-1, -1, 0, 0, 1, 1, 0} \right \}
$
& $\cases { \null \cr
X^{a_{12}} =
2 \left ( B^{\alpha_{5,7} } +
B^{\alpha_{5,8} } - B^{\alpha_{6,7} } + B^{\alpha_{6,8} } \right )
\cr
Y^{a_{12}} =
2 \left ( B^{\alpha_{5,5} } -
B^{\alpha_{5,6} } + B^{\alpha_{6,9} } + B^{\alpha_{6,10} } \right )
\cr \null \cr}$ \\ \hline
9 & ${\vec a}_{23}$ & $
\left \{{1, 1, 0, 0, 1, 1, 0} \right \}
$
& $\cases { \null \cr
X^{a_{23}} =
2 \left ( B^{\alpha_{5,7} } +
B^{\alpha_{5,8} } + B^{\alpha_{6,7} } - B^{\alpha_{6,8} } \right )
\cr
Y^{a_{23}} =
2 \left ( -B^{\alpha_{5,5} } +
B^{\alpha_{5,6} } + B^{\alpha_{6,9} } + B^{\alpha_{6,10} } \right )
\cr \null \cr}$ \\ \hline
10 & ${\vec a}_{34}$ & $
\left \{{0, 1, 0, 1, -1, 0, -1} \right \}
$
& $\cases { \null \cr
X^{a_{34}} =
2 \left ( B^{\alpha_{3,6} } +
B^{\alpha_{4,8} } - B^{\alpha_{6,24} } - B^{\alpha_{6,25} } \right )
\cr
Y^{a_{34}} =
2 \left ( -B^{\alpha_{3,5} } +
B^{\alpha_{4,9} } - B^{\alpha_{6,23} } + B^{\alpha_{6,26} } \right )
\cr \null \cr}$ \\ \hline
11 & ${\vec a}_{45}$ & $
\left \{{0, -1, -1, 0, 0, -1, -1} \right \}
$
& $\cases { \null \cr
X^{a_{45}} =
2 \left ( B^{\alpha_{5,11} }
+ B^{\alpha_{5,15} } - B^{\alpha_{6,17} } + B^{\alpha_{6,21} } \right )
\cr
Y^{a_{45}} =
-2 \left ( B^{\alpha_{5,12} } -
B^{\alpha_{5,16} } + B^{\alpha_{6,16} } + B^{\alpha_{6,20} } \right )
\cr \null \cr}$ \\ \hline
\hline
\end{tabular}
\end{center}
\end{table}
\par
\begin{table}[ht]\caption{{\bf
The step operators of the $SU(8)$...}{\sl continued $2^{nd}$}:}
\label{su8rutte2}
\begin{center}
\begin{tabular}{||ccl|l||}
\hline
\hline
  $\#$ & Root & Root & SU(8) step operator in\\
       & name & vector & terms of $E_{7(7)}$ step oper. \\
\hline
12 & ${\vec a}_{56}$ & $
\left \{ {1, -1, 1, -1, 0, 0, 0} \right \}
$
& $\cases { \null \cr
X^{a_{56}} =
2 \left ( B^{\alpha_{3,1} } +
B^{\alpha_{3,2} } - B^{\alpha_{4,5} } + B^{\alpha_{4,6} } \right )
\cr
Y^{a_{56}} =
-2 \left ( B^{\alpha_{3,3} } -
B^{\alpha_{3,4} } + B^{\alpha_{4,3} } + B^{\alpha_{4,4} } \right )
\cr \null \cr}$ \\ \hline
13 & ${\vec a}_{67}$ & $
\left \{{-1, 1, 1, -1, 0, 0, 0} \right \}
$
& $\cases { \null \cr
X^{a_{67}} =
2 \left ( B^{\alpha_{3,1} } +
B^{\alpha_{3,2} } + B^{\alpha_{4,5} } - B^{\alpha_{4,6} } \right )
\cr
Y^{a_{67}} =
-2 \left ( -B^{\alpha_{3,3} } + B^{\alpha_{3,4} }
+ B^{\alpha_{4,3} } + B^{\alpha_{4,4} } \right )
\cr \null \cr}$ \\ \hline
14 & ${\vec a}_{123}$ & $
\left \{{0, 0, 1, 1, 1, 1, 0} \right \}
$
& $\cases { \null \cr
X^{a_{123}} =
2 \left ( B^{\alpha_{5,3} } + B^{\alpha_{5,4} }
+ B^{\alpha_{6,3} } - B^{\alpha_{6,4} } \right )
\cr
Y^{a_{123}} =
2 \left ( -B^{\alpha_{5,1} } + B^{\alpha_{5,2} }
+ B^{\alpha_{6,5} } + B^{\alpha_{6,6} } \right )
\cr \null \cr}$ \\ \hline
15 & ${\vec a}_{234}$ & $
\left \{{0, 1, -1, 0, 0, 1, -1} \right \}
$
& $\cases { \null \cr
X^{a_{234}} =
2 \left ( B^{\alpha_{5,12} } + B^{\alpha_{5,16} }
+ B^{\alpha_{6,16} } - B^{\alpha_{6,20} } \right )
\cr
Y^{a_{234}} =
2 \left ( -B^{\alpha_{5,11} } + B^{\alpha_{5,15} }
+ B^{\alpha_{6,17} } + B^{\alpha_{6,21} } \right )
\cr \null \cr}$ \\ \hline
16 & ${\vec a}_{345}$ & $
\left \{{1, 0, 0, 1, 0, -1, -1} \right \}
$
& $\cases { \null \cr
X^{a_{345}} =
2 \left ( B^{\alpha_{5,9} } + B^{\alpha_{5,13} }
+ B^{\alpha_{6,15} } - B^{\alpha_{6,19} } \right )
\cr
Y^{a_{345}} =
-2 \left ( -B^{\alpha_{5,10} } + B^{\alpha_{5,14} }
+ B^{\alpha_{6,14} } + B^{\alpha_{6,18} } \right )
\cr \null \cr}$ \\ \hline
17 & ${\vec a}_{456}$ & $
\left \{{0, -1, 0, -1, -1, 0, -1} \right \}
$
& $\cases { \null \cr
X^{a_{456}} =
2 \left ( B^{\alpha_{3,6} } + B^{\alpha_{4,8} }
+ B^{\alpha_{6,24} } + B^{\alpha_{6,25} } \right )
\cr
Y^{a_{456}} =
2 \left ( B^{\alpha_{3,5} } - B^{\alpha_{4,9} }
- B^{\alpha_{6,23} } + B^{\alpha_{6,26} } \right )
\cr \null \cr}$ \\ \hline
18 & ${\vec a}_{567}$ & $
\left \{{0, 0, 1, -1, 1, -1, 0} \right \}
$
& $\cases { \null \cr
X^{a_{567}} =
2 \left ( B^{\alpha_{5,1} } + B^{\alpha_{5,2} }
- B^{\alpha_{6,5} } + B^{\alpha_{6,6} } \right )
\cr
Y^{a_{567}} =
-2 \left ( B^{\alpha_{5,3} } - B^{\alpha_{5,4} }
+ B^{\alpha_{6,3} } + B^{\alpha_{6,4} } \right )
\cr \null \cr}$ \\ \hline
\hline
\end{tabular}
\end{center}
\end{table}
\begin{table}[ht]\caption{{\bf
The step operators of the $SU(8)$...}{\sl continued $3^{rd}$}:}
\label{su8rutte3}
\begin{center}
\begin{tabular}{||ccl|l||}
\hline
\hline
  $\#$ & Root & Root & SU(8) step operator in\\
       & name & vector & terms of $E_{7(7)}$ step oper. \\
\hline
19 & ${\vec a}_{1234}$ & $
\left \{{-1, 0, 0, 1, 0, 1, -1} \right \}
$
& $\cases { \null \cr
X^{a_{1234}} =
2 \left ( B^{\alpha_{5,10} } +
B^{\alpha_{5,14} } + B^{\alpha_{6,14} } - B^{\alpha_{6,18} } \right )
\cr
Y^{a_{1234}} =
2 \left ( -B^{\alpha_{5,9} } +
B^{\alpha_{5,13} } + B^{\alpha_{6,15} } + B^{\alpha_{6,19} } \right )
\cr \null \cr}$ \\ \hline
20 & ${\vec a}_{2345}$ & $
\left \{{1, 0, -1, 0, 1, 0, -1} \right \}
$
& $\cases { \null \cr
X^{a_{2345}} =
2 \left ( -B^{\alpha_{1,1} } +
B^{\alpha_{4,10} } - B^{\alpha_{6,22} } + B^{\alpha_{6,27} } \right )
\cr
Y^{a_{2345}} =
-2 \left ( B^{\alpha_{2,3} } -
B^{\alpha_{4,7} } + B^{\alpha_{6,12} } + B^{\alpha_{6,13} } \right )
\cr \null \cr}$ \\ \hline
21 & ${\vec a}_{3456}$ & $
\left \{{1, 0, 1, 0, -1, 0, -1} \right \}
$
& $\cases { \null \cr
X^{a_{3456}} =
2 \left ( B^{\alpha_{2,3} } +
B^{\alpha_{4,7} } + B^{\alpha_{6,12} } - B^{\alpha_{6,13} } \right )
\cr
Y^{a_{3456}} =
2 \left ( -B^{\alpha_{1,1} } -
B^{\alpha_{4,10} } + B^{\alpha_{6,22} } + B^{\alpha_{6,27} } \right )
\cr \null \cr}$ \\ \hline
22 & ${\vec a}_{4567}$ & $
\left \{{-1, 0, 0, -1, 0, -1, -1} \right \}
$
& $\cases { \null \cr
X^{a_{4567}} =
2 \left ( B^{\alpha_{5,9} } +
B^{\alpha_{5,13} } - B^{\alpha_{6,15} } + B^{\alpha_{6,19} } \right )
\cr
Y^{a_{4567}} =
-2 \left ( B^{\alpha_{5,10} } -
B^{\alpha_{5,14} } + B^{\alpha_{6,14} } + B^{\alpha_{6,18} } \right )
\cr \null \cr}$ \\ \hline
23 & ${\vec a}_{12345}$ & $
\left \{{0, -1, 0, 1, 1, 0, -1} \right \}
$
& $\cases { \null \cr
X^{a_{12345}} =
2 \left ( B^{\alpha_{3,5} } +
B^{\alpha_{4,9} } + B^{\alpha_{6,23} } + B^{\alpha_{6,26} } \right )
\cr
Y^{a_{12345}} =
-2 \left ( B^{\alpha_{3,6} } -
B^{\alpha_{4,8} } - B^{\alpha_{6,24} } + B^{\alpha_{6,25} } \right )
\cr \null \cr}$ \\ \hline
24 & ${\vec a}_{23456}$ & $
\left \{{1, 0, 0, -1, 0, 1, -1} \right \}
$
& $\cases { \null \cr
X^{a_{23456}} =
2 \left ( B^{\alpha_{5,10} } +
B^{\alpha_{5,14} } - B^{\alpha_{6,14} } + B^{\alpha_{6,18} } \right )
\cr
Y^{a_{23456}} =
2 \left ( B^{\alpha_{5,9} } -
B^{\alpha_{5,13} } + B^{\alpha_{6,15} } + B^{\alpha_{6,19} } \right )
\cr \null \cr}$ \\ \hline
25 & ${\vec a}_{34567}$ & $
\left \{{0, 1, 1, 0, 0, -1, -1} \right \}
$
& $\cases { \null \cr
X^{a_{34567}} =
2 \left ( B^{\alpha_{5,11} } + B^{\alpha_{5,15} }
+ B^{\alpha_{6,17} } - B^{\alpha_{6,21} } \right )
\cr
Y^{a_{34567}} =
-2 \left ( -B^{\alpha_{5,12} } +
B^{\alpha_{5,16} } + B^{\alpha_{6,16} } + B^{\alpha_{6,20} } \right )
\cr \null \cr}$ \\ \hline
26 & ${\vec a}_{123456}$ & $
\left \{{0, -1, 1, 0, 0, 1, -1} \right \}
$
& $\cases { \null \cr
X^{a_{123456}} =
2 \left ( B^{\alpha_{5,12} } +
B^{\alpha_{5,16} } - B^{\alpha_{6,16} } + B^{\alpha_{6,20} } \right )
\cr
Y^{a_{123456}} =
2 \left ( B^{\alpha_{5,11} } -
B^{\alpha_{5,15} } + B^{\alpha_{6,17} } + B^{\alpha_{6,21} } \right )
\cr \null \cr}$ \\ \hline
27 & ${\vec a}_{234567}$ & $
\left \{{0, 1, 0, -1, 1, 0, -1} \right \}
$
& $\cases { \null \cr
X^{a_{234567}} =
2 \left ( -B^{\alpha_{3,5} } -
B^{\alpha_{4,9} } + B^{\alpha_{6,23} } + B^{\alpha_{6,26} } \right )
\cr
Y^{a_{234567}} =
2 \left ( -B^{\alpha_{3,6} } +
B^{\alpha_{4,8} } - B^{\alpha_{6,24} } + B^{\alpha_{6,25} } \right )
\cr \null \cr}$ \\ \hline
28 & ${\vec a}_{1234567}$ & $
\left \{{-1, 0, 1, 0, 1, 0, -1} \right \}
$
& $\cases { \null \cr
X^{a_{1234567}} =
2 \left ( -B^{\alpha_{1,1} } + B^{\alpha_{4,10} } +
B^{\alpha_{6,22} } - B^{\alpha_{6,27} } \right )
\cr
Y^{a_{1234567}} =
-2 \left ( -B^{\alpha_{2,3} } + B^{\alpha_{4,7} } +
B^{\alpha_{6,12} } + B^{\alpha_{6,13} } \right )
\cr \null \cr}$ \\ \hline
\hline
\end{tabular}
\end{center}
\end{table}
\begin{table}[ht]\caption{{\bf
Weights of the $28$ representation of $SU(8)$}:}
\label{28weights}
\begin{center}
\begin{tabular}{||cl|c|cl||}
\hline
\hline
   Weight & Weight & \null & Weight & Weight  \\
   name & vector & \null & name & vector  \\
\hline
\null & \null & {\bf (1,1,1)} & \null & \null \\
$ {\vec \Lambda}^{\prime(1)} \, = \, $ & $\{ 1,2,2,2,2,2,1\} $   & \null &
\null & \null \\
\hline
\null & \null & {\bf (1,1,15)} & \null & \null \\
$ {\vec \Lambda}^{\prime(2)} \, = \, $ & $\{ 0,0,0,1,1,2,1\} $   & \null &
$ {\vec \Lambda}^{\prime(3)} \, = \, $ & $\{ 0,0,0,1,1,1,1\} $    \\
$ {\vec \Lambda}^{\prime(4)} \, = \, $ & $\{ 0,0,0,1,1,1,0\} $   & \null &
$ {\vec \Lambda}^{\prime(5)} \, = \, $ & $\{ 0,0,1,2,2,2,1\} $     \\
$ {\vec \Lambda}^{\prime(6)} \, = \, $ & $\{ 0,0,0,1,2,2,1\} $   & \null &
$ {\vec \Lambda}^{\prime(7)} \, = \, $ & $\{ 0,0,0,0,0,1,0\} $    \\
$ {\vec \Lambda}^{\prime(8)} \, = \, $ & $\{ 0,0,0,0,0,0,0\} $   & \null &
$ {\vec \Lambda}^{\prime(9)} \, = \, $ & $\{ 0,0,0,0,0,1,1\} $     \\
$ {\vec \Lambda}^{\prime(10)} \, = \, $ & $\{ 0,0,1,1,1,2,1\} $   & \null &
$ {\vec \Lambda}^{\prime(11)} \, = \, $ & $\{ 0,0,1,1,1,1,1\} $     \\
$ {\vec \Lambda}^{\prime(12)} \, = \, $ & $\{ 0,0,1,1,1,1,0\} $   & \null &
$ {\vec \Lambda}^{\prime(13)} \, = \, $ & $\{ 0,0,0,0,1,1,1\} $     \\
$ {\vec \Lambda}^{\prime(14)} \, = \, $ & $\{ 0,0,0,0,1,2,1\} $   & \null &
$ {\vec \Lambda}^{\prime(15)} \, = \, $ & $\{ 0,0,0,0,1,1,0\} $     \\
$ {\vec \Lambda}^{\prime(16)} \, = \, $ & $\{ 0,0,1,1,2,2,1\} $   & \null &
\null & \null \\
\hline
\null & \null & {\bf (1,2,6)} & \null & \null \\
$ {\vec \Lambda}^{\prime(17)} \, = \, $ & $\{ 1,1,1,2,2,2,1\} $   & \null &
$ {\vec \Lambda}^{\prime(18)} \, = \, $ & $\{ 0,1,1,2,2,2,1\} $     \\
$ {\vec \Lambda}^{\prime(19)} \, = \, $ & $\{ 1,1,1,1,1,2,1\} $   & \null &
$ {\vec \Lambda}^{\prime(20)} \, = \, $ & $\{ 1,1,1,1,1,1,1\} $     \\
$ {\vec \Lambda}^{\prime(21)} \, = \, $ & $\{ 1,1,1,1,1,1,0\} $   & \null &
$ {\vec \Lambda}^{\prime(22)} \, = \, $ & $\{ 0,1,1,1,1,2,1\} $    \\
$ {\vec \Lambda}^{\prime(23)} \, = \, $ & $\{ 0,1,1,1,1,1,0\} $   & \null &
$ {\vec \Lambda}^{\prime(24)} \, = \, $ & $\{ 0,1,1,1,1,1,1\} $     \\
$ {\vec \Lambda}^{\prime(25)} \, = \, $ & $\{ 1,1,2,2,2,2,1\} $   & \null &
$ {\vec \Lambda}^{\prime(26)} \, = \, $ & $\{ 0,1,2,2,2,2,1\} $    \\
$ {\vec \Lambda}^{\prime(27)} \, = \, $ & $\{ 1,1,1,1,2,2,1\} $   & \null &
$ {\vec \Lambda}^{\prime(28)} \, = \, $ & $\{ 0,1,1,1,2,2,1\} $     \\
\hline
\end{tabular}
\end{center}
\end{table}
\begin{table}[ht]\caption{{\bf
Weights of the ${\bar 28}$ representation of $SU(8)$}:}
\label{28bweights}
\begin{center}
\begin{tabular}{||cl|c|cl||}
\hline
\hline
   Weight & Weight & \null & Weight & Weight  \\
   name & vector & \null & name & vector  \\
\hline
\null & \null & \null & \null & \null \\
\null & \null & ${\bf  {\overline {(1,1,1) }} }$  & \null & \null \\
$ {-\vec \Lambda}^{\prime(1)} \, = \, $ & $\{ 0,0,0,0,0,0,0\} $ & \null &
\null & \null \\
\hline
\null & \null & \null & \null & \null \\
\null & \null &  ${\bf  {\overline {(1,1,15)} } }$ & \null & \null \\
$ {-\vec \Lambda}^{\prime(2)} \, = \, $ & $\{ 1,2,2,1,1,0,0\} $ & \null &
$ {-\vec \Lambda}^{\prime(3)} \, = \, $ & $\{ 1,2,2,1,1,1,0\} $  \\
$ {-\vec \Lambda}^{\prime(4)} \, = \, $ & $\{ 1,2,2,1,1,1,1\} $ & \null &
$ {-\vec \Lambda}^{\prime(5)} \, = \, $ & $\{ 1,2,1,0,0,0,0\} $   \\
$ {-\vec \Lambda}^{\prime(6)} \, = \, $ & $\{ 1,2,2,1,0,0,0\} $ & \null &
$ {-\vec \Lambda}^{\prime(7)} \, = \, $ & $\{ 1,2,2,2,2,1,1\} $   \\
$ {-\vec \Lambda}^{\prime(8)} \, = \, $ & $\{ 1,2,2,2,2,2,1\} $ & \null &
$ {-\vec \Lambda}^{\prime(9)} \, = \, $ & $\{ 1,2,2,2,2,1,0\} $   \\
$ {-\vec \Lambda}^{\prime(10)} \, = \, $ & $\{ 1,2,1,1,1,0,0\} $ & \null &
$ {-\vec \Lambda}^{\prime(11)} \, = \, $ & $\{ 1,2,1,1,1,1,0\} $   \\
$ {-\vec \Lambda}^{\prime(12)} \, = \, $ & $\{ 1,2,1,1,1,1,1\} $ & \null &
$ {-\vec \Lambda}^{\prime(13)} \, = \, $ & $\{ 1,2,2,2,1,1,0\} $  \\
$ {-\vec \Lambda}^{\prime(14)} \, = \, $ & $\{ 1,2,2,2,1,0,0\} $ & \null &
$ {-\vec \Lambda}^{\prime(15)} \, = \, $ & $\{ 1,2,2,2,1,1,1\} $   \\
$ {-\vec \Lambda}^{\prime(16)} \, = \, $ & $\{ 1,2,1,1,0,0,0\} $ & \null &
\null & \null \\
\hline
\null & \null & \null & \null & \null \\
\null & \null & ${\bf  {\overline {(1,2,6)} } }$  & \null & \null \\
$ {-\vec \Lambda}^{\prime(17)} \, = \, $ & $\{ 0,1,1,0,0,0,0\} $ & \null &
$ {-\vec \Lambda}^{\prime(18)} \, = \, $ & $\{ 1,1,1,0,0,0,0\} $   \\
$ {-\vec \Lambda}^{\prime(19)} \, = \, $ & $\{ 0,1,1,1,1,0,0\} $ & \null &
$ {-\vec \Lambda}^{\prime(20)} \, = \, $ & $\{ 0,1,1,1,1,1,0\} $   \\
$ {-\vec \Lambda}^{\prime(21)} \, = \, $ & $\{ 0,1,1,1,1,1,1\} $ & \null &
$ {-\vec \Lambda}^{\prime(22)} \, = \, $ & $\{ 1,1,1,1,1,0,0\} $   \\
$ {-\vec \Lambda}^{\prime(23)} \, = \, $ & $\{ 1,1,1,1,1,1,1\} $ & \null &
$ {-\vec \Lambda}^{\prime(24)} \, = \, $ & $\{ 1,1,1,1,1,1,0\} $   \\
$ {-\vec \Lambda}^{\prime(25)} \, = \, $ & $\{ 0,1,0,0,0,0,0\} $ & \null &
$ {-\vec \Lambda}^{\prime(26)} \, = \, $ & $\{ 1,1,0,0,0,0,0\} $   \\
$ {-\vec \Lambda}^{\prime(27)} \, = \, $ & $\{ 0,1,1,1,0,0,0\} $ & \null &
$ {-\vec \Lambda}^{\prime(28)} \, = \, $ & $\{ 1,1,1,1,0,0,0\} $   \\
\hline
\hline
\end{tabular}
\end{center}
\end{table}

\end{document}